\documentclass{emulateapj}

\begin{document}

\title{A diversity of progenitors and histories for isolated spiral galaxies}

\author{Marie Martig}
\affil{Centre for Astrophysics \& Supercomputing, Swinburne University of Technology, P.O. Box 218, Hawthorn, VIC 3122, Australia. mmartig@astro.swin.edu.au}

\author{Fr\'ed\'eric Bournaud}
\affil{Laboratoire AIM Paris-Saclay, CEA/IRFU/SAp -- CNRS -- Universit\'e Paris Diderot, 91191 Gif-sur-Yvette Cedex, France.}

\author{Darren J. Croton}
\affil{Centre for Astrophysics \& Supercomputing, Swinburne University of Technology, P.O. Box 218, Hawthorn, VIC 3122, Australia.}

\author{Avishai Dekel}
\affil{Racah Institute of Physics, The Hebrew University, Jerusalem 91904, Israel.}

\author{Romain Teyssier}
\affil{Laboratoire AIM Paris-Saclay, CEA/IRFU/SAp -- CNRS -- Universit\'e Paris Diderot, 91191 Gif-sur-Yvette Cedex, France.}

\begin{abstract}

We analyze a suite of 33 cosmological simulations of the evolution of Milky Way-mass galaxies in low-density environments. Our sample spans a broad range of Hubble types at $z=0$, from nearly bulgeless disks to bulge-dominated galaxies. Despite the fact that a large fraction of the bulge is typically in place by $z\sim 1$, we find no significant correlation between the morphology at $z=1$ and at $z=0$. The $z=1$ progenitors of disk galaxies span a range of morphologies, including smooth disks, unstable disks, interacting galaxies and 
bulge-dominated systems. By $z\sim 0.5$, spiral arms and bars are largely in place and the progenitor morphology is correlated with the final morphology. 
We next focus on late-type galaxies with a bulge-to-total ratio B/T$<0.3$ at $z=0$.  These show a correlation between B/T at $z=0$ and the mass ratio of the largest merger at $z<2$, as 
well as with the gas accretion rate at $z>1$.  We find that the galaxies with the lowest B/T tend to have a quiet baryon input history, with no 
major mergers at $z<2$, and with a low and constant gas accretion rate that keeps a stable 
angular-momentum direction.  More violent merger or gas accretion histories lead to galaxies with more prominent bulges.  
Most disk galaxies have a bulge S\'{e}rsic index $n<2$. The galaxies with the highest bulge S\'{e}rsic index tend to have histories 
of intense gas accretion and disk instability rather than active mergers.

\end{abstract}

\section{Introduction}

Spiral galaxy formation is shaped by a number of key dynamical processes, including galaxy-galaxy mergers, gas accretion from cosmic filaments and internal processes driven by disk instabilities. Understanding how these mechanisms combine to produce the range of morphologies observed in the Local Universe is one of the great challenges for galaxy formation theory.

Mergers have been shown to be important for the formation of elliptical galaxies and for the growth of bulges in spirals. In particular, major mergers can transform disks into ellipticals \citep{Toomre1972,Barnes1992,Naab1999,Naab2003,Cox2006a}, especially when they involve two relatively gas-poor galaxies, which can often be the case at low redshift. The end-product of gas-rich mergers is more complicated, with early studies arguing that the merger remnant can still contain a large rotating disk \citep{Springel2005,Robertson2006,Hopkins2009d}, while recent simulations resolving the star-forming regions and capturing supersonic turbulence suggest that disk survival is very unlikely \citep{Bournaud2011}.

While major mergers are violent events strongly impacting a galaxy's morphology, they are relatively rare in a $\Lambda$CDM context. Minor mergers (with mass ratios lower than 4:1) are much more frequent and can also induce bulge formation \citep{Bournaud2004,Bournaud2005a, Cox2008}, either by adding satellite stars to the center of the galaxy \citep{Aguerri2001} or by triggering gas inflows followed by central starbursts \citep{Eliche-Moral2006}. However, the bulge fraction and S\'{e}rsic index of the remnant depend strongly on the detailed properties of the merger.
Mass ratio is one of the key parameters: mergers with mass ratios between 4:1 and 10:1 are the most efficient at heating the galactic disks \citep{Bournaud2005a}. Smaller mergers only slightly perturb the disks, except if they are frequent enough \citep{Bournaud2007}.

In the end, the effect of mergers on disk galaxies is still debated, and even if they temporarily destroy disks, new disks could reform from the infall of fresh gas from cosmic filaments. At low redshift, mergers become less frequent and gas accretion can then play an important part in slowly building stellar disks around  pre-existing bulges. 

Gas accretion however is not always slow and steady, particularly at high redshift where cold flows can rapidly accumulate large amounts of gas in galactic disks and make them violently unstable \citep{Dekel2009}. These gas-rich and unstable disks fragment and are observed as having a clumpy morphology in high-redshift galaxy surveys \citep{Elmegreen2004a,Elmegreen2007}. The clumps migrate toward the center of the galaxy under the effect of dynamical friction (although it is debated if they survive the intense star formation they host and the associated feedback -- \citealp{Krumholz2010, Murray2010}), where they participate in building the bulge, thus providing another channel for bulge formation at high redshift \citep{Noguchi1999,Immeli2004,Bournaud2007a,Genzel2008,Elmegreen2008,Agertz2009,Dekel2009a, Ceverino2010}.

The picture becomes even more complex with the realization that local spiral galaxies host bulges with a whole range of morphologies. These are usually broken into two types, classical bulges and pseudo-bulges. Classical bulges have a high S\'{e}rsic index (meaning their mass distribution is highly concentrated) and a spheroidal shape. A mechanism explaining their formation is the violent relaxation of stars during mergers, either major or frequent enough minor ones \citep{Bournaud2007}. 
Furthermore, \cite{Elmegreen2008}  argue that clump coalescence in gas rich disks is actually similar to a merger in terms of orbit mixing and also produces classical bulges (although see \citealp{Inoue2011} showing it might not be the case).

On the other hand, pseudo-bulges can be either flattened, with disk-like profile and kinematics or have a boxy-peanut shape (see the review by \citealp{Kormendy2004}). 
They are most likely the result of secular processes re-arranging disk material on timescales of a few billion years. Bars are in particular thought to play a central part in pseudo-bulge formation since they trigger gas inflows, resulting in increased star formation in the central regions \citep{Athanassoula2002, Athanassoula2005,Heller2007,Fisher2009}. As bars grow older, vertical instabilities can give them the boxy/peanut shape observed in many edge-on spirals \citep{Combes1981,Combes1990, Raha1991,Bureau1999, Debattista2006}. Finally, minor mergers themselves are also a possible mechanism for triggering gas inflows and producing bulges with a low S\'{e}rsic index \citep{Eliche-Moral2006,Mendez-Abreu2010}.

In the local Universe, pseudo-bulges are mostly found in late-type galaxies, especially in isolated environments \citep{Durbala2008}. The Milky-Way itself hosts a pseudo-bulge, with the contribution of a classical bulge limited to 8\% of the total stellar mass \citep{Shen2010}. Even more stricking is the fact that 80\% of local galaxies (i.e. within the local 11 Mpc sphere)  more massive than 10$^9$ M$_{\odot}$ are either bulgeless or only have a pseudo-bulge (\citealp{Fisher2011}, see also \citealp{Kautsch2006} and \citealp{Kormendy2010}). The widespread existence of galaxies without  a classical bulge is extremely puzzling given the variety of mechanisms expected to produce them, and the frequency of these mechanisms in our standard cosmological model.

The formation of realistic spiral galaxies has indeed  been a major challenge for cosmological simulations in the last 20 years. Early simulations suffered from extreme angular momentum loss during mergers, giving birth to galaxies with overly concentrated mass distributions and massive bulges (e.g., \citealp{Navarro1991,Navarro1994,Abadi2003a}). The increased resolution of recent simulations, as well as the inclusion of additional physics, in particular feedback from supernovae, has considerably improved the situation \citep{Governato2007a,Scannapieco2008,Piontek2011}.
Zoom cosmological simulations can now produce Milky-Way mass  galaxies with a bulge fraction of 0.2--0.3 \citep{Agertz2011, Guedes2011, Brook2011}, closer to that observed in our own Galaxy. In such simulations, the formation of galaxies with a small bulge relies on the use of a low star formation efficiency, a high threshold for star formation and/or gas recycling with galactic fountains. Although no clear consensus has been reached on how to model star formation and feedback, a picture emerges in which more disk-dominated galaxies form if their star formation histories are delayed significantly by the chosen subgrid recipe \citep{Scannapieco2011b}. However, no simulation has ever been able to produce a Milky-Way mass bulgeless galaxy (note that this problem is partially solved for dwarf galaxies, where supernova feedback is much more efficient at removing baryons with a low angular momentum---see \citealp{Governato2010}).

More generally, the intrinsic limitations of zoom simulations make it hard for them to study galaxy formation from a statistical point of view. 
In this type of simulations, a given sub-volume of the box is simulated at high resolution, while the rest of the box is kept at a very coarse resolution. The high-resolution sub-volume is centered on the galaxy that is being studied, and this volume needs to encompass all particles ending-up in the galaxy at $z=0$. This can lead to a very complex, and possibly large, zoom volume, often making  the technique highly inefficient, so that not all galaxies can be simulated this way. It is in particular difficult to simulate galaxies with high accretion and merger rates at all redshifts, although more isolated galaxies or massive galaxies with early assemblies are more easily modeled.

In any case, gathering a larger sample of such simulations is a challenge given today's supercomputers.
Hence, it has been difficult to test the formation of galaxies of different Hubble types (at fixed physical recipes), to link their morphology with their history or to study the connection between high reshift galaxy populations and their $z=0$ counterparts.

In this paper we present  a suite of 33 simulations performed with an alternative zoom re-simulation technique (presented in \citealp{Martig2009}). This technique consists of extracting the merger and gas accretion histories of dark matter haloes in a large-scale cosmological simulation and then re-simulating these histories at much higher resolution. Its main advantage is its low computational cost that makes it possible to gather a large sample of simulated galaxies and to follow their evolution from $z=5$ to $z=0$. 

The simulated sample is made of 33 isolated galaxies with a halo mass between  $2.7 \times 10^{11} $ and $2 \times 10^{12}$ M$_{\odot}$. At $z=0$, we find galaxies with a large range of Hubble types, from bulgeless to bulge-dominated galaxies. Most of the galaxies host pseudo-bulges, with a S\'{e}rsic index lower than 2. 

We study the evolution of their morphology with redshift. We do not find any correlation between the morphology at $z=1$ and at $z=0$,  with a whole range of possibilities for the $z=1$ progenitors of spiral galaxies (interacting galaxies, bulge-dominated systems, pure disks, unstable disks...). By contrast, there is a much better morphological correlation between $z=0.5$ and 0, with spiral arms and bars being mostly in place at $z=0.5$.

Focussing on the formation histories of galaxies with B/T$<0.3$ (typically Sb and later types, and corresponding to 16 galaxies of our sample), we find that the most disk-dominated of these galaxies have both an extremely quiet merger and gas accretion history. By contrast, more violent merger or gas accretion histories give birth to galaxies with more prominent bulges. We find that the galaxies with the highest bulge S\'{e}rsic index at $z=0$ are not those with mergers but those with intense gas accretion at $z=1$ and either early bar formation or other disk instabilities.

In Section 2 we describe the simulation technique and discuss the selection of the sample. In Section 3 we present the techniques we used to perform morphological decompositions, while in Section 4 we discuss the general properties of the sample at $z=0$. Section 5 is devoted to the high redshift progenitors of spiral galaxies, where we present the evolution of their mass, star formation rate, size and morphology with redshift. Finally, Section 6 is focused on our most disk-dominated galaxies (with a bulge fraction lower than 0.3) and characterizes their merger and gas accretion histories.

\medskip

\section{Simulations}
We study the evolution of 33 simulated galaxies from $z=5$ to $z=0$ using the zoom-in technique described in Appendix A of \cite{Martig2009}.
This technique consists of extracting merger and accretion histories (and geometry) for a given halo in a $\Lambda$-CDM cosmological simulation, and then re-simulating these histories at much higher resolution, replacing each halo by a realistic galaxy containing gas, stars and dark matter. 

In this Section, we will briefly recall the main details of the simulation technique, present the characteristics of the model galaxies, and explain how the sample of 33 re-simulated galaxies was chosen.

\subsection{Simulation technique}

The collisionless cosmological simulation we use as a first step in this work has been performed with the Adaptive Mesh Refinement code RAMSES \citep{Teyssier2002}, in a box of comoving length equal to  20~h$^{-1}$~Mpc. It contains 512$^3$ dark matter particles with each a mass of $6.9 \times 10^6$~M$_{\sun}$. The cosmology is set to $\Lambda$-CDM with $\Omega_m$=0.3, $\Omega_{\Lambda}$=0.7, H$_0$=70~km~s$^{-1}$~Mpc$^{-1}$ and $\sigma_8$=0.9.

The dark matter halos are detected in the cosmological simulation using the HOP algorithm \citep{Eisenstein1998}.
Particles that do not belong to a gravitationally bound halo are also  taken into account; we will refer to them as ``diffuse'' mass accretion when they are accreted by the studied halo.

We extract the merger and diffuse accretion histories of a halo by tracking halos and particles crossing a fixed spherical boundary drawn around the target halo. We record the position, velocity and spin of each incoming satellite as well as the date of the interaction. We then perform a new simulation, in which we now replace each halo of the cosmological simulation by a galaxy made up of gas, star and dark matter particles, and each diffuse particle of the cosmological simulation with a blob of lower-mass, higher-resolution gas and dark matter particles (see  a detailed explanation of how this is done in Appendix A of \citealp{Martig2009}). They start interacting with the main galaxy following the orbital and spin parameters given by the cosmological simulation. The re-simulation starts at $z=5$ inside a 800~kpc-large zoom area, and follows the evolution of the main galaxy down to $z=0$. 

The code we use for the re-simulation is the Particle-Mesh code described in \cite{Bournaud2002,Bournaud2003}. The density is computed thanks to a Cloud-In-Cell interpolation and the Poisson equation is then solved using Fast Fourier Transforms. Time integration is made using a leapfrog algorithm with a time step of 1.5~Myr.
The spatial resolution (gravitational softening) is 150~pc and the mass resolution (particle mass) is $1.5\times 10^4$~M$_{\sun}$ for gas, $7.5\times 10^4$~M$_{\sun}$ for stars, and  $3\times 10^5$~M$_{\sun}$ for dark matter. In Appendix A we present a resolution test, with the spatial and mass resolutions higher by a factor of two and six respectively. We find very little difference between the high-resolution and standard runs. 

Gas dynamics is modeled with a sticky-particle scheme (with a similar resolution of  150~pc) and star formation is computed with a Schmidt-Kennicutt law \citep{Kennicutt1998b} with an exponent of 1.5 and an efficiency of 2\%. The threshold for star formation is set at 0.03 M$_{\sun}$pc$^{-3}$ (i.e. one atom per cubic centimeter), which corresponds to the minimal density for diffuse atomic cloud formation \citep{Elmegreen2002}. We include the energy feedback from supernovae explosions, and use a kinetic feedback scheme: 20\% of the energy of the supernovae is distributed to neighbouring gas particles (within a radius of 70 pc) in the form of a radial velocity kick.  We also include the continuous gas return from high-, intermediate- and low-mass stars following the scheme proposed by \cite{Jungwiert2001} and used in \cite{Martig2010}. We indeed found  stellar mass loss to be an important ingredient  for the formation of realistic late-type galaxies in cosmological simulations \citep{Martig2010}.

\subsection{Model galaxies}

Each halo recorded in the dark matter-only cosmological simulation is replaced with a realistic galaxy, made of a disk (gas and stars), a stellar bulge and a dark matter halo. The disks have a Toomre profile, the bulge is a Plummer sphere and the halo follows a Burkert profile \citep{Burkert1995,Salucci2000}.
The core radii of our dark matter halos are chosen according to the scaling relations proposed by \cite{Salucci2000} and the halos are truncated at their virial radius.

The stellar mass of a galaxy is set according to Table 7 of \cite{Moster2010a} as a function of its halo mass and of the redshift at which the galaxy is introduced in the simulation. All model galaxies are assumed to be disk galaxies without a bulge, except for massive galaxies (M$_{\rm halo}>10^{11}\rm M_{\sun}$) at $z<1$, in which case the mass of the bulge is set to 20\% of the stellar mass.

The gas content in the disk also varies according to redshift and total mass: at $z<1$, the gas fraction (with respect to total baryonic mass) is 0.3 for small galaxies ($\rm M_{\rm halo}<10^{11}\rm M_{\sun}$) and 0.15 for massive galaxies. At higher redshifts, the gas fraction is chosen independently of the total mass, and is set to 0.5 for $1<z<3$ and 0.7 for $z>3$.
No hot gas halo is included, which should be a sensible approximation given that the mass of the model galaxies is always lower than the critical mass for virial shocks to be stable (\mbox{$\sim 10^{12} \rm M_{\sun}$}, see \citealp{Birnboim2003} and \citealp{Dekel2006}).

\bigskip

\subsection{Advantages and limitations of the simulation technique}

\begin{figure}
\centering 
\includegraphics[width=0.35\textwidth]{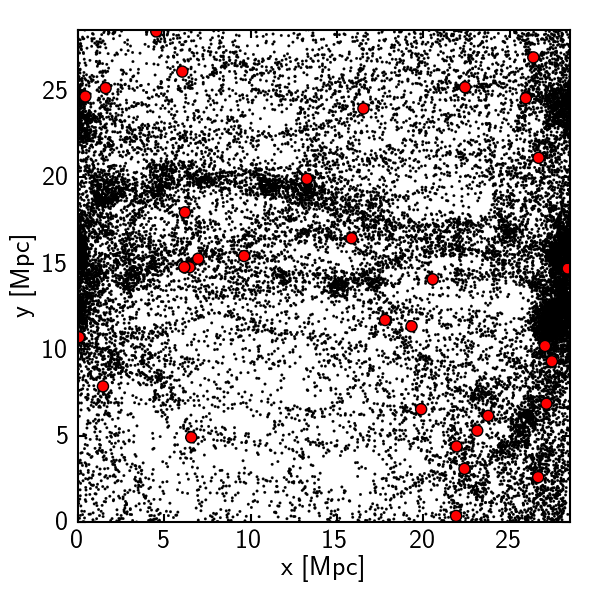}
\includegraphics[width=0.35\textwidth]{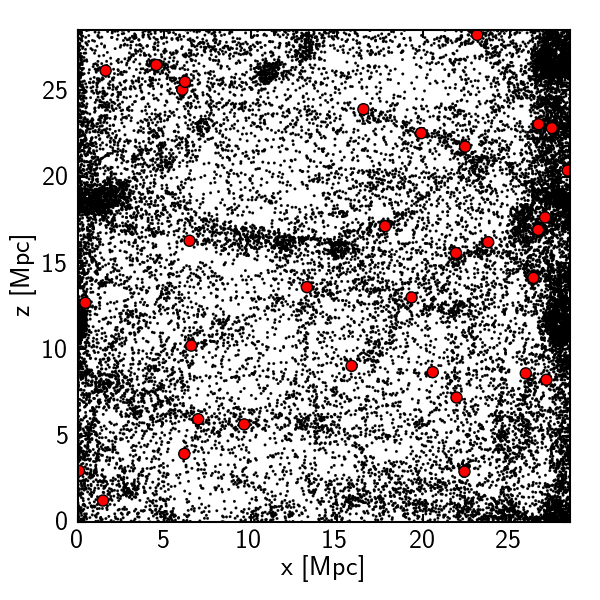}
\caption{Location of the chosen galaxies within the cosmological box. The two panels show two different projections (in the xy and xz planes), each black dot corresponds to a dark matter halo (only halos more massive than $10^9$ M$_{\odot}$ are shown) and the red dots correspond to the 33 galaxies chosen for the re-simulation. The chosen galaxies sample the whole volume but  avoid the densest regions. }
\label{fig:loc_halos}
\end{figure}

The technique we developed was designed to overcome some of the limitations of standard zoom simulations. 
Such a re-simulation technique has already been used by \cite{Kazantzidis2007}, \cite{Read2008} and \cite{Villalobos2008} to study the stability of a stellar disk undergoing a series of collisionless minor mergers. A major difference with our work is that these simulations include no gas component (and consider mergers only above a given mass ratio) so that their scope is limited and they cannot be used to study most mechanisms governing galaxy evolution.

The main advantage of the re-simulation technique is the low computation time \footnote{around 5 days for each simulation presented in this paper} compared to zoom cosmological simulations, so that running a large number of simulations is easier and statistical studies are possible.
Another advantage is that it is possible to simulate galaxies with all types of merger histories, in contrast to zoom simulations that need all the progenitors to be in the sub-volume treated at high resolution. Some merger histories are thus for now impossible to simulate with a standard zoom technique, in particular if they involve a merger at $z \sim 0$ with a massive galaxy, in which case the sub-volume on which one has to zoom becomes very large.
Finally, the fact that we decouple the evolution of our main galaxy from the expansion of the Universe keeps the physical resolution constant as a function of time at no additional cost (all positions are expressed in physical units rather than in comoving units). This is not necessarily the case in cosmological simulations, that often have a constant resolution in comoving coordinates, so that the physical resolution decreases with redshift unless (in the case of AMR simulations) additional levels of refinement are added, with a large computational cost.

 \begin{figure*}
 \centering 
 \includegraphics[width=\textwidth]{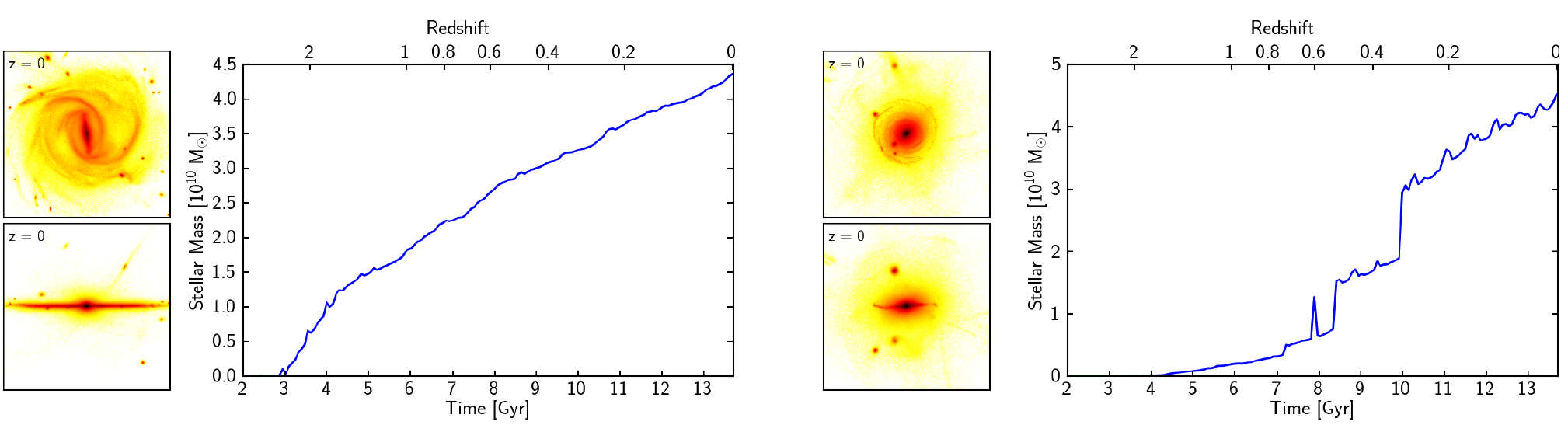}
 \caption{Examples of i-band surface brightness maps (face-on and edge-on, 70x70 kpc, color scale from 16 to 27~mag~arcsec$^{-2}$) for 2 simulated galaxies, and the corresponding stellar mass growth histories (stellar mass computed within the optical radius). The two cases shown here represent two extremes: a disk-dominated galaxy with an extremely quiet history, and a bulge-dominated one with a much more violent history (two major mergers at $z<1$). }
 \label{fig:examples_z0}
 \end{figure*}
 
 \begin{figure}
 \centering 
 \includegraphics[width=0.4\textwidth]{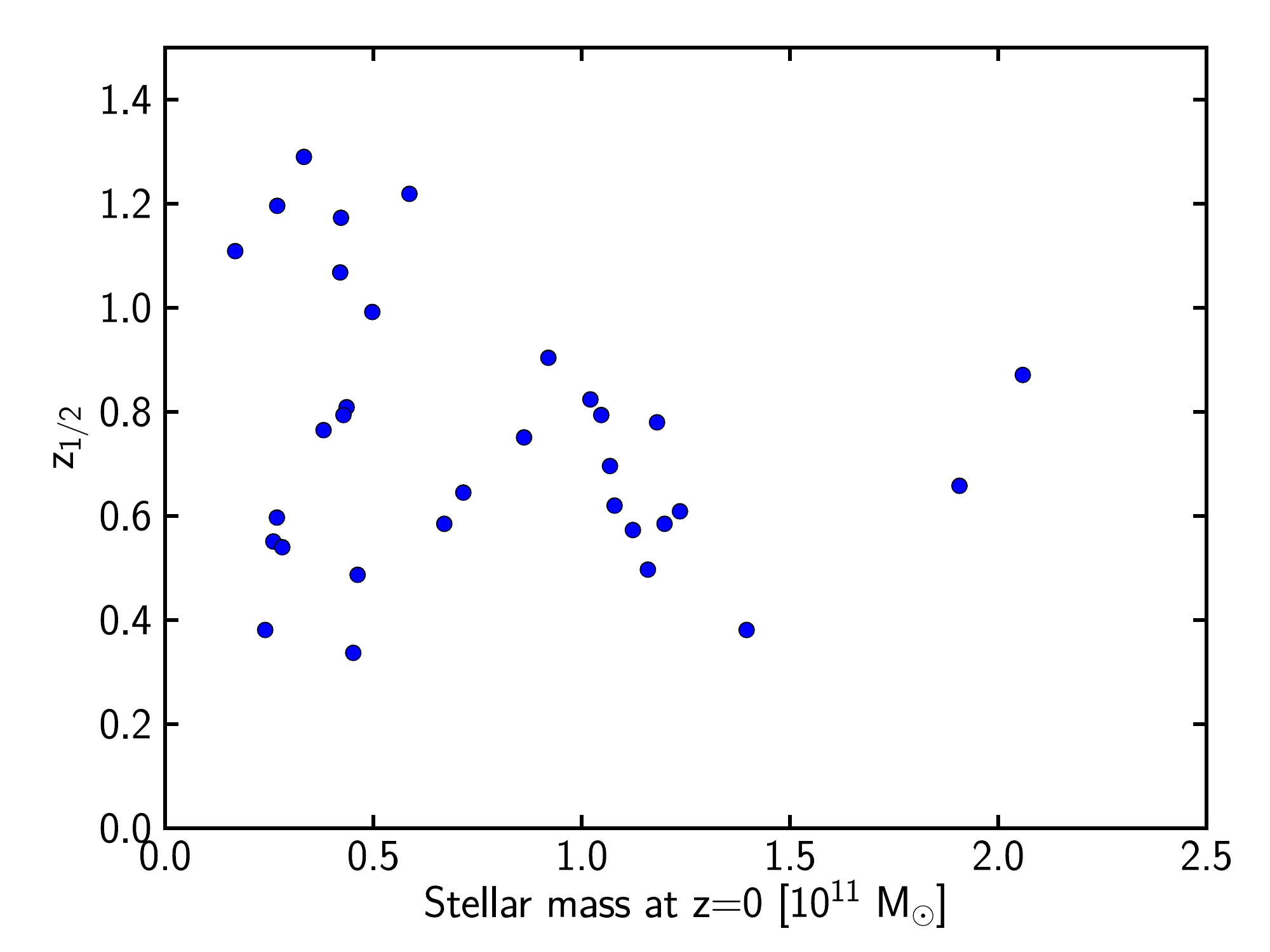}
 \caption{Redshift at which galaxies reach half of  their final stellar mass ( $z_{1/2}$, or formation redshift) as a function of their stellar mass at $z=0$. The formation redshift of most galaxies is between 0.5 and 1, suggesting a late assembly of their stellar content.}
 \label{fig:z_half}
 \end{figure}

Of course there are also some major drawbacks. The most important is probably the large number of free parameters for the model galaxies. Even if we carefully select these parameter to be close to the observed galaxies, we will never reach the level of diversity found both in observations and in fully cosmological simulations. The case of our $z = 5$ galaxies is even more complicated, with very few data both from observations and previous simulations. However, at $z=5$ only a minor fraction of the final baryonic mass of a simulated galaxy is already in place: on average, we find that only 5\% of the $z=2$ baryons are in place at $z=5$. The mergers and instabilities happening between $z=5$ and $z=2$ redistribute these initial baryons, so that the initial assumptions do not make a big difference for our purpose here (see a test of a change of initial conditions in Appendix A of \citealp{Martig2009}). We are thus careful to ``discard'' the early phase of the simulations and only analyse them after $z=2$ to erase the memory of the initial conditions.

Finally, an important limitation is linked with the sticky-particle model, that poorly treats the hot gas phase (see also the discussion in Appendix B of \citealp{Martig2009}). This technique is thus unable to treat the case of massive halos (above the critical mass of $\sim 10^{12}$~M$_{\sun}$  for virial shocks to be stable), for instance massive early-type galaxies, but also groups and clusters. We are limited to galaxies below the threshold for virial shocks, for which we know the cold accretion mode is predominant. In that case, the feeding of cold gas to the galaxy is most easily tied to the properties of the cosmic-web filaments to which the galaxy is connected (e.g., \citealp{Dekel2009}).

\subsection{Sample selection}

We select 33 halos with a mass between  $2.7 \times 10^{11} $ and $2 \times 10^{12}$ M$_{\odot}$ and that are relatively isolated at $z=0$: no halo more massive than half of their mass can be found within 2 Mpc, and they are at least 6 Mpc away from one of the four most massive halos in the simulation box. The chosen halos are distributed across the whole volume of the simulation box but avoid the densest regions (see Figure \ref{fig:loc_halos}).
We computed the local density contrast at $z=0$ in spheres of radius 8 Mpc around each halo, $\delta=\frac{\rho-\bar{\rho}}{\bar{\rho}}$, and find values ranging from -0.29 to -0.01, with all but 5 galaxies in the range [-0.29,-0.22]. This confirms that the simulated galaxies are in underdense environments, with densities intermediate between the Local Group and the center of voids \citep{Romano-Diaz2007}.

No additional criterion is set, in particular there is no constraint on the merger history of the chosen halos. The chosen halos represent 26\% of the total number of halos in the simulation box within this mass range, and 69\% of the halos satisfying both the mass and environment criteria. Note that in the mass range that we are considering the abundance and internal properties of halos are unaffected by the relatively small box size of the cosmological simulation \citep{Power2006}.

Appendix \ref{app:sample} shows for each simulated galaxy several i-band surface brightness maps (computed using PEGASE, with a Kroupa IMF, and without dust extinction) at various redshifts, the i-band surface brightness profile at $z=0$ as well as the stellar mass evolution with time. The mass growth histories are usually very quiet for these galaxies, resulting in a disk+bulge (and possibly bar) morphology for  all of them. Figure \ref{fig:examples_z0} shows examples of i-band images and stellar mass evolution plots for two extreme cases: a disk-dominated galaxy (left panels) with an extremely quiet history, and a bulge-dominated one that has undergone several mergers at $z<1$.

We find that at $z=2$ the simulated galaxies have at most reached 20\% of their final stellar mass, at $z=1$ they have typically between 30 and 60 \% of their final mass, and between 50 and 80\% for $z=0.5$. 

The redshift at which half of the final stellar mass is in place ($z_{1/2}$) is shown in Figure \ref{fig:z_half}.  While $z_{1/2}$ is greater than 1 for  22\% of the simulated galaxies, the majority (69\%) of galaxies reach half of their final stellar mass between $z=1$ and $z=0$.5. These results suggest a late assembly of our sample galaxies. We find one extreme case of a galaxy with only 0.001\% of its mass in place at $z=2$, and 4\% at $z=1$; this is a galaxy undergoing several major mergers after $z=1$ and having a bulge-dominated morphology at $z=0$ (this is the galaxy shown on the right in Figure \ref{fig:examples_z0}).

\section{Bulge/Disk/Bar decomposition}
We use two different methods to characterize the morphology of the simulated galaxies. One of them is based on stellar kinematics. It is usually robust but fails at discriminating bars and bulges. The other technique consists in using GALFIT \citep{Peng2002,Peng2010} to fit mock images of the galaxies with a sum of bulge, bar and disk profiles.  This technique is less robust, especially for galaxies that are difficult to fit with smooth profiles, which is often the case at high redshift and which is the reason why we will only use this technique for the galaxies at $z=0$. On the other hand,  GALFIT has the advantage of giving measures that are easy to compare to observations. It also allows us to discriminate bars and bulges, and to measure their S\'{e}rsic index. A detailed comparison of the relative advantages and limitations of the two techniques can be found in \cite{Scannapieco2010}.
\subsection{Kinematic decomposition}\label{sec:kinematics}

\begin{figure*}
\centering 
\includegraphics[width=\textwidth]{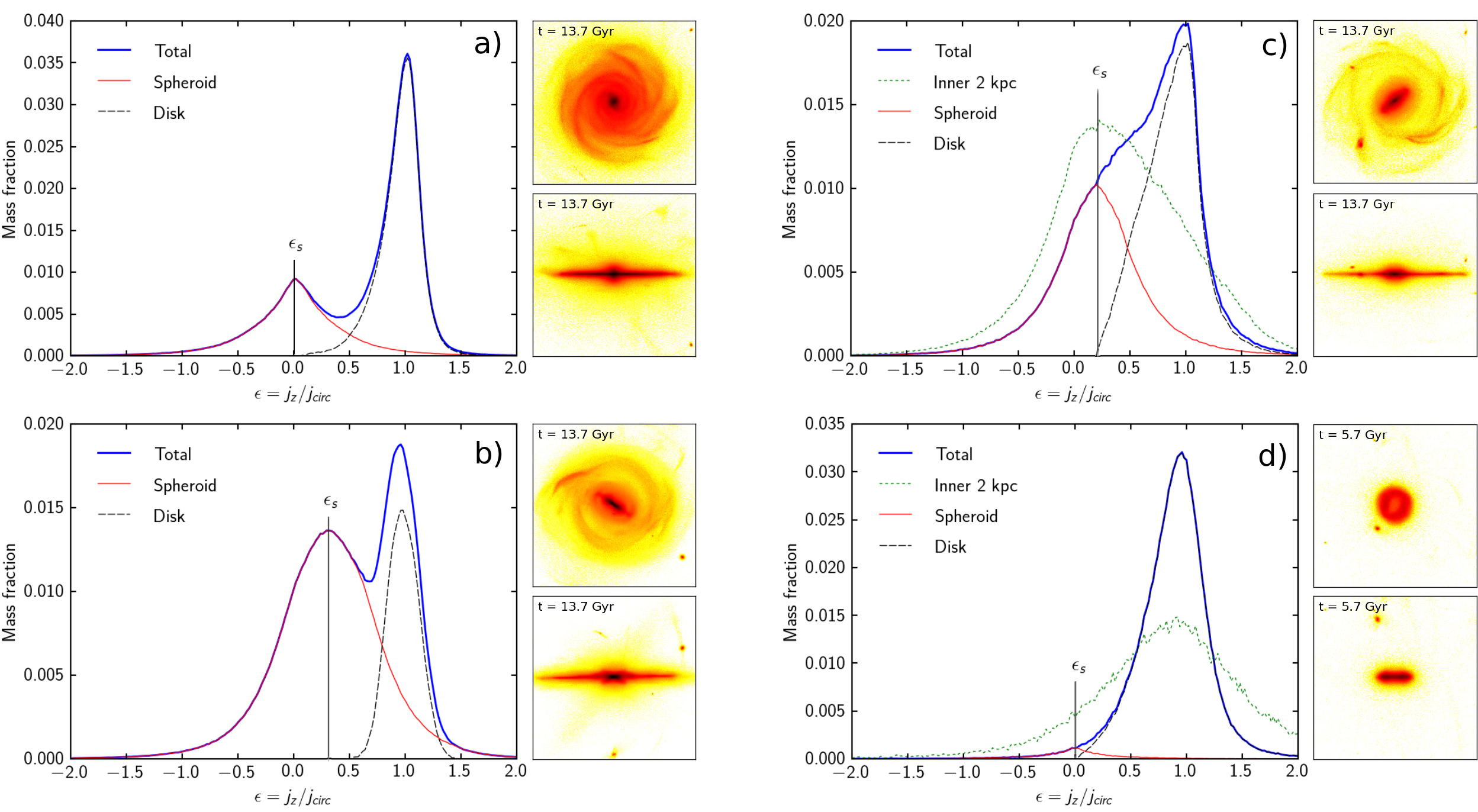}
\caption{Examples of bulge/disk decompositions based on kinematics. Each panel illustrates a different possible case; for each galaxy the distribution of $\epsilon = j_z/j_{circ}$ is shown (together with the resulting spheroid/disk decomposition), as well as i-band face-on and edge-on images (50 kpc x 50 kpc images). Panels a and b correspond to cases where two peaks are easily identified in the distribution of $\epsilon$ so that $\epsilon_s$  is straightforward to measure (note that Panel b shows a galaxy with a bar, for which $\epsilon_s \sim 0.3$). Panel c and d correspond to galaxies for which the spheroidal component is harder to identify. In this situation, we either set  $\epsilon_s$ to the value corresponding to the peak of the distribution of $\epsilon$ restricted to stars within the central inner 2 kpc of the galaxy (Panel c), or if this does not help  $\epsilon_s$ is set to 0  (Panel d). }
\label{fig:eps}
\end{figure*}

  We set the z-axis to the direction of the total angular momentum of all stars within the optical radius  (R$_{\mathrm{25}}$ in g band).
We compute the circularity parameter $\epsilon = \frac{j_z}{j_{circ}(R)}$ for each of the stellar particles within the optical radius, where $j_{circ}(R)$ is the angular momentum a particle would have if it were on a circular orbit at its current radius.
In a standard spiral galaxy, the distribution of $\epsilon$ shows two peaks, one at $\epsilon_d \sim 1$
corresponding to the rotating disk component, and one at 
$\epsilon_s \sim 0$ corresponding to the non-rotating spheroidal component (see Panel a in Figure \ref{fig:eps}). We identify the spheroidal component by assuming it corresponds to a distribution of $\epsilon$ symmetric around $\epsilon_s$ so that its mass is $M_s=2 \times M(\epsilon<\epsilon_s)$. Note that in some cases, particularly if the galaxy hosts a bar, $\epsilon_s$ is not 0, but most often between 0.2 and 0.4, which means the central regions of the galaxy have a net rotation movement (see Panel b in Figure \ref{fig:eps}).

In some cases however, $\epsilon_s$ can be difficult to measure if the two peaks are not clearly separated (see Panels c and d  in Figure \ref{fig:eps}). In this situation, we either set  $\epsilon_s$ to the value corresponding to the peak of the distribution of $\epsilon$ restricted to stars within the central inner 2 kpc of the galaxy (see Panel c), or if this does not help  $\epsilon_s$ is set to 0  (see Panel d).

\subsection{Photometric decomposition with GALFIT}
\begin{figure}
\centering 
\includegraphics[width=0.45\textwidth]{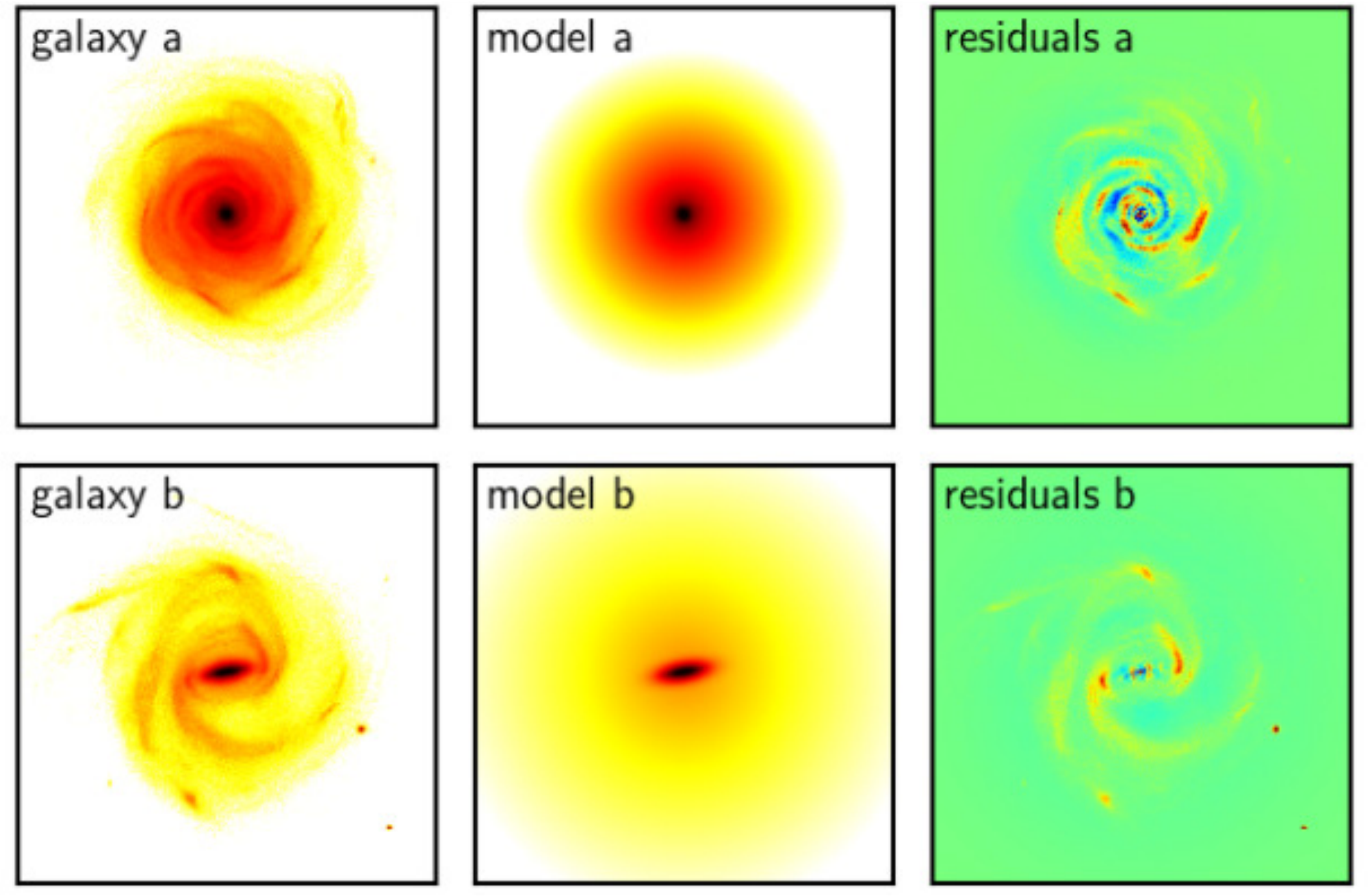}
\caption{Examples of GALFIT decompositions. The top panels show a bulge/disk decomposition (left: the simulated galaxy, middle: the GALFIT model, right: the residuals), while the bottom panels correspond to a bulge/bar/disk decomposition.}
\label{fig:galfit_examples}
\end{figure}

We create i-band images of the $z=0$ galaxies, extending out to 1.5 times the optical radius. We mask out of the image satellite galaxies as well as any bright clump that would make the identification of the main components more difficult. We use GALFIT in two steps: first we try a bulge+disk model and then a bulge+bar+disk model (see some examples of decompositions in Figure \ref{fig:galfit_examples}). Bulges and bars are assumed to follow a S\'{e}rsic profile while disks are assumed to have an exponential profile. In a bulge+bar+disk decomposition, the bar is chosen to be the S\'{e}rsic component with the lowest S\'{e}rsic index, largest effective radius and smallest axis ratio.

Between a bulge+disk and a bulge+bar+disk  decomposition, $\chi^2$ is often not sufficiently different to clearly identify the best model. In this case we add two additional criteria: the amplitude of the residuals both in the 2D image and in the radial profile. We then set the total bulge and/or bar luminosity to the corresponding S\'{e}rsic fits. As far as the disk is concerned, taking the luminosity of the smooth exponential profile is not always relevant, especially in the case of rings or lopsided features. We then measure the disk luminosity by subtracting the bulge and/or bar profiles to the total light distribution.

Note that 5 galaxies present such a complex structure that no satisfying GALFIT decomposition can be achieved, we remove them from studies involving photometric decompositions.

\subsection{Comparing the two techniques}

\begin{figure}
\centering 
\includegraphics[width=0.35\textwidth]{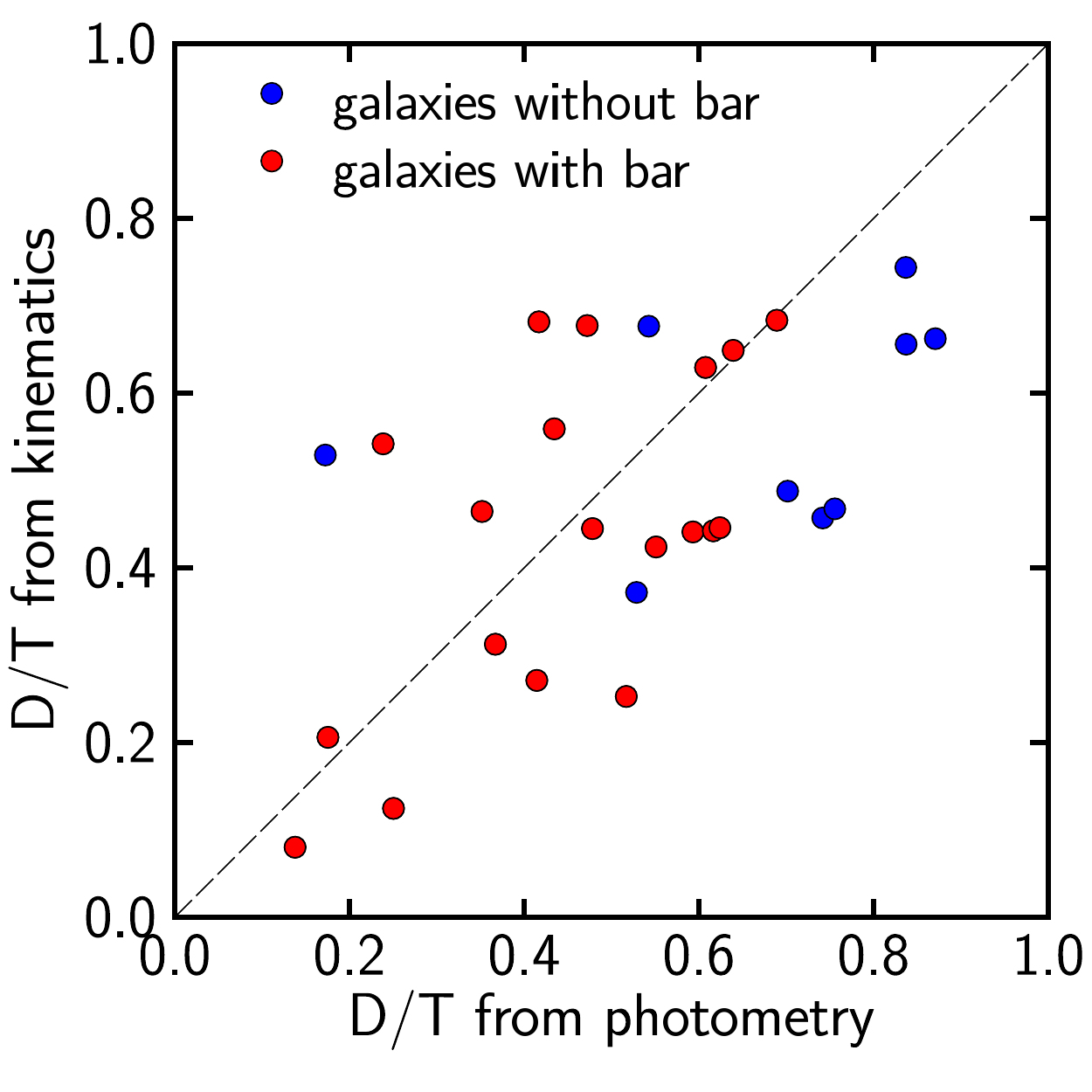}
\caption{Comparison of photometric and kinematic decompositions. We show the disk-to-total ratio (D/T) obtained with the kinematics as a function of that obtained with GALFIT. Barred galaxies correspond to the red dots  (we define barred galaxies as having a bar-to-total ratio greater than 0.1 according to GALFIT), non-barred galaxies to the blue dots.  }
\label{fig:compar_morpho}
\end{figure}

Figure \ref{fig:compar_morpho} compares the disk-to-total (D/T) ratio obtained for each galaxy with the kinematic and photometric techniques. There does not seem to be any strong  systematic offset between the values obtained by the two methods, although the scatter is quite large. For galaxies without a bar however, we find that GALFIT tends to give higher values of D/T compared with the kinematic decomposition. This trend was also found by \cite{Scannapieco2010}. They argue that the most likely reason for this difference is the fact that the photometric technique assumes that the disk exponential profile extends all the way to the center of the galaxy, while the stars in the center most likely are dispersion-dominated and are instead attributed to the bulge component by the kinematic decomposition. Additional differences can be attributed to measuring the morphology from the mass distribution instead of the luminosity distribution.

The fact that there is no systematic offset between the disk fraction obtained by the two methods suggests however that the ``spheroid'' component identified kinematically roughly corresponds to the bulge and bar components identified by GALFIT.

\section{Properties of the sample at $z=0$}

\begin{figure}
\centering 
\includegraphics[width=0.4\textwidth]{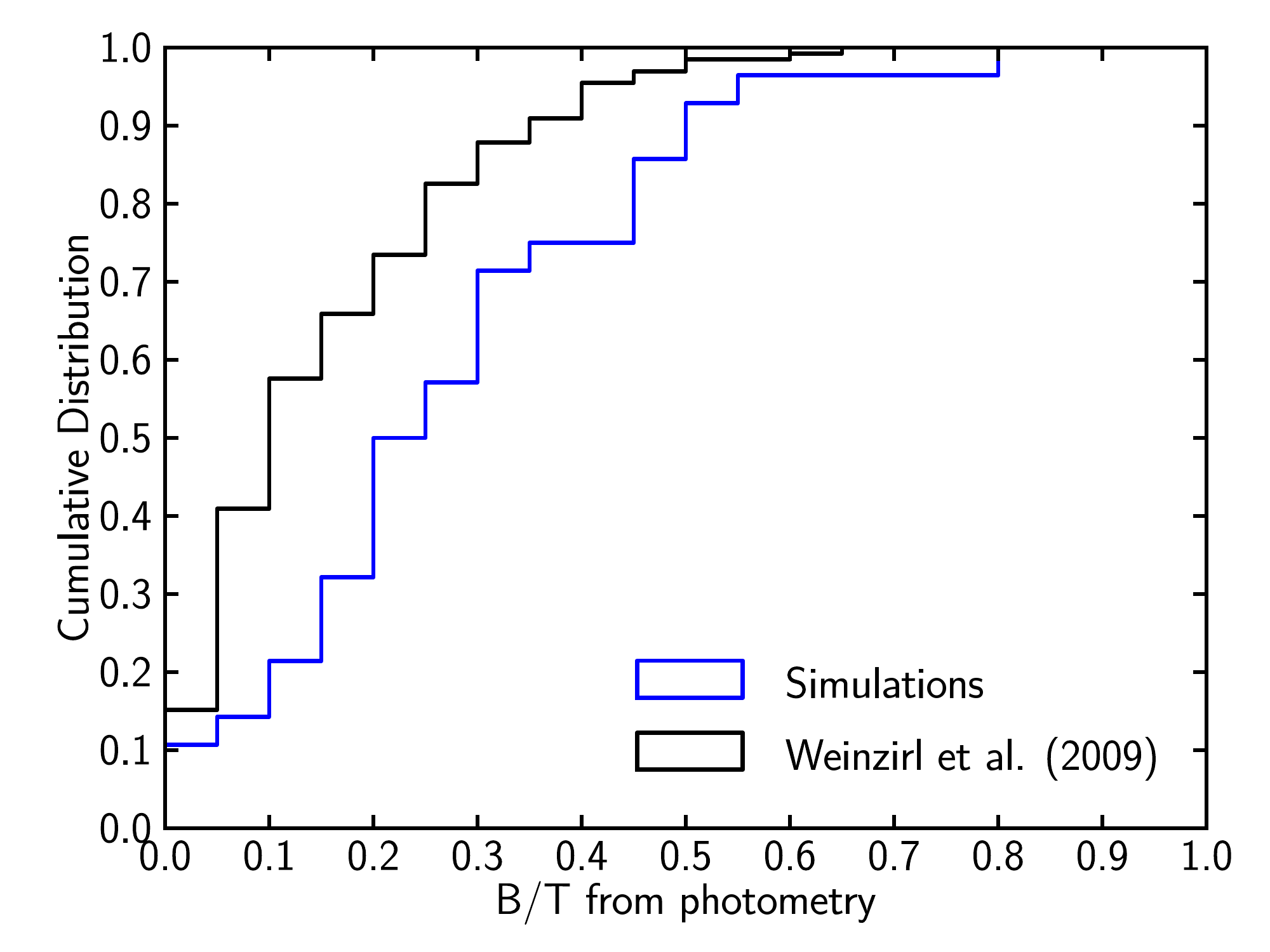}
\caption{Cumulative distribution of the bulge-to-total ratio measured with GALFIT, compared to observations by \cite{Weinzirl2009}. The simulations explore a large range of bulge fractions, from nearly bulgeless to bulge-dominated galaxies. However, the fraction of galaxies with B/T$\leq 0.2$ is only 0.34 for our sample of simulations, vs a fraction of 0.66 for the sample of local massive spirals observed by \cite{Weinzirl2009}. This is an important discrepancy, even if our sample is small and statistical comparisons are hard to perform. }
\label{fig:histo_bulge}
\end{figure}

We study the global properties of our simulated galaxies at $z=0$. We use in this section the photometric measurements to facilitate comparisons with observations, especially with \cite{Weinzirl2009}, who studied a sample of local massive spirals with a GALFIT decomposition technique similar to the one adopted here. We show in Figure \ref{fig:histo_bulge} the cumulative distribution of the bulge-to-total ratio as measured with GALFIT

The simulated galaxies span a large range of bulge fractions, from nearly bulgeless galaxies with B/T of the order of 0.05 to bulge-dominated galaxies with B/T close to 0.8. However, the distribution of the bulge fractions differs significantly from the observed one:  B/T tends to be too high in the simulations. Figure \ref{fig:histo_bulge} shows that only 32\% of our simulated galaxies have B/T$<0.2$, a value much lower than the 66\% found by \cite{Weinzirl2009}. Similarly, 40\% of the observed sample is found at B/T$<0.1$, a value that is only reached by 14\% of our sample (these nearly bulgeless galaxies are shown in Figure \ref{fig:AppA1}). In spite of the success of the simulations at producing galaxies with small bulges, there is thus strong evidence that not enough simulated galaxies are produced with a low B/T. Although the images used by \cite{Weinzirl2009} are in H-band while ours are in i-band and we do not include dust extinction, and while these differences will affect the measured bulge fractions \citep{Graham2008}, this is probably not enough to reconcile the simulations with the observations.

In addition, even if we probe similar stellar mass ranges, our simulated galaxies are likely more isolated than the observed sample, that is only magnitude-limited, with no restriction on environment. However, if we compare the simulations to samples of isolated galaxies, the discrepancy becomes even greater: the observed  population of isolated galaxies is strongly dominated by late-type spirals \citep{Sulentic2006,Durbala2008}. 

Note also that we do not find any pure elliptical galaxies, but this is not so worrying since they are rather rare, especially in isolated environments (e.g. \citealp{Dressler1980,Hogg2003,Croton2005}). \cite{Tasca2011} perform bulge/disk decompositions of SDSS galaxies and also show that very few galaxies actually have bulge fractions greater than 0.8.

\begin{figure}
\centering 
\includegraphics[width=0.4\textwidth]{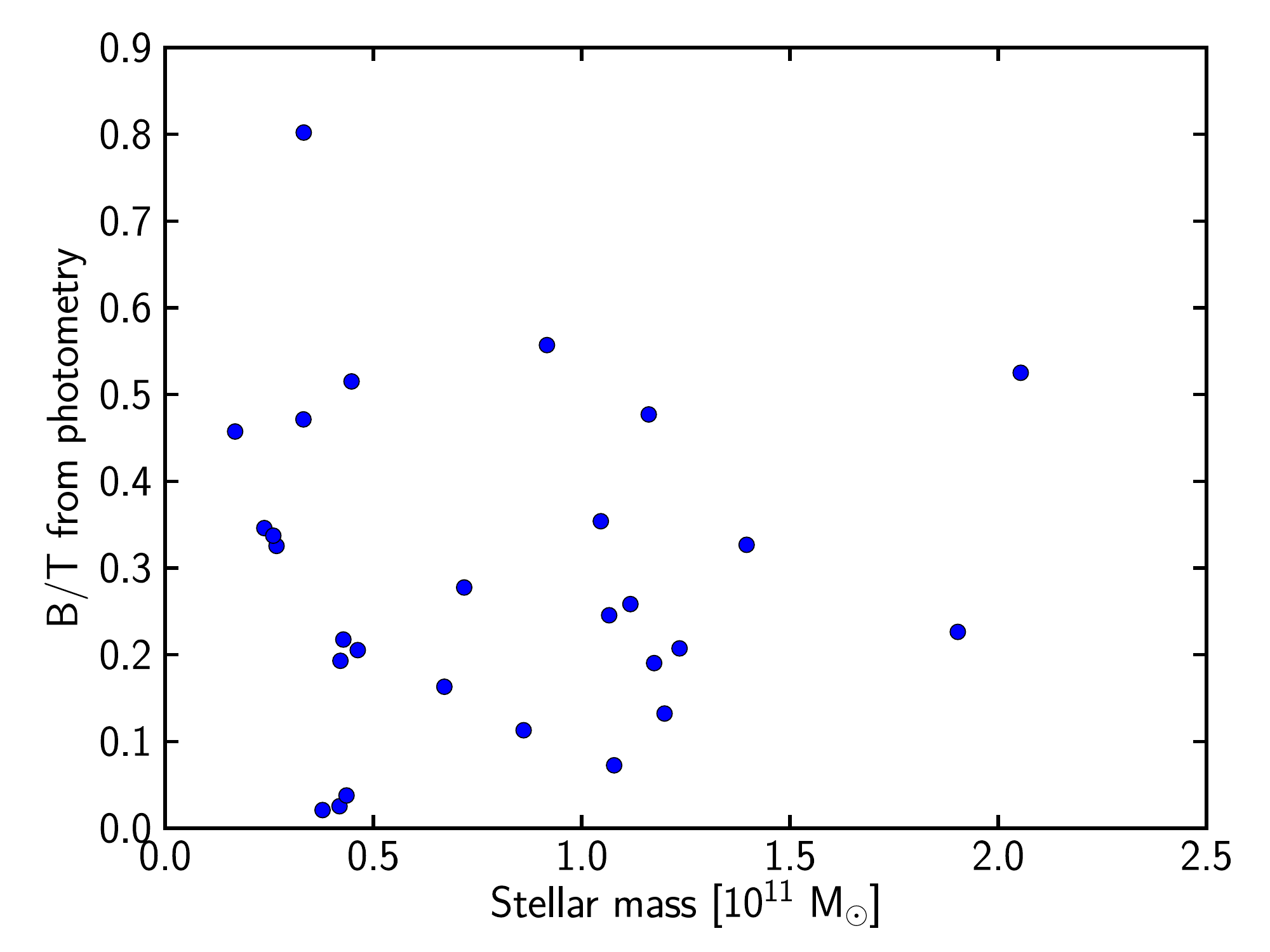}
\caption{Bulge-to-total ratio (measured with GALFIT) as a function of stellar mass at $z=0$ for the simulated galaxies. We do not find any trend between bulge content and stellar mass, contrary to observations showing that more massive galaxies tend to be more bulge-dominated (e.g., \citealp{Weinzirl2009}). \hspace{5cm} }
\label{fig:mstar_bulge}
\end{figure}

Another difference with observations is that we do not find any correlation between stellar mass and bulge content (Figure \ref{fig:mstar_bulge}), contrary for instance to \cite{Weinzirl2009} that show a trend to higher bulge fractions with higher stellar masses. This difference might be insignificant due to our small sample and due to different methods used to measure stellar masses in observations and simulations (\cite{Weinzirl2009}  estimate the stellar mass of a galaxy from its $B-V$ color, they estimate typical errors to be within a factor of 2--3), but it could also reflect a more profound discrepancy between our simulations and observations.

\begin{figure}
\centering 
\includegraphics[width=0.4\textwidth]{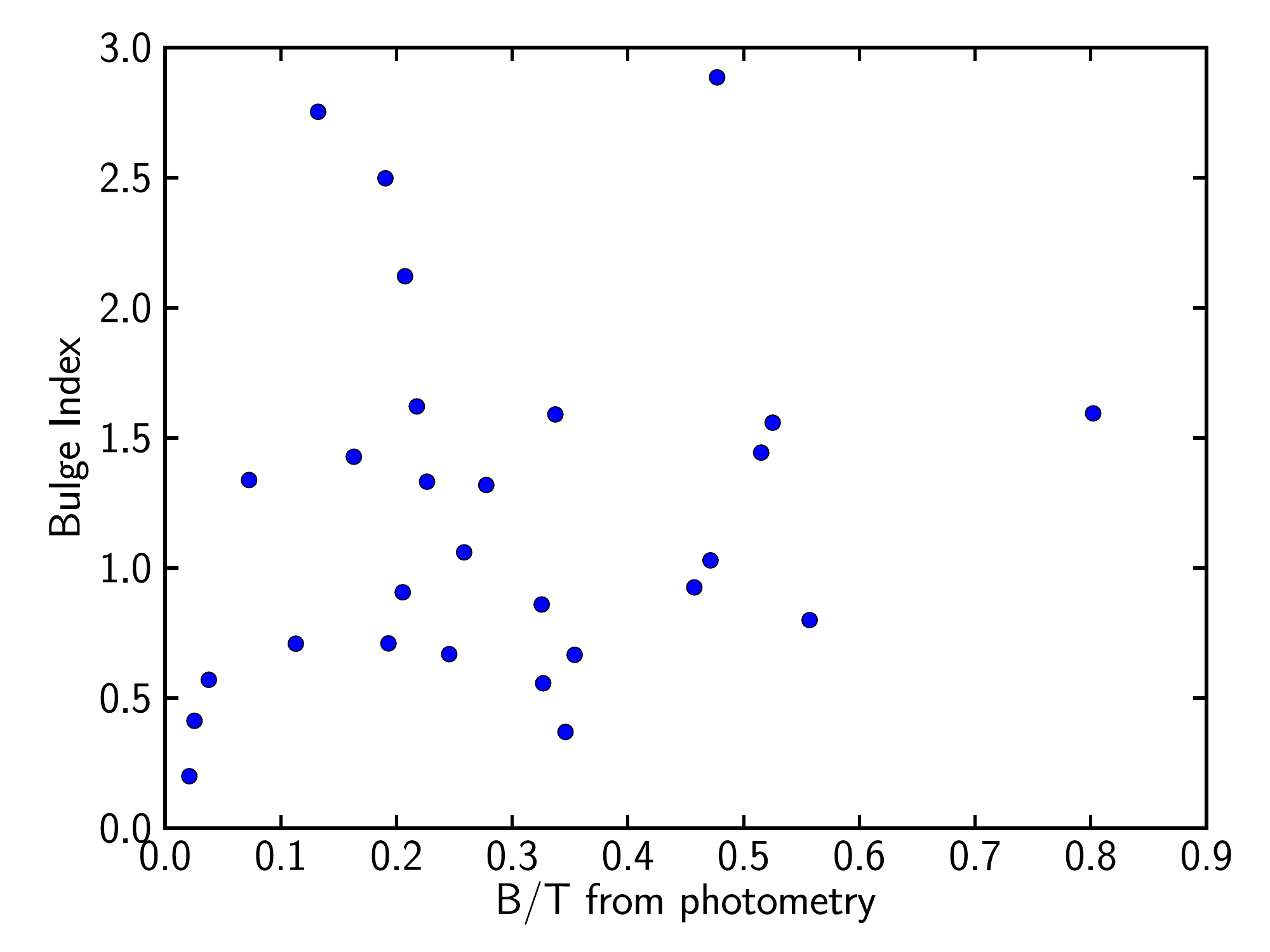}
\caption{Bulge S\'{e}rsic index as a function of  the bulge-to-total ratio measured with GALFIT. Most of the simulated galaxies have a bulge S\'{e}rsic index lower than 2, suggesting a high incidence of pseudo-bulges, which is consistent with the observed properties of local spirals  \citep{Laurikainen2007,Durbala2008,Weinzirl2009}. }
\label{fig:Sersic}
\end{figure}

Figure \ref{fig:Sersic}  shows the bulge S\'{e}rsic index as a function of the bulge fraction. Most of our simulated galaxies have a low S\'{e}rsic index, the majority lower than 2, indicating a large fraction of pseudo-bulges within our sample. This is consistent with the observed properties of local disk galaxies; \cite{Laurikainen2007} find for instance that the S\'{e}rsic index of bulges is on average lower than 2 for all morphological types (see also \citealp{Durbala2008,Weinzirl2009}).

The bar content of our galaxies also appears roughly consistent with observations: 70\% of our galaxies host a bar, to be compared to values of 60--70\% for local galaxies when observed in the near-infrared \citep{Eskridge2000,Marinova2007}. The simulated galaxies span a large range of bar fractions, from galaxies with no bar to galaxies where the bar luminosity amounts to nearly half of the total luminosity. 
In a companion paper, we examine the properties of bars in our sample, including bar formation and the redshift evolution of the fraction of barred galaxies.

To summarize our results thus far:
\begin{itemize}
\item Simulated Milky Way-mass spiral galaxies show at $z=0$  a wide range of bulge fractions, from bulgeless to bulge-dominated galaxies, though the bulgeless disks are under-represented.
\item These galaxies also display a diverse range of bar fractions
\item The majority of bulges in our sample would be classified as pseudo-bulges and not classical bulges based on their S\'{e}rsic index.
\end{itemize}

Our simulations  fail to reproduce the observed distribution of bulge fractions (with too few bulgeless galaxies within the simulated sample), as well as the observed correlation between bulge fraction and stellar mass. In spite of these failures, the sample of simulated galaxies remains relevant to study the redshift evolution of galaxy morphologies and the physical mechanisms building bulges.
The next sections will investigate the nature of the high-redshift progenitors of the simulated galaxies, and will also focus on the most disk-dominated cases at $z=0$ and study their formation histories.

\section{The high-redshift progenitors of spiral galaxies}
\begin{figure}
\centering 
\includegraphics[width=0.45\textwidth]{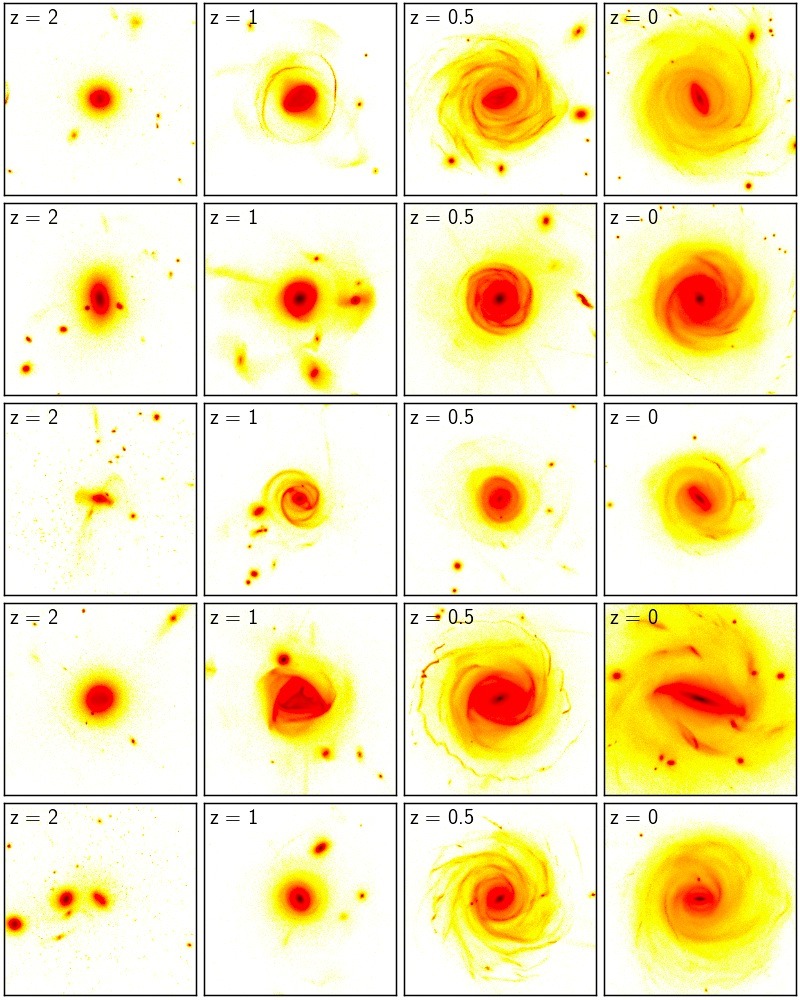}
\caption{Morphological evolution from $z=2$ to 0 for a subset of simulated galaxies. We show  i-band surface brightness maps (face-on projection, 70x70 kpc) at z= 2, 1, 0.5 and 0. }
\label{fig:evol_maps}
\end{figure}

In this section, we study the evolution with redshift of the star formation rates, gas fractions, sizes and morphologies of the simulated galaxies.
Figure \ref{fig:evol_maps} shows i-band surface brightness maps for a subset of simulated galaxies at  $z=2$, 1, 0.5 and 0. The maps for the whole sample can be found in Appendix \ref{app:sample}.
\subsection{Mass growth and star formation histories}

\begin{figure}
\centering 
\includegraphics[width=0.40\textwidth]{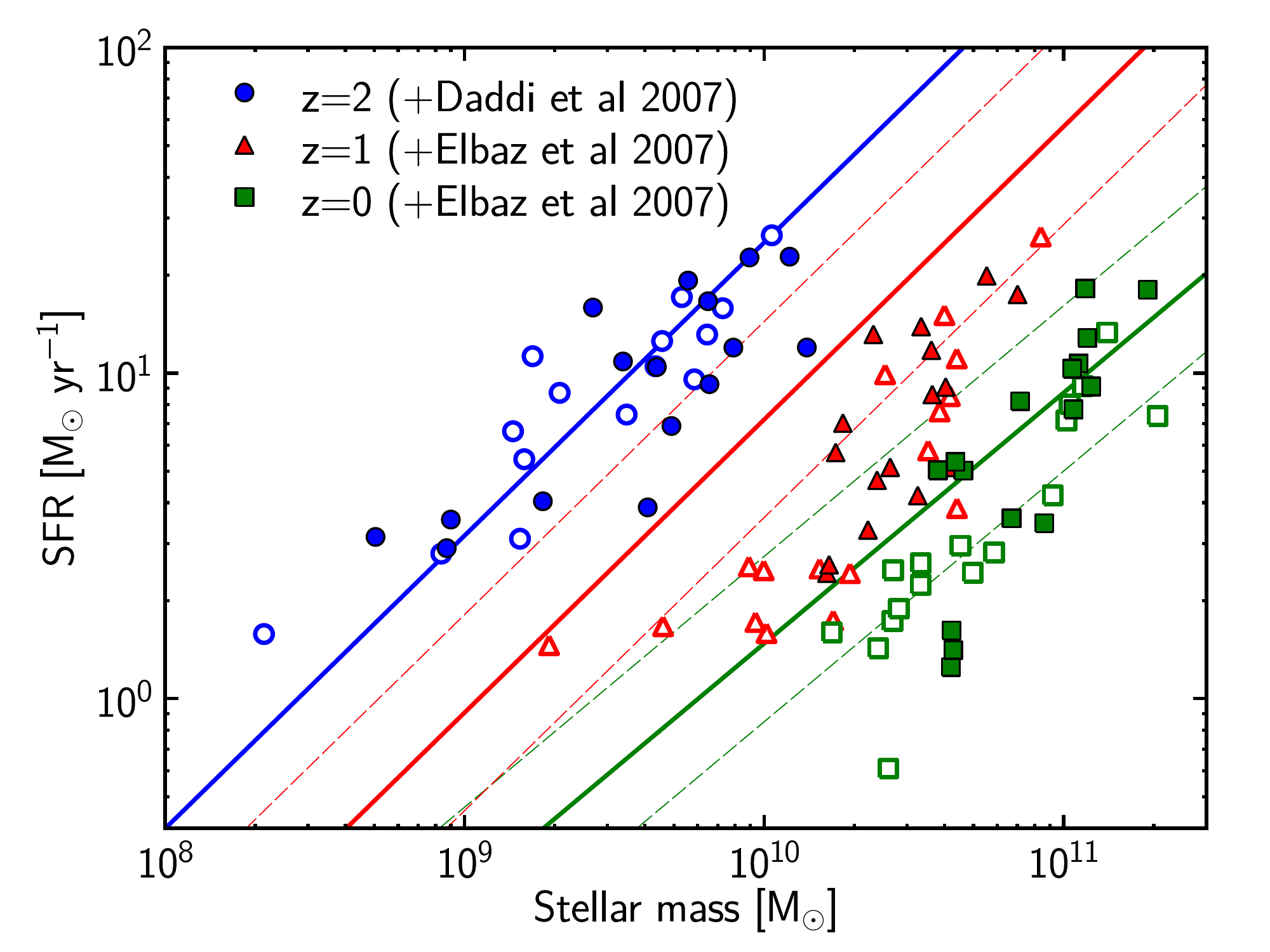}
\caption{Star formation rate as a function of stellar mass for the simulated galaxies at $z=2$ (blue dots), 1 (red triangles) and 0 (green squares), compared to the correlations observed for star forming galaxies ($z=2$, \citealp{Daddi2007}, blue line; $z=1$, \citealp{Elbaz2007}, red line; $z=0$, \citealp{Elbaz2007}, green line). The filled symbols correspond to galaxies that are disk-dominated at $z=0$ (i.e. with a photometric B/T lower than 0.3), the empty symbols correspond to the rest of the sample.}
\label{fig:SFR_mstar}
\end{figure}

Figure \ref{fig:SFR_mstar} shows the star formation rate of the simulated galaxies as a function of their stellar mass at redshift 2, 1 and 0. The simulations are compared to observations of star-forming galaxies \citep{Elbaz2007,Daddi2007}. We find an agreement between simulations and observations at $z=2$ and 0, while at $z=1$ the simulated galaxies have systematically lower SFRs than the observed ones. This disagreement could be (at least partially) due to selection effects, for instance it could be that  $z=0$ spirals were not on the star-forming sequence at $z=1$, i.e. that the galaxies on the star forming sequence at $z=1$  have evolved into bulge-dominated galaxies at $z=0$. However, another explanation is a too short gas consumption timescale in the simulation between $z=1$ and 2 (maybe linked with an overly weak feedback implementation), which would result in low gas fractions at $z=1$, and thus low SFRs.

\begin{figure}
\centering 
\includegraphics[width=0.40\textwidth]{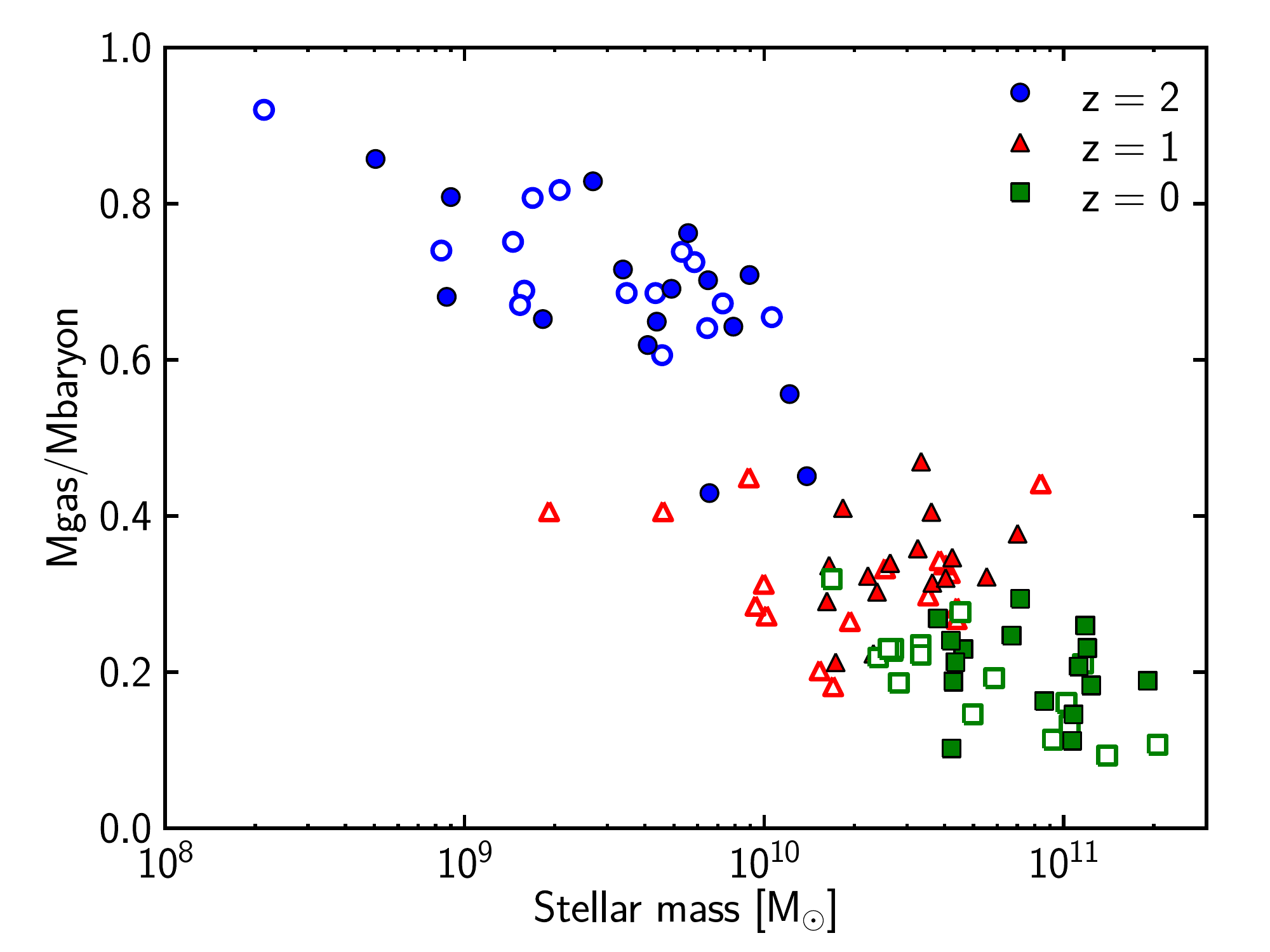}
\caption{Gas fraction as a function of stellar mass for the simulated galaxies at $z=2$, 1 and 0. The gas fraction is defined as the total gas mass within R25 divided by the total baryonic mass within R25. The filled symbols correspond to galaxies that are disk-dominated at $z=0$ (i.e. with a photometric B/T lower than 0.3), the empty symbols correspond to the rest of the sample.}
\label{fig:mstar_gas_frac}
\end{figure}

To investigate this issue, Figure \ref{fig:mstar_gas_frac} shows the evolution with redshift of the gas fraction, defined as the total gas mass within R25 divided by the total mass of gas and stars within R25. While the values for $z=2$ seem in agreement with observations, the $z=1$ gas fractions are indeed probably a bit too low \citep{Daddi2010a,Tacconi2010,Geach2011}, especially since they reflect the whole gas content of the galaxies and not only their molecular gas content (precise comparisons are thus difficult, in addition to observational uncertainties in the CO-to-H$_2 $ conversion factor). It is interesting to notice a tight sequence of decreasing gas fraction both with decreasing redshift and increasing stellar mass.

\begin{figure}
\centering 
\includegraphics[width=0.40\textwidth]{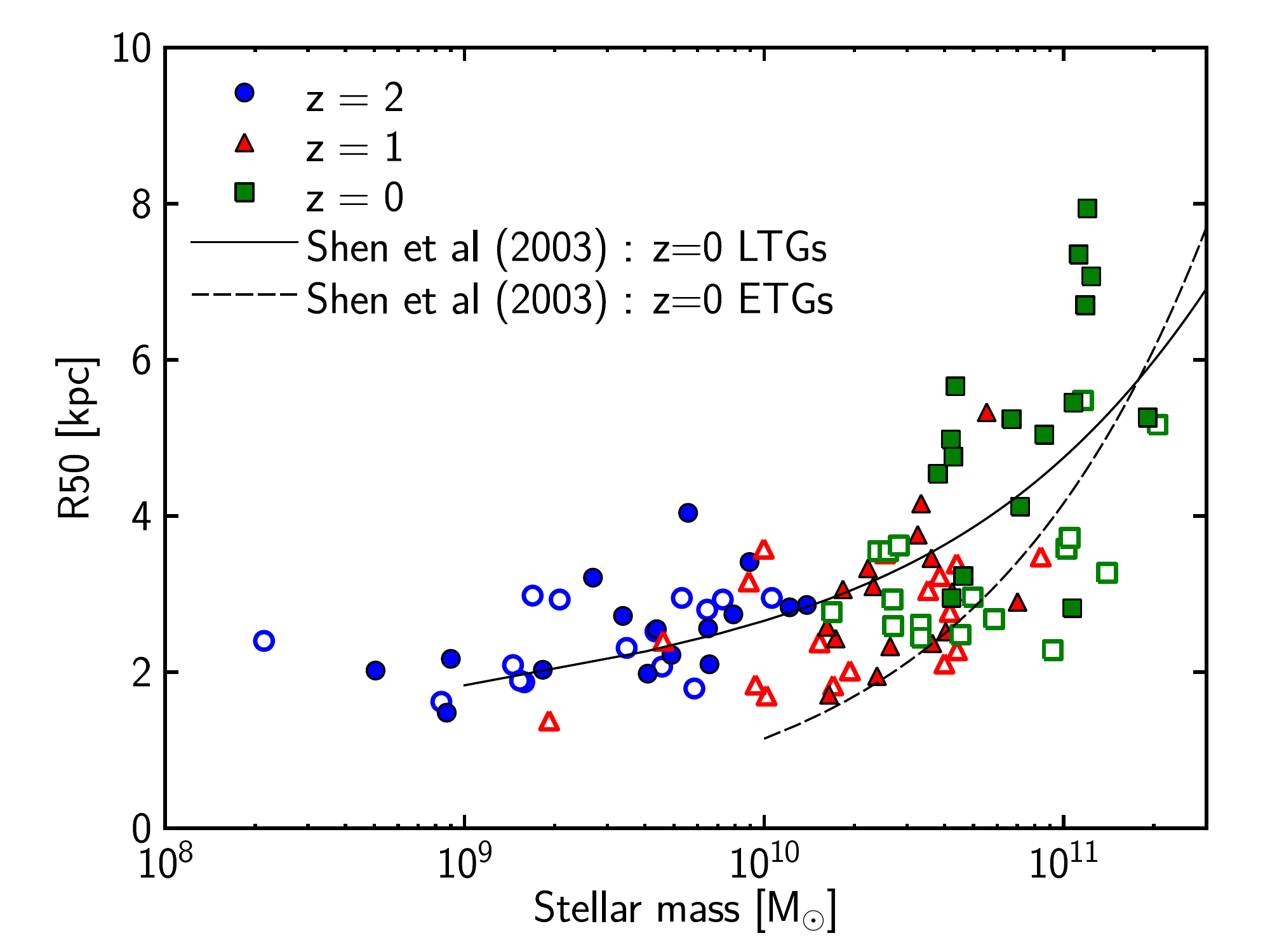}
\caption{Half-mass radius as a function of stellar mass for the simulated galaxies at $z=2$, 1 and 0. The filled symbols correspond to galaxies that are disk-dominated at $z=0$ (i.e. with a photometric B/T lower than 0.3), the empty symbols correspond to the rest of the sample. The $z=0$ observed relations for late-type and early-type galaxies \citep{Shen2003} are also plotted. }
\label{fig:Radius_Mass}
\end{figure}

\subsection{Sizes}
 
We study the half-mass radius of the galaxies as a function of their stellar mass. Figure \ref{fig:Radius_Mass} shows the results at $z=2$, 1 and 0, compared with observations by \cite{Shen2003} for $z=0$.  \cite{Shen2003} actually compute R50 in the r-band, which is different from using the mass distribution. We find however that our simulations agree nicely with their measures. The simulated galaxies have reasonable sizes, a fact that had been an issue for galaxy formation simulations in the last few decades but not so much now \citep{Brooks2011,Guedes2011,Brook2011b}.  We note that at fixed stellar mass our disk-dominated galaxies tend to have larger values of R50 with respect to the bulge-dominated ones, which is also found in $z=0$ observations.

A difficulty when trying to compare with observations at higher redshifts comes from the variety of morphological types and the limited mass range that we explore. For instance studies of the sizes in the zCOSMOS sample at $z\sim1$ \citep{Sargent2007,Maier2009} focus on galaxies with a stellar mass greater than $5\times 10^{10}\rm M_{\sun}$, a threshold that is only reached by 3 of our simulated at this redshift. Our results are thus not necessarily in contradiction with \cite{Sargent2007}, who find as many galaxies with a radius between 5 and 7 kpc at $z=1$ and $z=0$, whereas we find many more of these galaxies at $z=0$ than at $z=1$. Additional discrepancies could come from using light profiles instead of mass profiles (and dust extinction could play a part: if extinction is underestimated in the central regions, the half-light radii could be overestimated).

\begin{figure*}[!ht]
\centering 
\includegraphics[height=5cm]{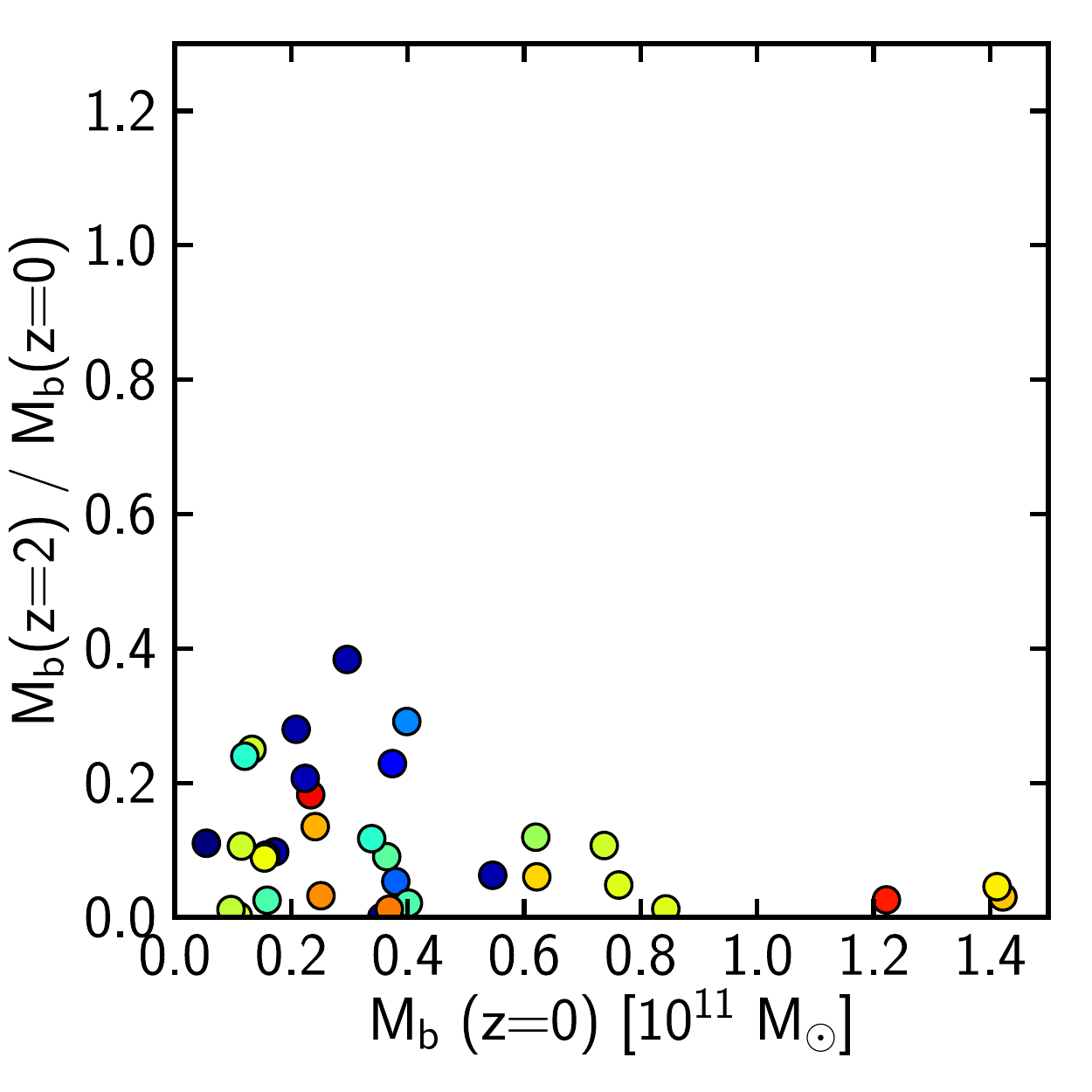}
\includegraphics[height=5cm]{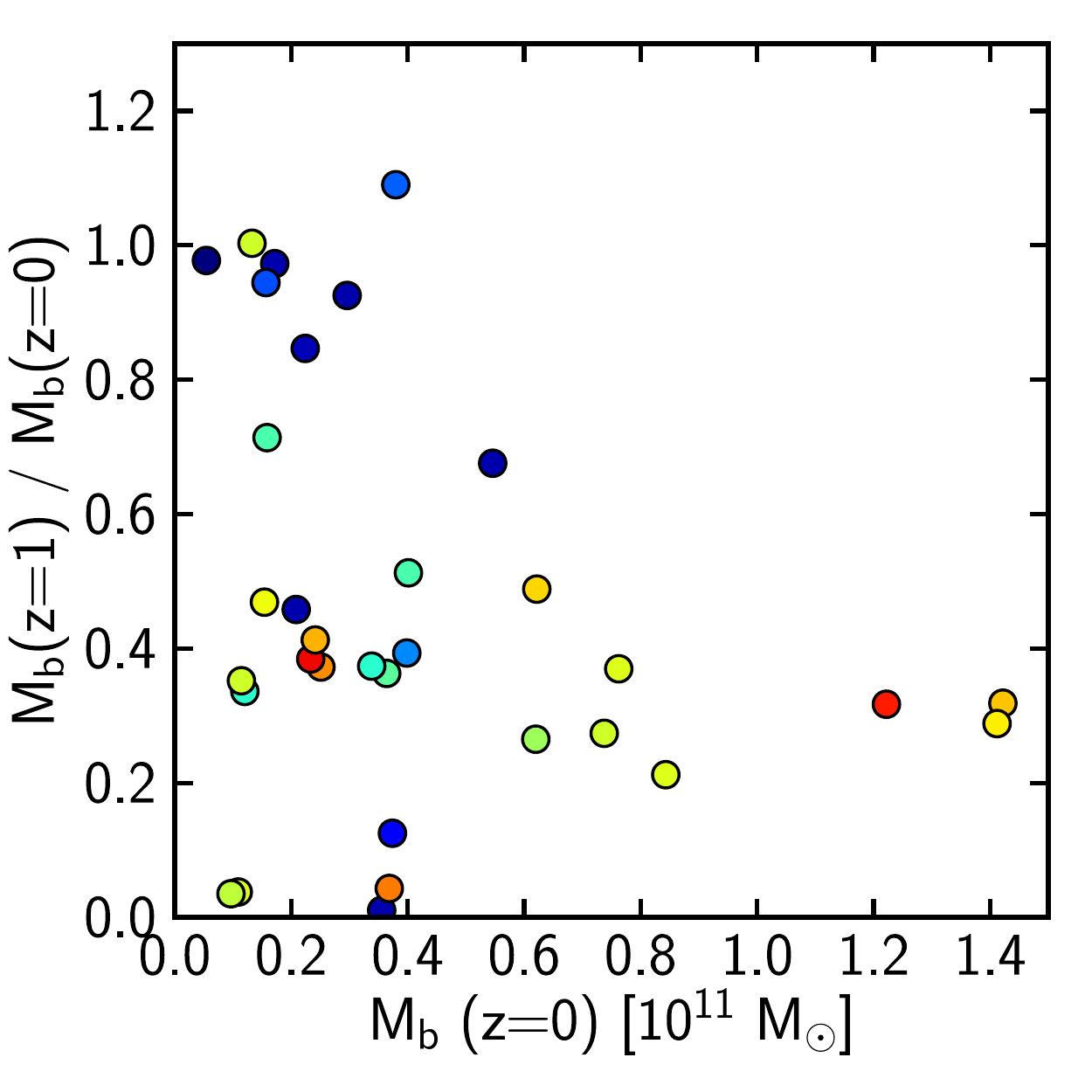}
\includegraphics[height=5cm]{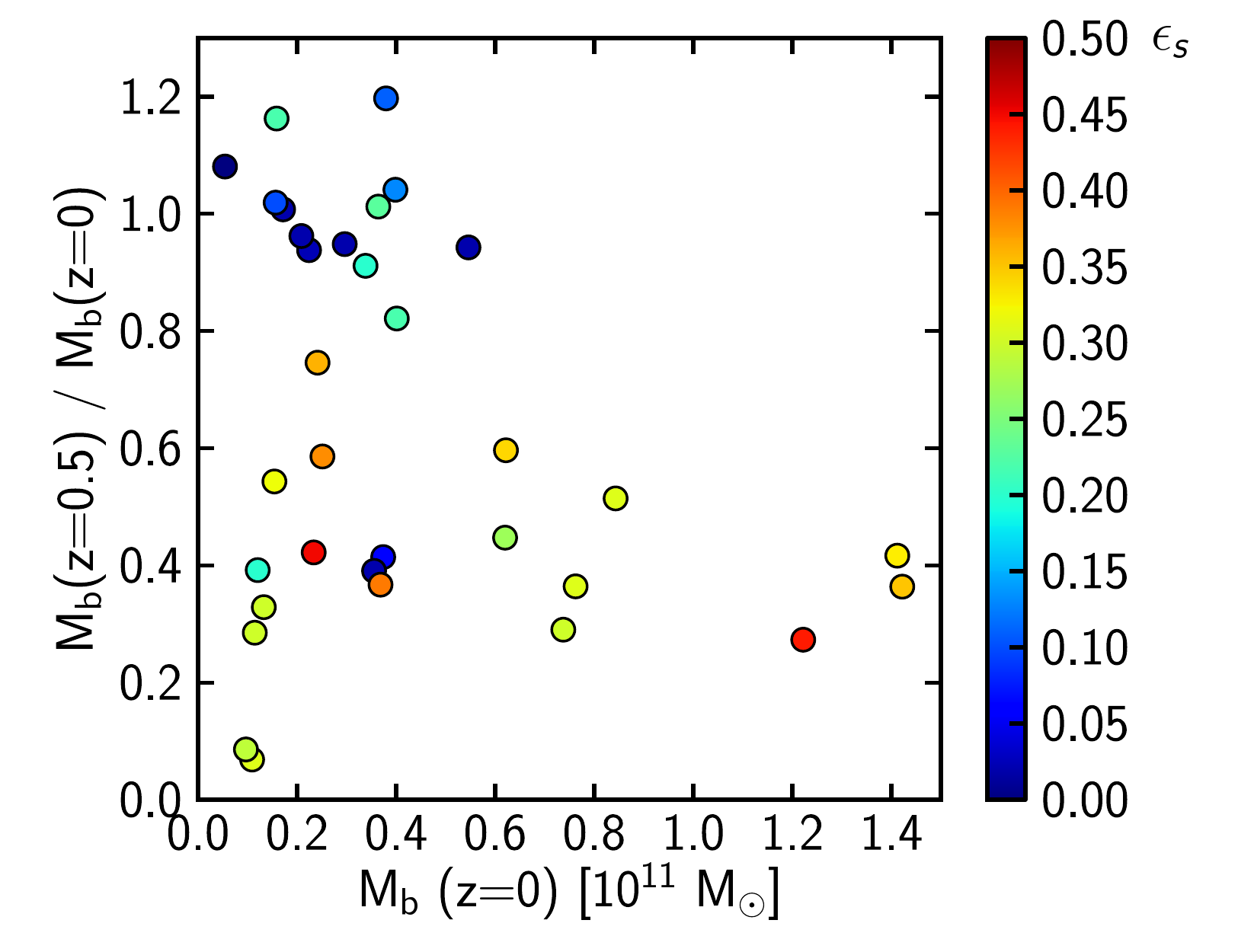}
\caption{Evolution of the mass of the dispersion-dominated component (bulge+bar as measured by kinematics technique) with redshift. The three panels show the fraction of the mass of this component already in place at high redshift (left panel: $z=2$; middle: $z=1$; right: $z=0.5$) as a function of the total bulge+bar mass at $z=0$. The color code indicates the value of $\epsilon_s$ at $z=0$ for each galaxy, it is thus related to the amount of rotation in the central regions, i.e. to the presence of a bar or a rotating bulge. }
\label{fig:Mb_evol}
\end{figure*}
\begin{figure*}[!ht]
\centering 
\includegraphics[height=5cm]{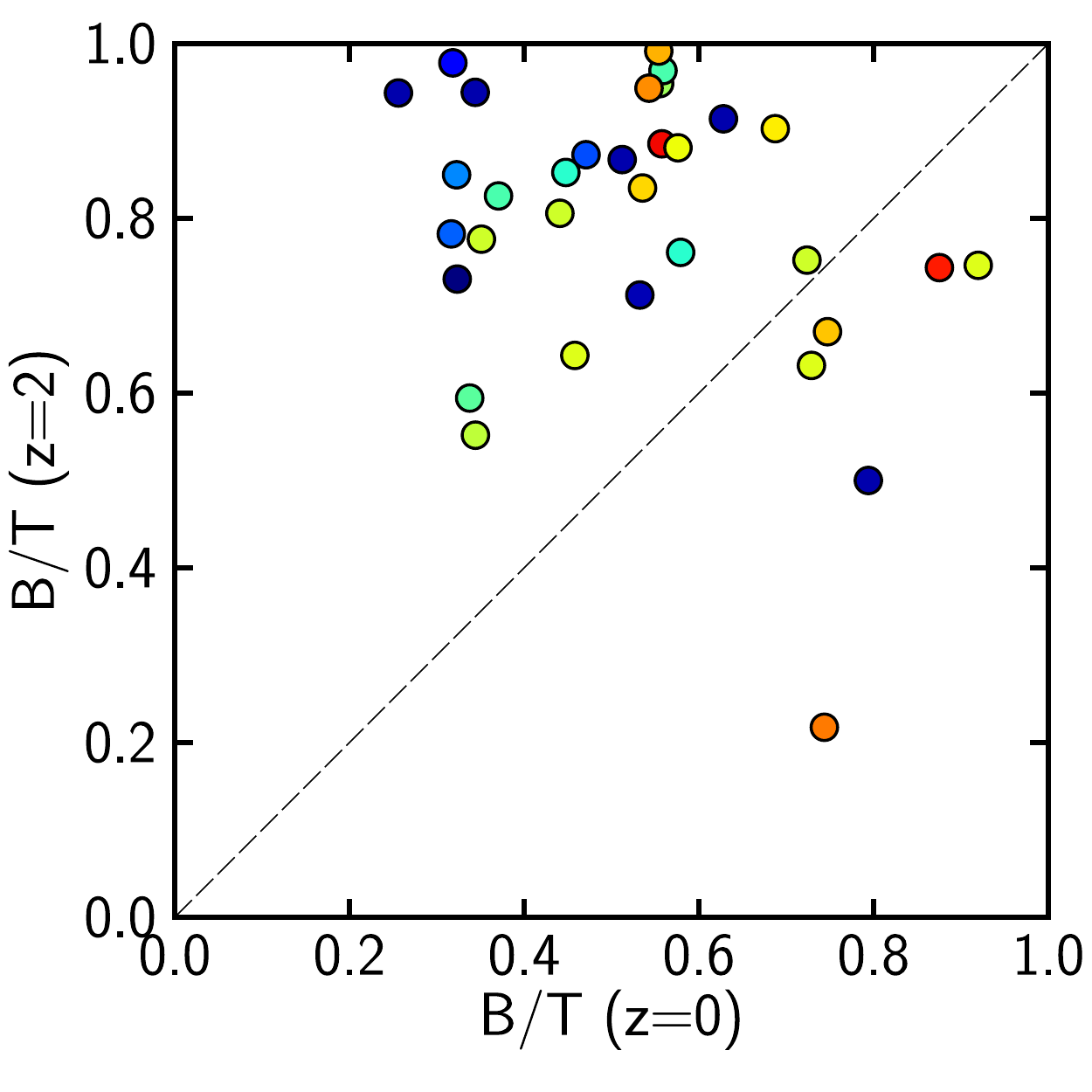}
\includegraphics[height=5cm]{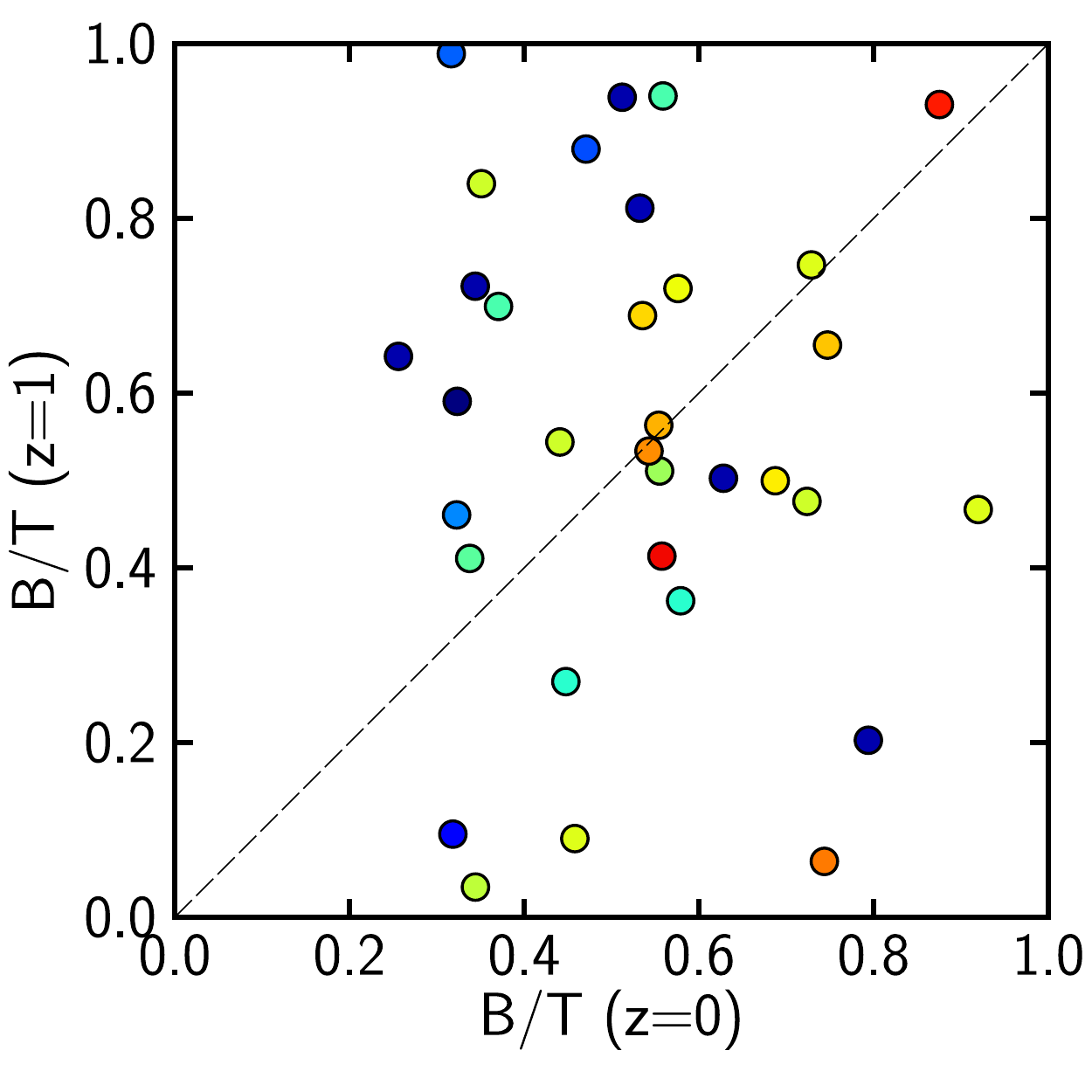}
\includegraphics[height=5cm]{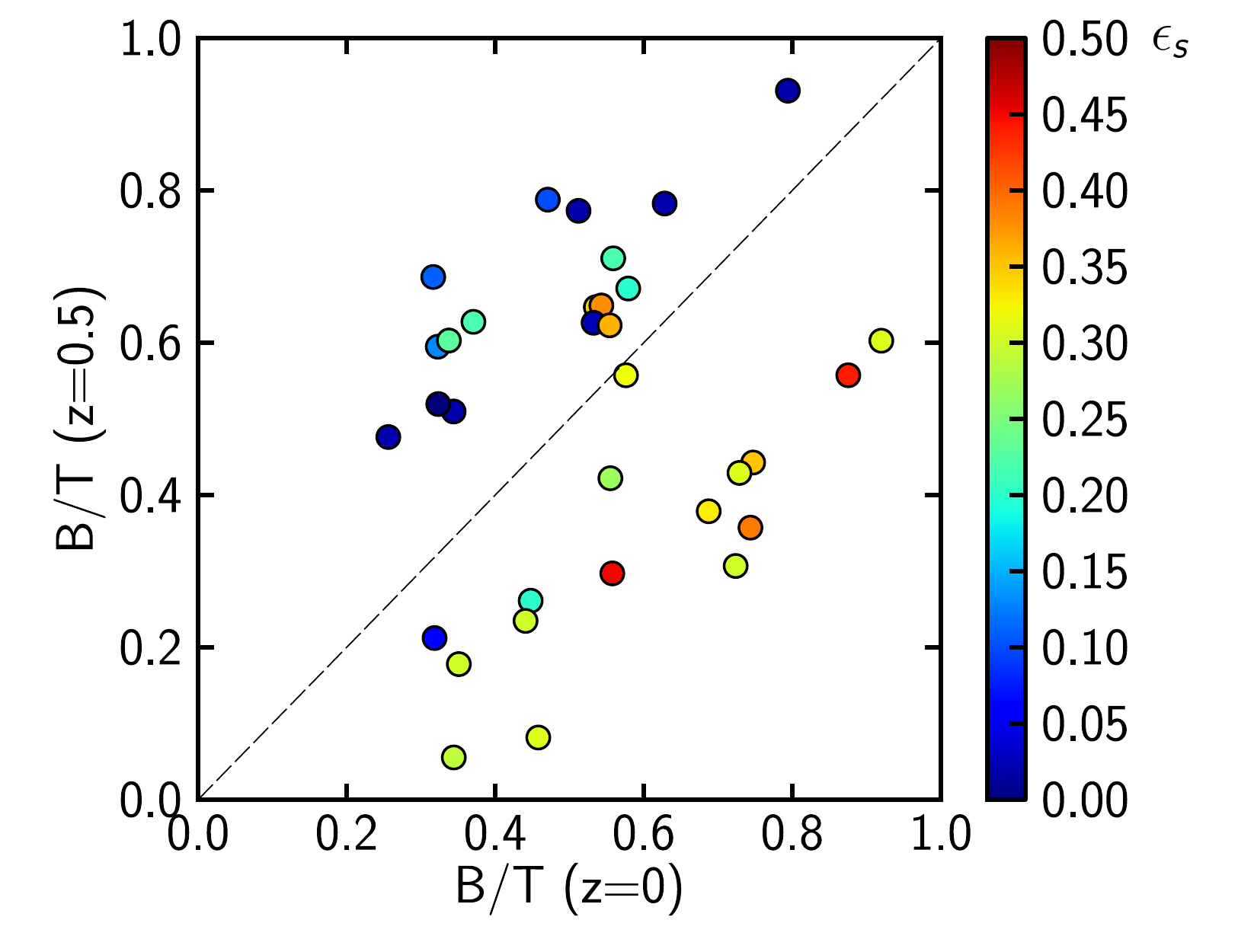}
\caption{Evolution of the (bulge+bar)-to-total ratio (bulge+bar as measured by kinematics) with redshift. The three panels show B/T at different redshifts (left panel: $z=2$; middle: $z=1$; right: $z=0.5$) as a function of B/T at $z=0$. The color code is the same as in Figure \ref{fig:Mb_evol}}
\label{fig:BoverT_evol}
\end{figure*}

Overall, we find that the simulated galaxies at $z=2$ and 1 seem to follow the same relation as the local galaxies (\citealp{Shen2003} for the SDSS): disk galaxies grow along a sequence in the radius-mass plane. This idea is supported by observations \citep{Barden2005, Maier2009}, and it has already been seen in other simulations \citep{Brooks2011} and semi-analytic models \citep{Somerville2008a, Firmani2009, Dutton2010b}.

None of our high redshift galaxies are particularly compact, which does not allow us to draw any conclusion on the nature and frequency of compact ellipticals. This suggests however that these galaxies are not found within the progenitors of isolated, Milky Way-mass galaxies (early-type galaxies are found in denser environments), and that instead they follow a different evolutionary track.

\subsection{Morphologies}\label{sec:morpho}

To characterize the morphological evolution of the simulated galaxies, we first study bulge formation as a function of redshift. Figure  \ref{fig:Mb_evol} shows, as a function of the final bulge mass, the fraction of the bulge already in place at $z=2$, 1 and 0.5. We use the kinematic decomposition so that ``bulge'' here actually means ``bulge+bar'' (as explained in Section 3, using GALFIT is extremely difficult on high redshift galaxies). The color coding in Figure  \ref{fig:Mb_evol}  indicates the value of $\epsilon_s$ at $z=0$ for each galaxy and is related to the amount of rotation in the central regions, i.e. to the presence of a bar or a rotating bulge  ($\epsilon_s$ is defined in Section \ref{sec:kinematics} as the average circularity for stars in the central spheroidal component, it is equal to 0 for a non-rotating component).

We find that only a minor fraction of the bulges were already in place at $z=2$: galaxies typically have at most 15\% of their bulge-bar formed at this redshift. A strong evolution occurs between $z=2$ and $z=1$, and our simulations show a few cases of galaxies that have their entire $z=0$ bulge in place by $z=1$; these tend to be the galaxies without bar at $z=0$  (blue colors on Figure  \ref{fig:Mb_evol} ). Galaxies with a strong bar at $z=0$ have by contrast still a lot of bulge-bar growth at $z<0.5$.
Note that Figure \ref{fig:Mb_evol} shows a few cases of galaxies with a greater bulge mass at $z=0.5$ or 1 than at $z=0$. This can be due either to errors in bulge mass determination or to a transfer of mass from the bulge to the disk (see \citealp{Martig2010} for an example of how stellar mass loss from bulge stars can transfer mass from a bulge to a disk).

\begin{figure}
\centering 
\includegraphics[width=0.4\textwidth]{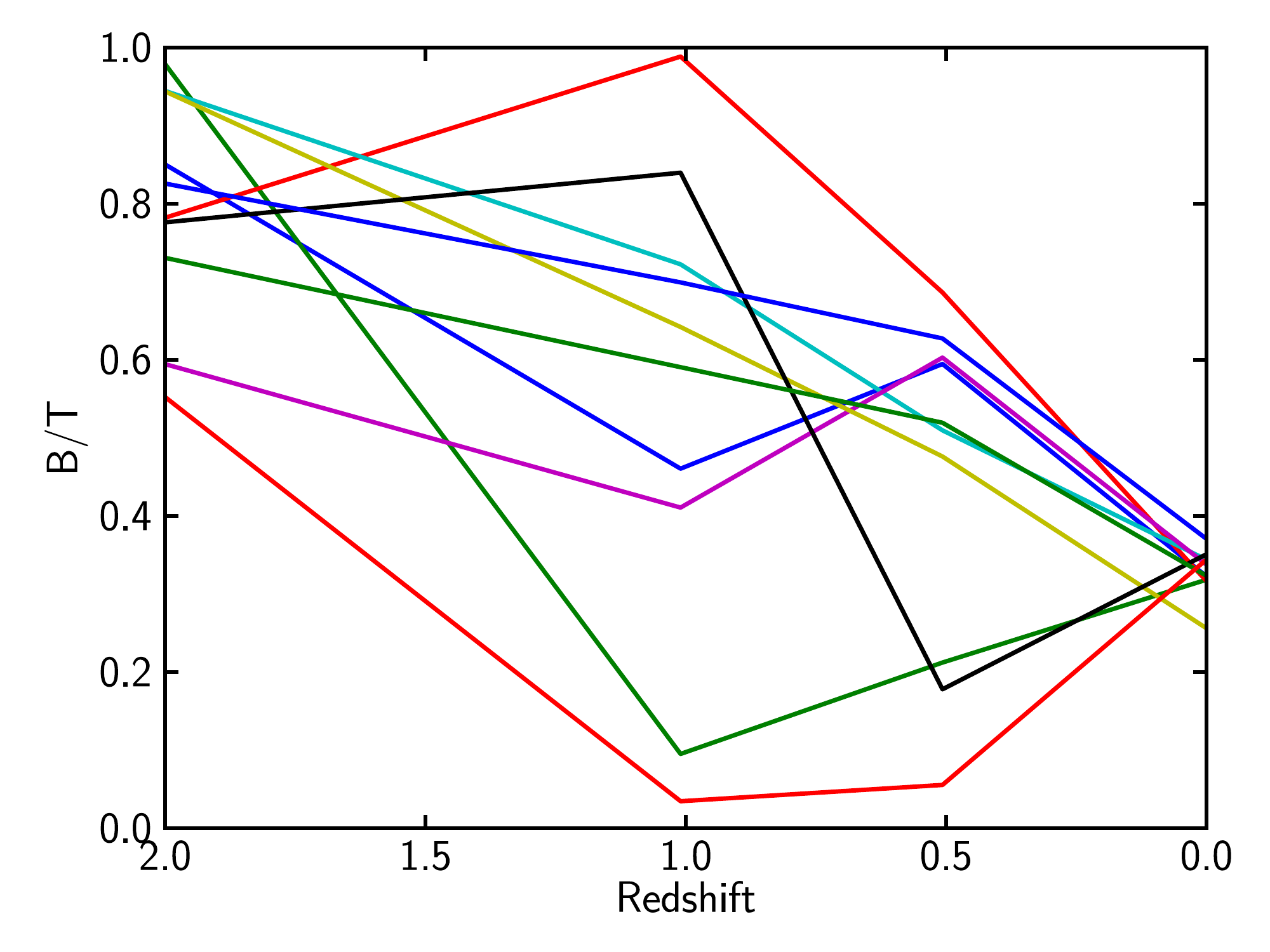}
\caption{Evolution of the mass of the (bulge+bar)-to-total ratio with redshift for galaxies with B/T$<0.4$ at $z=0$. A large diversity of histories can give birth to galaxies with a similar bulge-bar fraction at $z=0$. The two red lines correspond to galaxies with the same final B/T, but at $z=1$ one is a pure elliptical, the other is a pure disk. The general trend between $z=1$ and $z=0$ is a decrease in the bulge fraction, except for galaxies that are nearly bulgeless at $z=1$. }
\label{fig:Btracks}
\end{figure}

Instead of the total bulge mass, Figures \ref{fig:BoverT_evol} and \ref{fig:Btracks} show how the bulge fraction of the simulated galaxies evolves with redshift. Figure \ref{fig:BoverT_evol} studies the (potential) correlations between B/T at $z=2$, 1, 0.5 and 0 for the whole sample, while Figure  \ref{fig:Btracks} shows some examples of redshift evolution of B/T for a subset of galaxies.
We find that at $z=2$ (left panel of Figure \ref{fig:BoverT_evol}), most galaxies (except one) are bulge-dominated according to their kinematics (they all have a relatively low mass, however, so that the measures are noisy). Many changes happen between $z=2$ and $z=1$, and by $z=1$, the whole range of B/T is explored, from pure bulges to pure disks, with a few cases of mergers and of disks undergoing a phase of violent instabilities (see the i-band surface brightness maps in Figure \ref{fig:evol_maps} and in Appendix \ref{app:sample}). Interestingly enough, no correlation is found between the morphology at $z=1$ and 0: pure bulges and pure disks at $z=1$ can end up with the same bulge content at $z=0$. This is also illustrated by Figure  \ref{fig:Btracks} that shows the evolution of B/T for ten galaxies ending-up with a similar bulge fraction at $z=0$, and that shows the large diversity of possible evolutionary tracks.

\begin{figure}
\centering 
\includegraphics[width=0.4\textwidth]{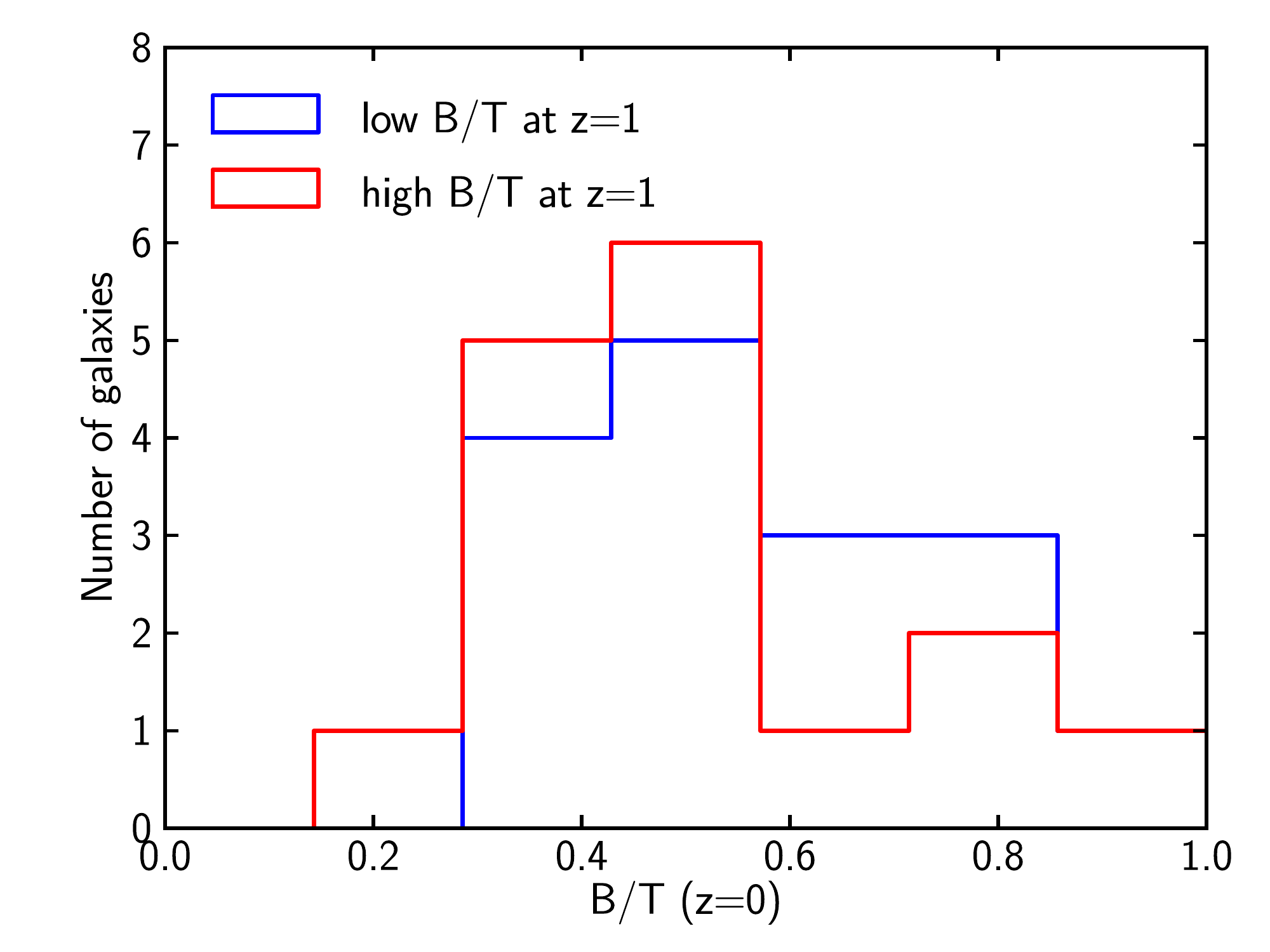}
\caption{Distribution of the final values of the  (bulge+bar)-to-total ratio for galaxies having at $z=1$ either a low or a high B/T (the distinction between low and high B/T is made with respect to the median B/T at $z=1$). Both bulge-dominated and disk-dominated $z=1$ galaxies have a large range of $z=0$ descendants. \hspace{6cm} } 
\label{fig:histoB}
\end{figure}

Finally, we divide the $z=1$ galaxies into bulge-dominated and disk-dominated samples, using the median B/T at that redshift (equal to 0.54) as the limit between the samples. For each sample, we show in Figure \ref{fig:histoB} the range of final values of B/T reached at $z=0$. Both bulge- and disk-dominated $z=1$ galaxies have a large (and similar) range of $z=0$ morphologies. We might even see a slight anti-correlation between the bulge content at $z=1$ and 0, with $z=1$ disk-dominated galaxies tending to be more bulge-dominated at $z=0$.

This is also related to the fact that pure disks at $z=1$ do not remain bulgeless until $z=0$: either a bar or a bulge grows in between. These pure disks are indeed very unstable, and even a small perturbation can destroy them. We find for instance the case of a galaxy that is bulgeless until $z=0.2$, when a fly-by happens and triggers bulge formation (this is the last galaxy of Figure \ref{fig:AppA8}).  This is consistent with the studies of minor mergers performed by \cite{Cox2008}, where bulges are shown to stabilize disks, and to suppress merger-driven inflows and associated star formation, so that bulgeless galaxies are much more fragile.

In contrast to the high redshift results, there is more correlation between B/T at $z=0.5$ and $z=0$ (right panel of  Figure \ref{fig:BoverT_evol}). This correlation is actually closer to a bimodal behaviour since  galaxies appear to be separated into two different sequences. For one sequence, the bulge-bar content decreases between $z=0.5$ and 0, these are the galaxies that have no (or nearly no) bar at $z=0$ (blue circles in Figure \ref{fig:BoverT_evol}). The galaxies on the other sequence show an increase of their bulge-bar content, and these are galaxies having a substantial bar at $z=0$ (yellow to red circles). This suggest a scenario for which, in galaxies without a bar, the bulge is in place by $z=1$--0.5, and some important disk growth happens at $z<0.5$. By contrast, in galaxies that develop a bar, there is some bar growth between $z=0.5$ and 0, accompanied with late bulge formation.

A visual inspection of the face-on images in Appendix \ref{app:sample} shows that bars and spiral arms are present in a few galaxies at $z=1$, but are rare. They become much more common at $z=0.5$.


To summarize this Section:
\begin{itemize}
\item simulated galaxies tend to follow the observed scaling relations between stellar mass, SFR, gas fraction and sizes. One exception could be a too strong gas consumption between $z=2$ and 1, leading to too low SFRs and gas fractions at $z=1$
\item there is no correlation between the morphology at $z=1$ and at $z=0$, and there is a whole range of possibilities for the  $z=1$ progenitors of spirals galaxies
\item the morphology at $z=0.5$ is much closer to the final morphology, with spiral arms and bars being mostly in place at $z=0.5$
\item the main epoch for bulge formation seems to be around $z=1$, and we find some late bulge growth that accompanies bar formation at $z<0.5$
\end{itemize}

\section{Formation histories of the most disk-dominated galaxies}
In this Section, we discuss the merger and accretion histories leading to the formation of the most disk-dominated galaxies of our sample. The reason to focus on these galaxies is that they are usually rare in cosmological simulations, and we wish to understand what is special about them.
We use GALFIT photometric decomposition throughout this Section to classify morphology because it is the only way we can discriminate between bulge and bar. We select galaxies with B/T$<0.3$ at $z=0$, which is usually used as a threshold for identifying galaxies of Hubble types later than Sb, and corresponds to 16 galaxies in our sample.

\subsection{Method}
We fix a boundary, which is a sphere centered on the main galaxy, and with a radius equal to $1.5\times R_{25}(z=0)$. We follow the inflow of gas and stars through this boundary with a time resolution of 37.5 Myr. We compute both the mass and the angular momentum of the inflowing material. We discriminate between particles belonging to a satellite (that then correspond to mergers or fly-bys) and diffuse accretion of gas.

We also compute this inflow only for particles that end-up within the optical radius at $z=0$. This allows to discriminate fly-bys (where a satellite passes through the boundary but only leaves an insignificant amount of stars and gas in the main galaxy) and mergers (where a significant amount of mass from the satellite ends-up in the final galaxy at $z=0$).

\subsection{Merger histories}

\begin{figure}[]
\centering 
\includegraphics[width=0.45\textwidth]{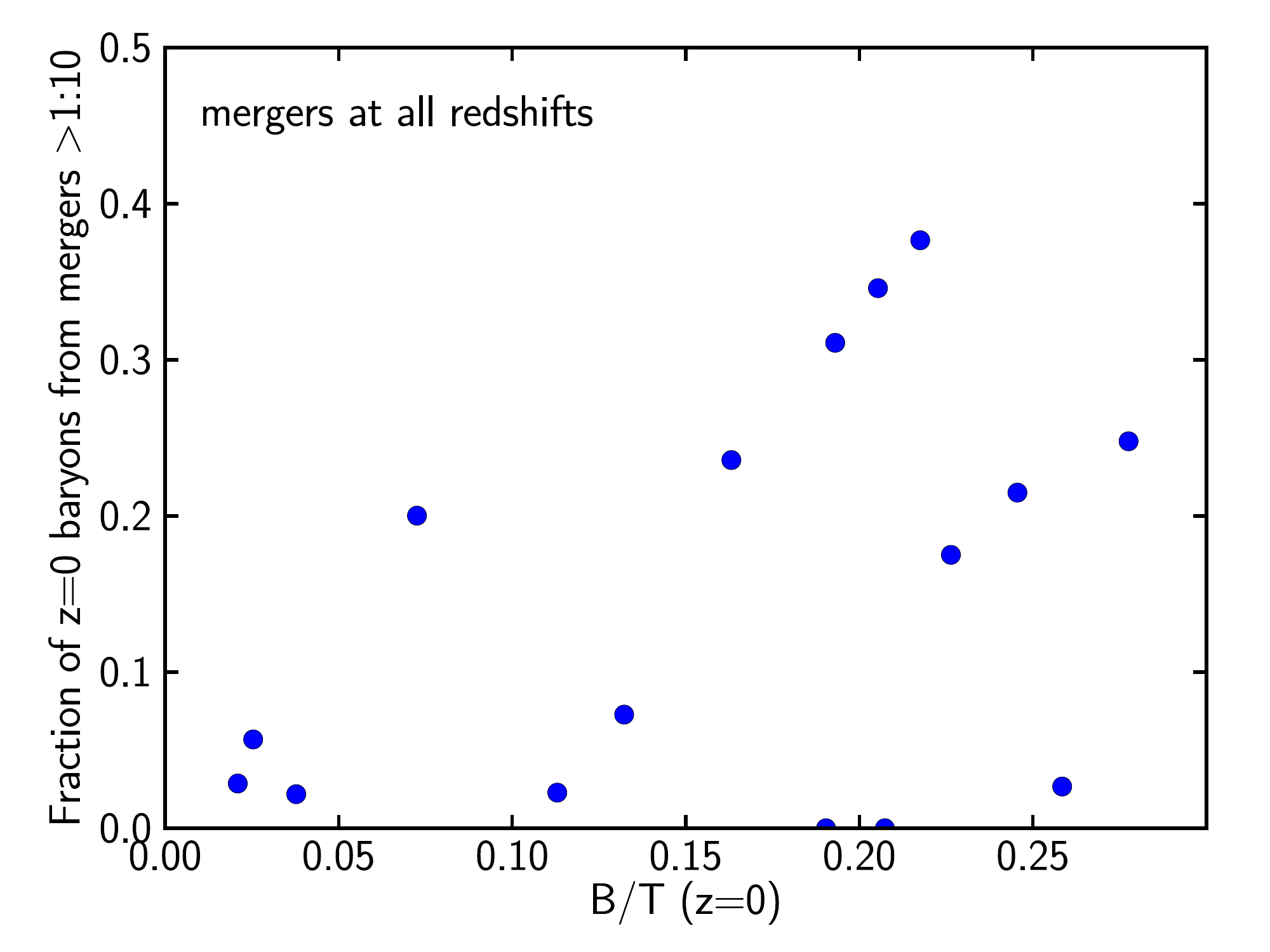}
\includegraphics[width=0.45\textwidth]{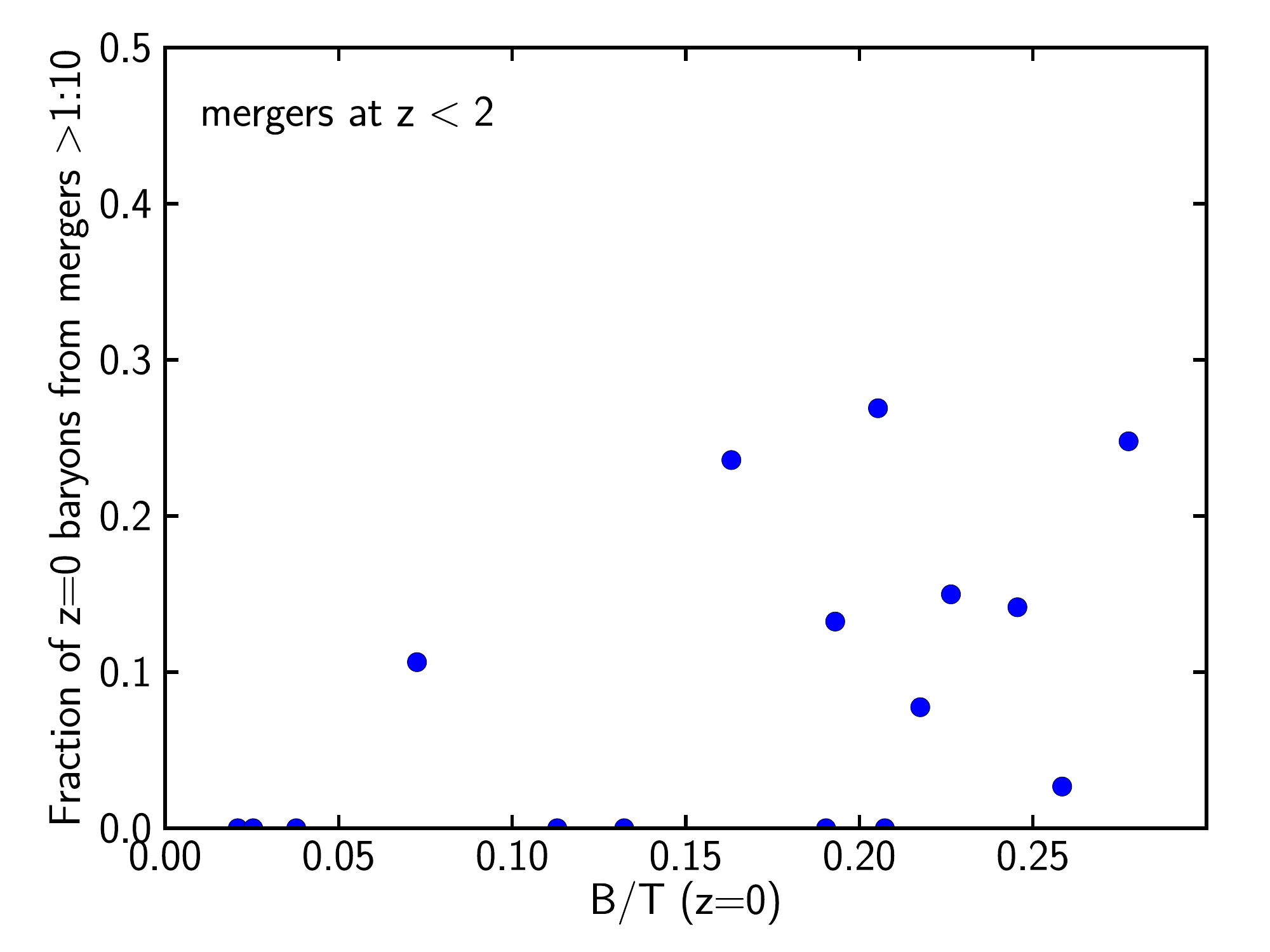}
\caption{Fraction of mass (gas and stars) that joined the galaxies through intermediate and major mergers as a function of their B/T at $z=0$. We  either consider mergers happening at all redshifts (top panel) or only at $z<2$ (bottom panel). }
\label{fig:Bulge_mu10}
\end{figure}

\begin{figure*}[]
\centering 
\includegraphics[height=5cm]{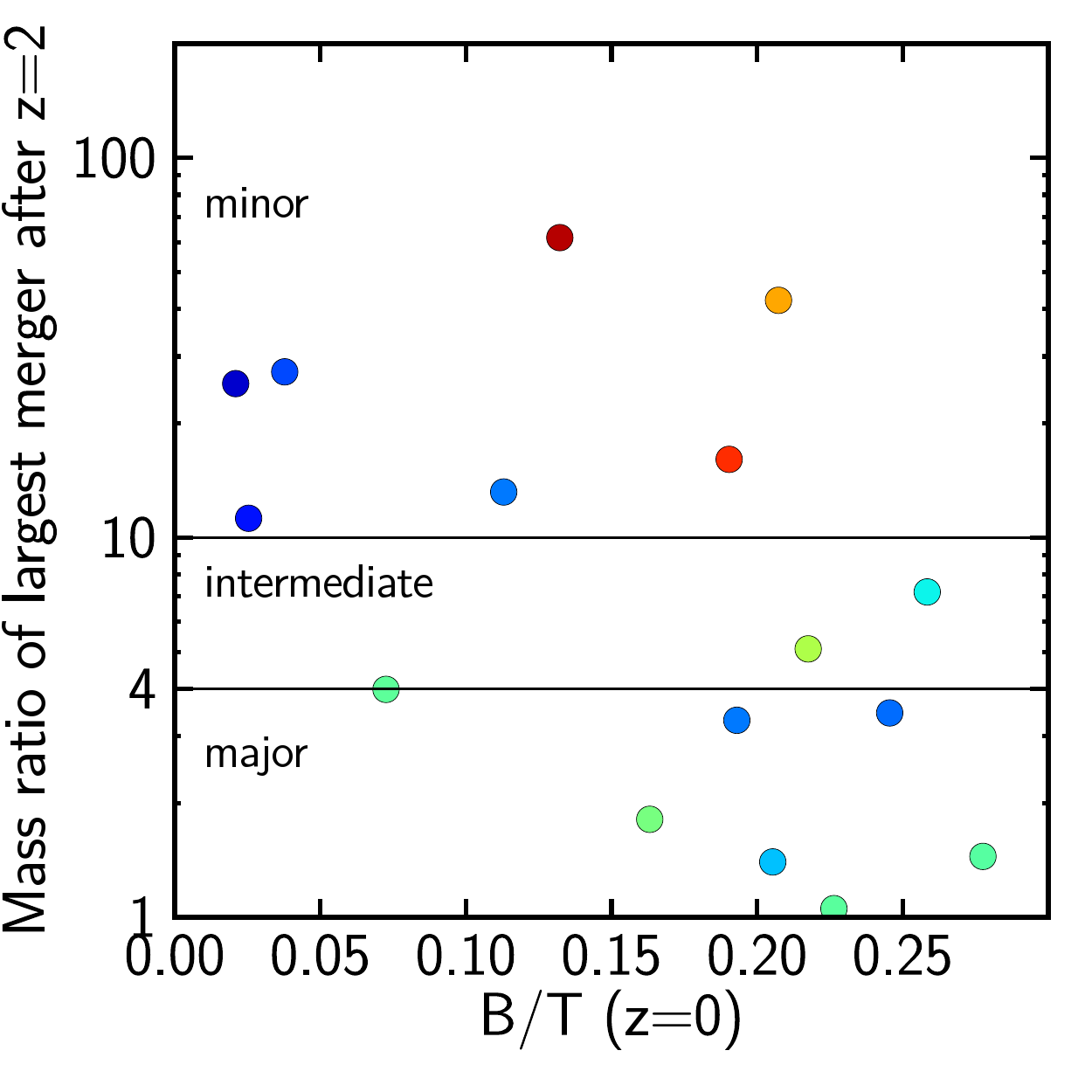}
\includegraphics[height=5cm]{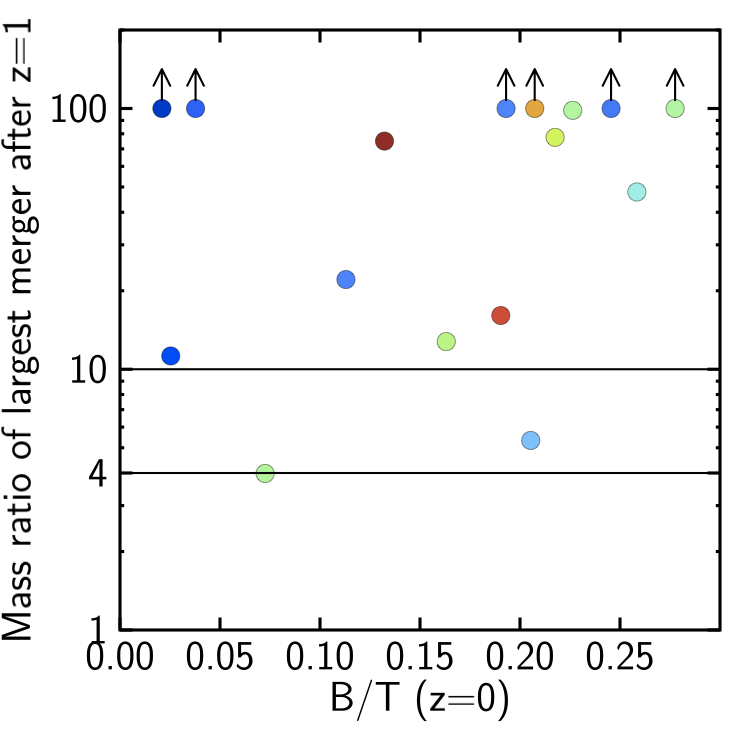}
\includegraphics[height=5cm]{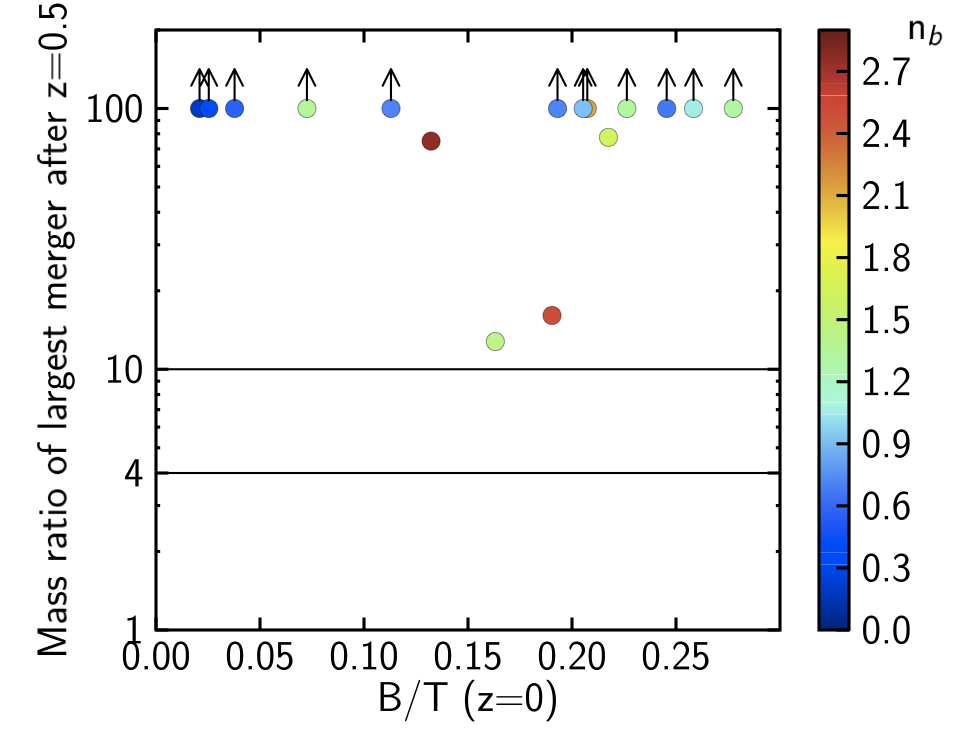}
\caption{Mass ratio of the largest merger undergone by the galaxies at $z<2$ (left panel), 1 (middle) and 0.5 (right) as a function of their B/T at $z=0$. The arrows mark lower limits. The color code indicates the S\'{e}rsic index of the bulge at $z=0$. }
\label{fig:ratio_mergers}
\end{figure*}

To study the merger histories of the simulated galaxies, we will make a distinction between major (mass ratios from 1:1 to 1:4), intermediate (1:4 to 1:10) and minor (smaller than 1:10) mergers. This distinction follows the work by \cite{Bournaud2004,Bournaud2005a} showing that intermediate mergers can significantly affect galaxies, producing remnants with spiral-like luminosity profiles but with elliptical-like kinematics.

Figure \ref{fig:Bulge_mu10} shows the contribution of major and intermediate mergers to the baryonic growth of the simulated galaxies (more exactly the fraction of gas and stars within the optical radius at $z=0$ that have been brought in by these mergers) as a function of their final bulge fraction. The top panel of this Figure takes into account mergers at all redshifts. We find an overall low contribution  (no more than 40\%) of major and intermediate mergers to the mass growth of these galaxies, and a slight trend of an increasing importance of mergers with increasing final B/T. The galaxies with the lowest $z=0$ bulge fractions all assemble only 10--20\% of their mass through major and intermediate mergers. Overall, the values we find are in rough agreement with other cosmological simulations studying the contributions of mergers and accretion to the growth of galaxies \citep{Murali2002,Semelin2005,Dekel2009,L'Huillier2011}.
The bottom panel of Figure shows only the contribution of mergers occurring at $z<2$ (this is when most of the mass growth happens). A large majority of galaxies assemble less than 20\% of their baryons through major and intermediate mergers at $z<2$, and all but one of the galaxies with B/T$<0.15$ undergo only minor mergers after $z=2$.

Figure \ref{fig:ratio_mergers} shows the mass ratio of the largest merger undergone by the galaxies after a given redshift (2, 1 and 0.5) as a function of their final B/T.  We find that some of the most bulge-dominated spirals of our sample have undergone at least one major merger between $z=2$ and $z=1$, but no galaxy has had a major merger at $z<1$. Only 2 of the 16 galaxies have at least one intermediate merger at $z<1$, and none of them has such a merger at $z<0.5$. Note also that 7 galaxies (nearly half the cases) only experience minor mergers at $z<2$. 

Focusing now on the left panel of this Figure, i.e., on the mass ratio of the largest merger after $z=2$, one stricking feature is the large range of possible final bulge fractions for a similar merger history. Galaxies undergoing only minor mergers at $z<2$ can either be nearly bulgeless or have a bulge fraction of 0.2.

The S\'{e}rsic index of the bulges at $z=0$ (shown by the color code in Figure \ref{fig:ratio_mergers}) can help to understand part of this messy situation. Three galaxies have a final bulge with a S\'{e}rsic index greater than 2, and they appear located in the same region on the left panel of Figure \ref{fig:ratio_mergers}). These three galaxies only undergo minor mergers at $z<2$, and have final bulge fractions of 0.10--0.2. They appear as a distinct population to galaxies who have also grown through minor mergers but instead have a lower B/T  by $z=0$. They are also distinct from bulge-rich galaxies with a more violent merger history. This actually suggests that mergers are not responsible for most of the bulge growth in these three galaxies, and we will get back to this in more detail in the next Section. Excluding these three galaxies, and only keeping in mind  those with a S\'{e}rsic index lower than 2, we find a clearer trend of increasing bulge fraction and increasing S\'{e}rsic index with increasing mass ratio of the largest merger undergone after $z=2$. A large scatter still remains however, and we do not find a one-to-one correlation between merger history and $z=0$ properties.

An additional fact to notice here is that the galaxies undergoing a major merger at high redshift are found to have final S\'{e}rsic indices in the intermediate range, with values between 0.6 and 1.5.  These are relatively low values for galaxies undergoing a major merger, and they do not fit in the standard picture of major mergers building classical bulges. We have studied the properties of these galaxies after the mergers (waiting for the remnant to be relaxed), and find that the bulges already have low S\'{e}rsic indices: the low $z=0$ values are not the result of bulge evolution following (for instance) disk re-growth, rather the structure of the bulge is  already in place after the merger. This might be due to the gas-rich nature of the mergers happening at high redshift.

An interesting case is a galaxy undergoing a merger with a mass ratio 1:4.1 just after $z=1$, but that ends up as extremely disk-dominated at $z=0$  (B/T=0.07, Bar/T=0.06). 
We find that the interaction increases the radius of the disk and triggers the formation of a spiral structure. 
This galaxy at $z=0$ has a very small bulge with a higher S\'{e}rsic index (n=1.3) than the other B/T$<0.1$ galaxies (for which n is between 0.2 and 0.6), and a lower bar fraction.

To summarize, we find that galaxies with B/T$<0.3$ have extremely quiet merger histories, both in terms of the mass ratio of most important merger they undergo and in terms of the fraction of baryons brought in by major and intermediate mergers. None of them have a major merger at $z<1$, and only two of them have an intermediate merger after $z=1$ (and these take place between $z=1$ and 0.5). The most disk-dominated cases (B/T$<0.1$) tend to have even quieter histories. If we exclude the three galaxies having a $z=0$ bulge with a S\'{e}rsic index greater than 2, we find a correlation (although a very weak one) between the bulge fraction at $z=0$ and the mass ratio of the largest merger undergone after $z=2$.
This suggests that mergers cannot be solely described by their mass ratio, but are highly complex events (mergers of a similar mass ratio can have very different consequences as a function of the orbit, gas fraction...). The existence of galaxies with quiet merger histories and large bulge fractions indicates that the morphology is partly determined by the gas accretion history, as we discuss in the next Section.

\subsection{Gas accretion histories}

\begin{figure*}[!ht]
\centering 
\includegraphics[height=5cm]{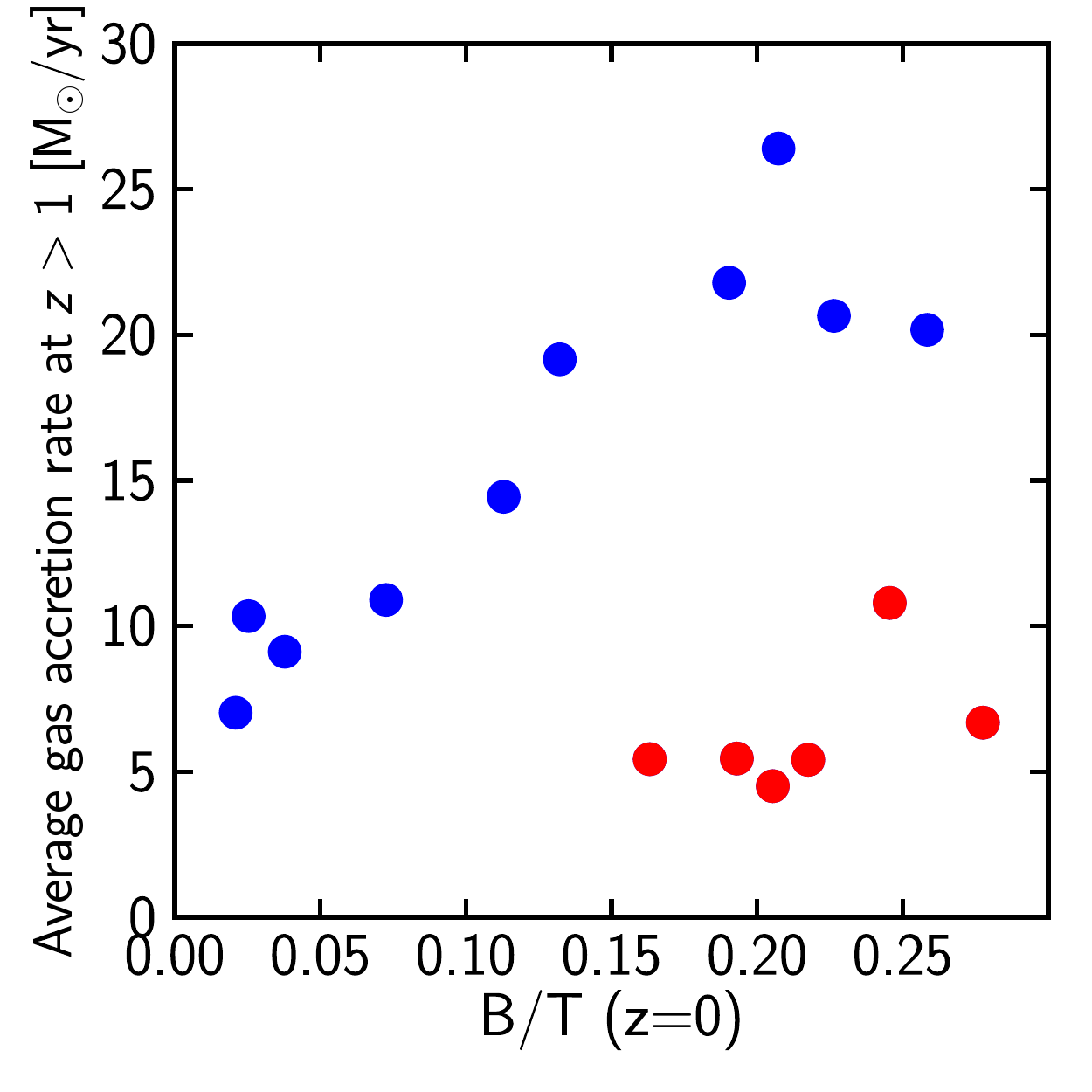}
\includegraphics[height=5cm]{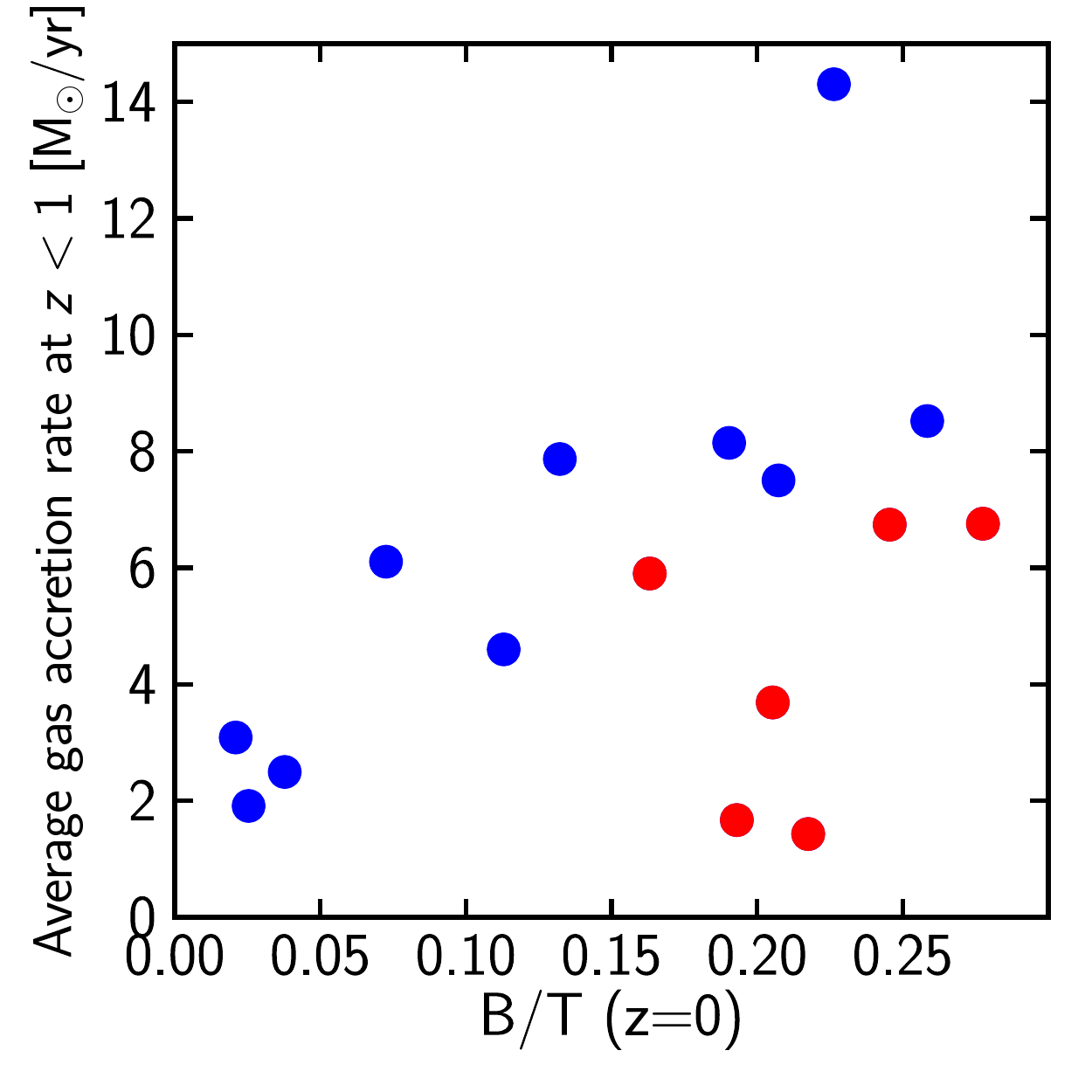}
\includegraphics[height=5cm]{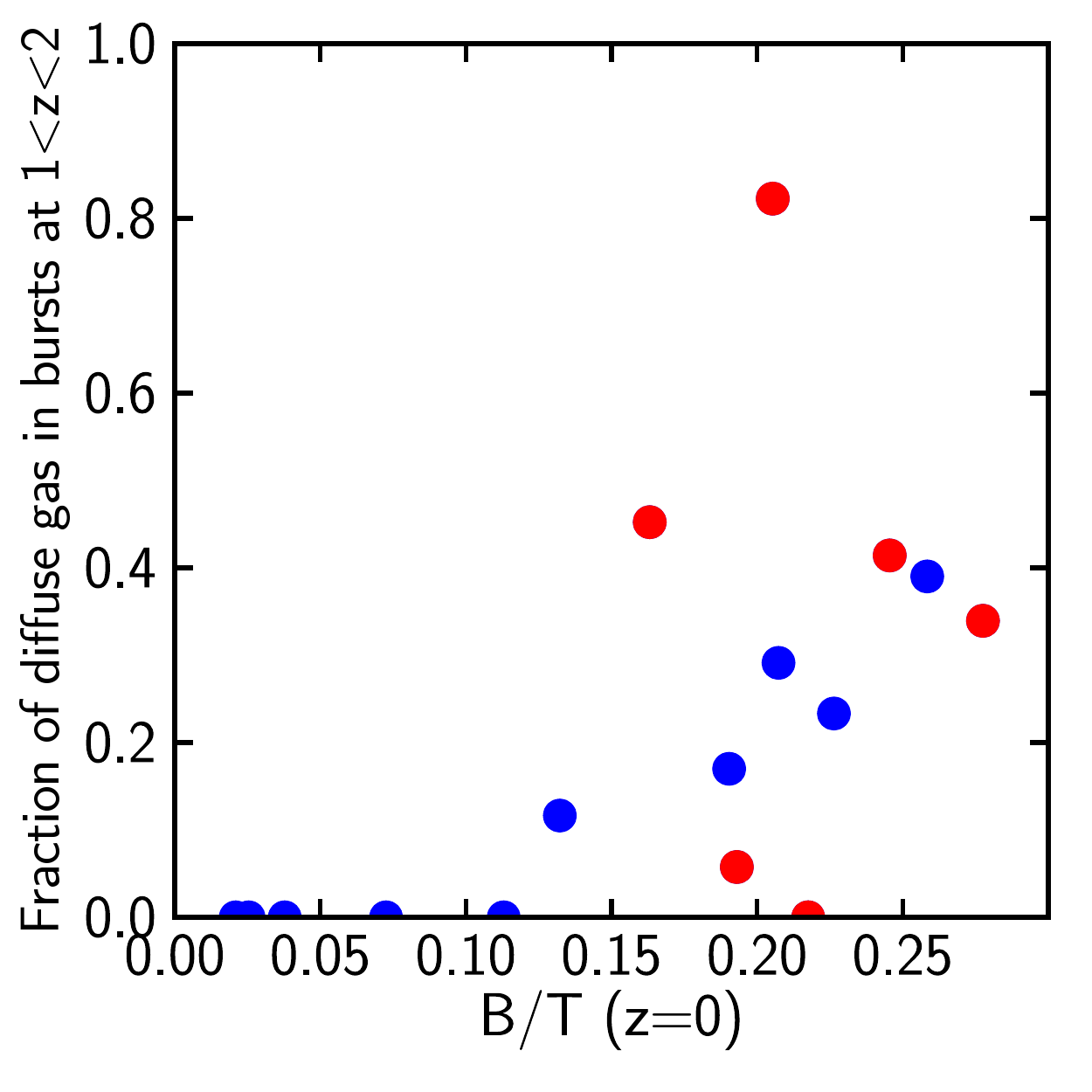}
\caption{Gas accretion rates and their relation with the bulge fraction of the simulated galaxies. The left panel shows the average gas accretion rate at high redshift ($z>1$), while the middle panel shows it for $z<1$. The right panel  shows for  $1<z<2$ the fraction of diffuse gas that is accreted in bursts, i.e. with a rate that would double (or more)  the current baryonic mass of the galaxy within 1 Gyr. The most disk-dominated galaxies do not undergo any burst of gas accretion at these redshifts. The red dots highlight the cloud of galaxies with a low gas accretion rate at $z>1$ but a high bulge fraction, that appear as outliers from the tight sequence found for the other galaxies. All these outliers are actually galaxies undergoing at least one merger with a mass ratio greater than 1:5 after $z=2$, so that most of their bulge content is probably not determined by their gas accretion history.}
\label{fig:rate_accretion}
\end{figure*}

\begin{figure*}[!ht]
\centering 
\includegraphics[height=5cm]{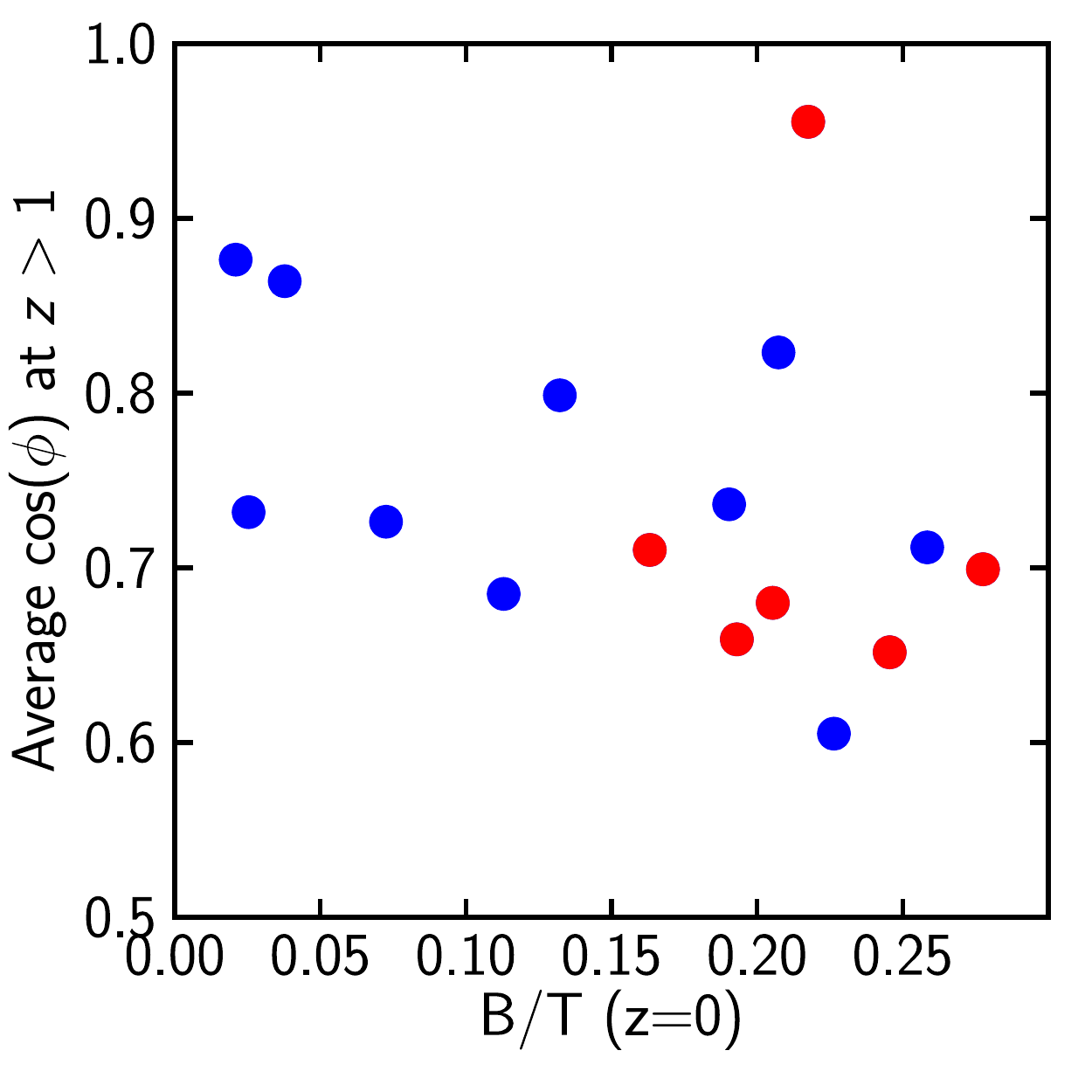}
\includegraphics[height=5cm]{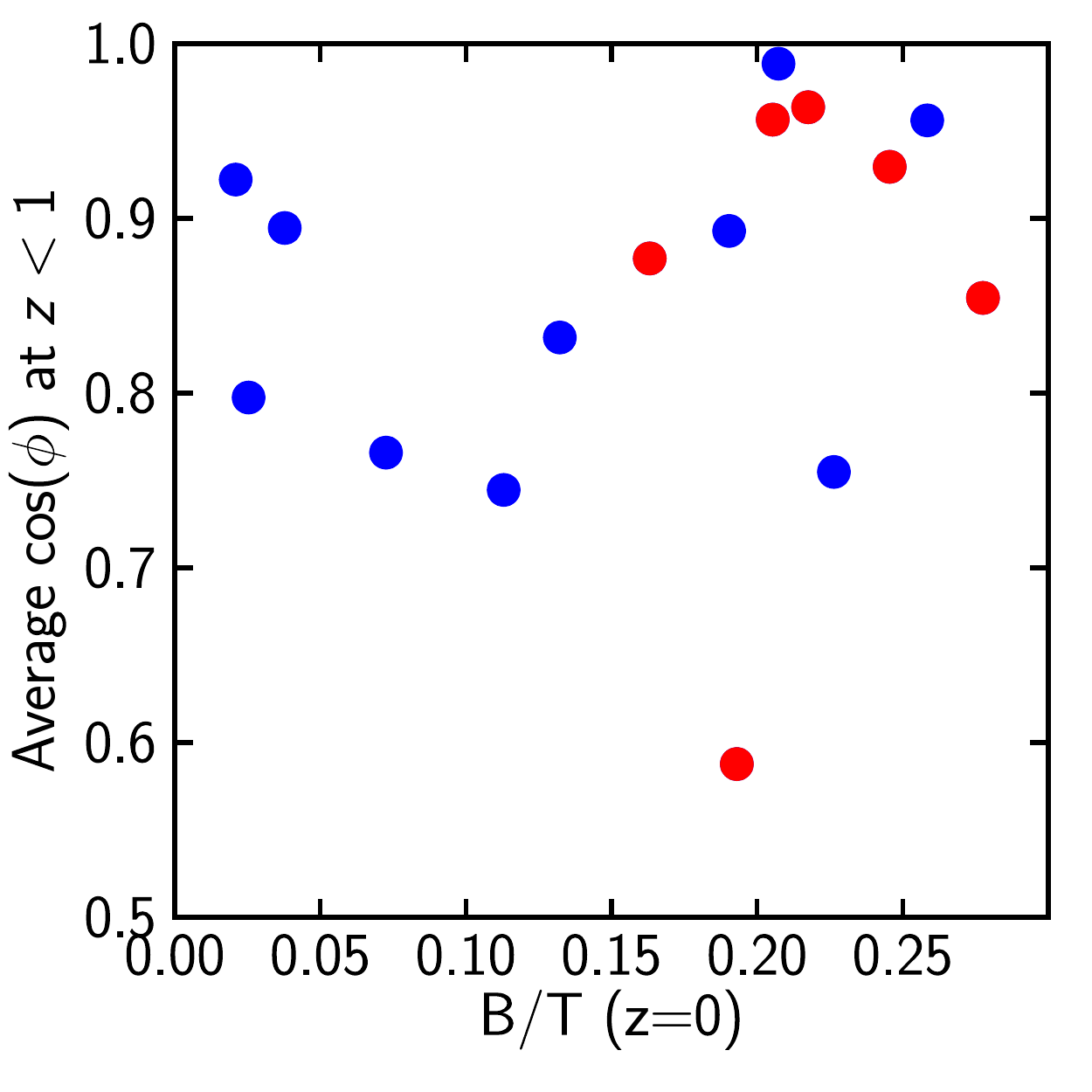}
\includegraphics[height=5cm]{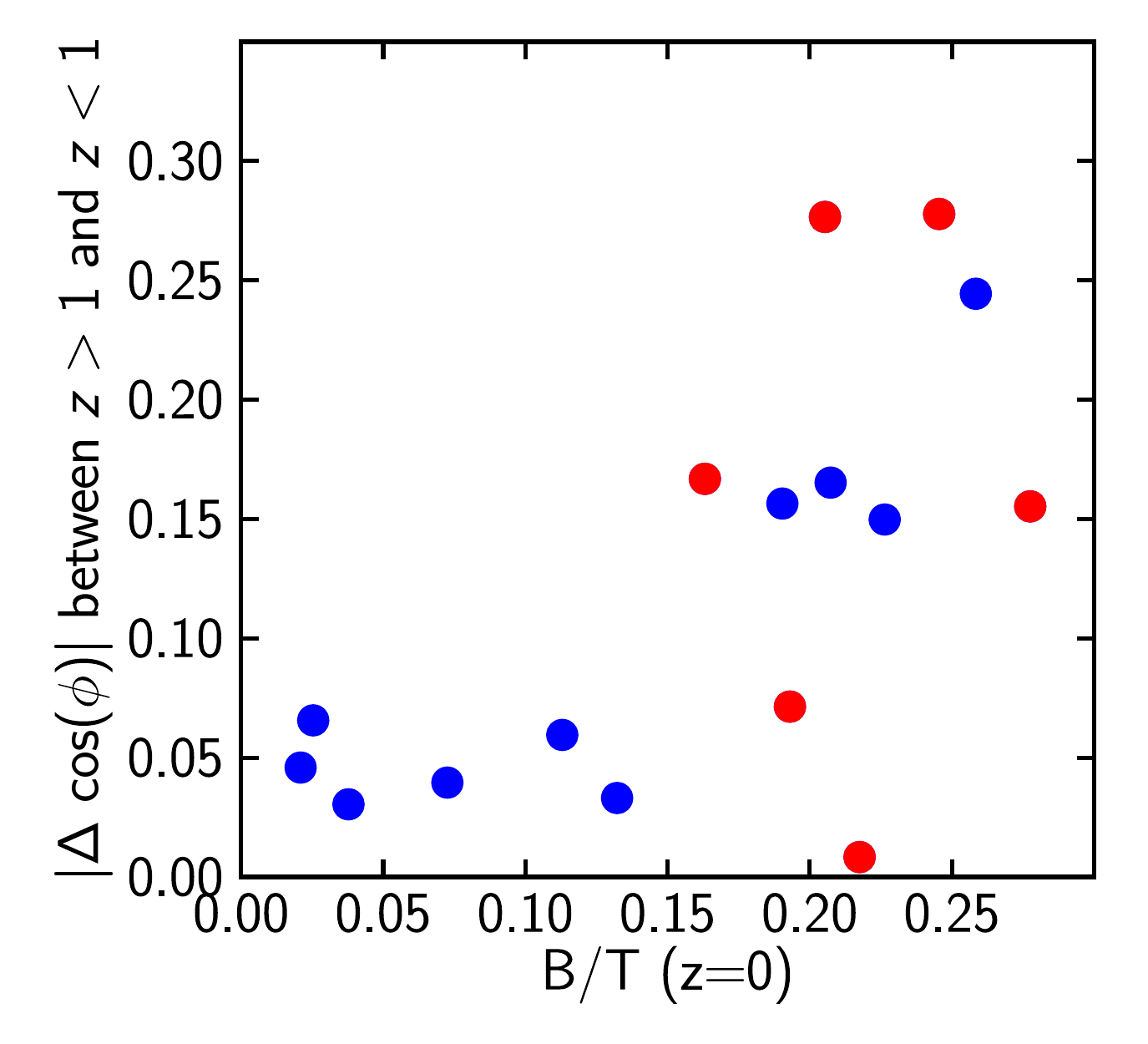}
\caption{Evolution with time of the direction of the angular momentum of the accreted gas. We study the evolution with time of  cos($\phi$), where $\phi$ is the angle between the angular momentum of the gas accreted at a given time and the final angular momentum of the galaxy. The left panel shows the mass-weighted average value of cos($\phi$) for $z>1$, the middle panels shows the mass-weighted  average cos($\phi$) for $z<1$, and the right panel shows the difference between these two values. The red dots correspond to the same galaxies as in Figure \ref{fig:rate_accretion}.}
\label{fig:angle_accretion}
\end{figure*}

We study the evolution with time of the average gas accretion rate (Figure \ref{fig:rate_accretion}) and angular momentum (Figure  \ref{fig:angle_accretion}), and look for potential correlations with their bulge content at $z=0$. Note that we only consider here the accretion of diffuse gas, and do not take into account gas brought in by satellite galaxies.

We show in Figure \ref {fig:rate_accretion} the average gas accretion rate for our simulated galaxies, separated in accretion at high ($z>1$ --- left panel) and low ($z<1$ --- middle panel) redshifts. The choice of $z=1$ as a threshold is motivated by the results of Section \ref{sec:morpho}, where we show that this is the typical epoch of bulge formation.  When studying the influence of the gas accretion rate at $z>1$ on the bulge fraction, we find that galaxies can be divided into two different populations: a tight sequence of increasing bulge fraction with increasing gas accretion rate, and a clearly distinct  cloud of galaxies with low gas accretion rates but high bulge fractions (marked with blue and red points respectively in Figures \ref{fig:rate_accretion} and \ref{fig:angle_accretion}). This cloud is actually made of galaxies that undergo at least a merger with a mass ratio greater than 1:5 after $z=2$. If mergers are responsible for most of their bulge formation, it is thus not surprising to find them as outliers on this Figure. These galaxies also have a low gas accretion rate at $z<1$ (middle panel of Figure \ref{fig:rate_accretion}), so that disk re-building is harder, thus increasing the chance for these galaxies to have a large bulge fraction at $z=0$.

Setting apart these 6 galaxies, we find a good correlation between the bulge content of a galaxy and its gas accretion history. The most disk-dominated cases have relatively low gas accretion rates, both at high and low redshift. The right panel in Figure \ref{fig:rate_accretion} studies gas accretion between $z=2$ and 1, and shows the fraction of this accretion that happens in bursts. We define a burst as a period of gas accretion at a rate that would double (or more) the baryonic mass of the galaxy in less than 1 Gyr. We find that the most disk dominated galaxies do not have any bursts of gas accretion between $z=1$ and 2, whereas galaxies that do have some bursts end up more bulge-dominated by $z=0$. Note that these bursts of gas accretion are not necessarily related to mergers: Figure \ref{fig:rate_accretion} shows cases of galaxies with a quiet merger history but a bursty gas accretion (the blue points with B/T$>0.15$ on the right panel).

Finally, we measure the angular momentum of the accreted gas, and how the direction of this angular momentum changes with time. We define $\phi$ as the angle between the angular momentum of the gas accreted at a given time (at a constant radius of $1.5\times R_{25}(z=0)$) and the final angular momentum of the stellar component. We show in Figure \ref{fig:angle_accretion} mass-weighted average values of cos($\phi$) at low and high redshift. We find that the galaxies with the highest bulge fraction tend to have a smaller cos($\phi$) at high redshift, corresponding to a larger offset in the direction of the angular momentum of the accreted gas. At $z<1$ most galaxies accrete gas with an angular momentum more closely aligned with the final $z=0$ momentum (which is not surprising since this gas is often that which gives birth to a large fraction of the galactic stellar disk).
The right panel of Figure \ref{fig:angle_accretion} shows the difference between cos($\phi$) at $z>1$ and at $z<1$. We find that the galaxies with the lowest bulge fraction undergo very little changes in cos($\phi$) as a function of time, and that large changes are linked with higher bulge fractions (note that a similar conclusion is reached by \citealp{Sales2011b}).
Indeed, for the gas which is accreted outside of the disk plane, when it finally settles into the disk it only keeps the vertical component of its angular momentum. It then settles into a smaller, more concentrated disk.

It thus seems that the formation of bulgeless galaxies requires an extremely calm gas accretion history, both in term of accretion rate and in terms of changes in the direction of the accreted angular momentum. The strongest correlation seems to be between the bulge content at $z=0$ and the gas accretion rate at high redshift.

\begin{figure*}
\centering 
\includegraphics[width=0.85\textwidth]{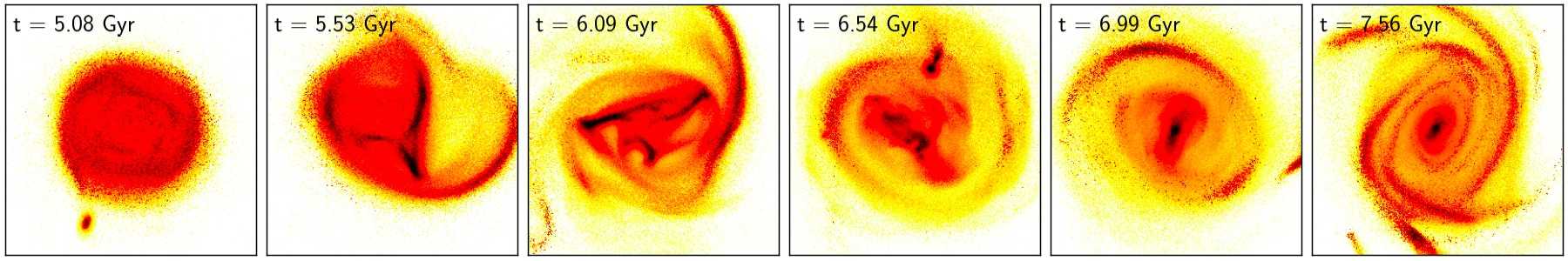}
\caption{Disk instabilities in one of the galaxies with a high gas accretion rate at $z>1$. The images show B-band surface brightness maps (30$\times$30~kpc, color scale from 18 to 25~mag~arcsec$^{-2}$) that are 500 Myrs apart, spanning a redshift range from 1.2 to 0.8. The initially smooth disk becomes unstable and fragmented, the fragments then migrate to the center of the galaxy and participate in bulge formation. }
\label{fig:unstable_disk}
\end{figure*}

To investigate the reasons why a more intense and more bursty accretion history gives birth to bulge-dominated galaxies, we study in more detail the five galaxies that have a high gas accretion rate at high redshift. Among these 5 galaxies, one has a major merger between $z=2$ and 1 so that gas accretion is probably not the main reason for the high bulge content. Among the four remaining galaxies, a visual inspection of i-band images at $z=1$ reveals that three are cases of very early bar formation, and the last one is undergoing violent disk instabilities, eventually resulting in the formation of a bulge and finally a small bar  a few 100 Myr after the end of the clumpy phase (this galaxy is shown in Figure \ref{fig:unstable_disk}).

Interestingly, face-on images of these 4 galaxies still show a bar at $z=0$. However, three of them have a low value of $\epsilon_s$ (which means a low amount of rotation in the central regions), and these are the three galaxies with a S\'{e}rsic index greater than 2. The fourth galaxy has a bulge with a lower S\'{e}rsic index ($n_b=1.1$), but shows significant rotation in the central regions.

The low values of $\epsilon_s$ could be due to a slow down of the bar with time (which has already been shown in simulations, see for instance \citealp{Combes1993,Athanassoula2003}). This slow down (and maybe partial dissolution) seems to be linked with the formation of a bulge with a high S\'{e}rsic index. This picture, although based on an extremely small number of cases, seems in agreement with observations by \cite{Durbala2008} showing that most of the local Sb-Sc galaxies with a bulge S\'{e}rsic index greater than 1.7 have a bar. Our simulations further suggest a link between old bars, formed in gas rich disks at $z \sim 1$ and bulges with a high S\'{e}rsic index.

\begin{figure}
\centering 
\includegraphics[width=0.45\textwidth]{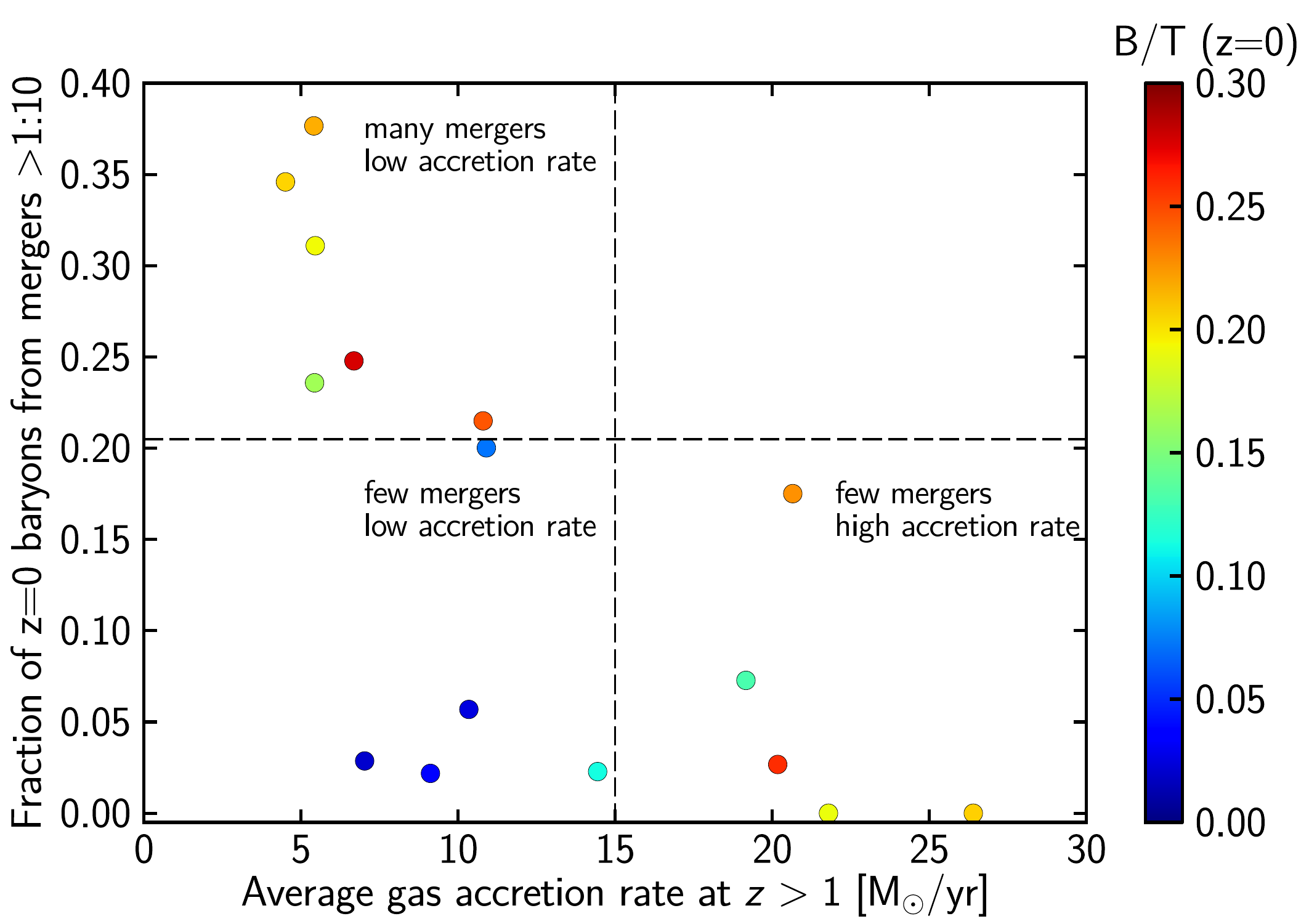}
\caption{Fraction of baryons brought in by intermediate and major mergers as a function of the average gas accretion rate at $z>1$, with the simulated galaxies color-coded as a function of their final B/T. We find that galaxies with a low B/T have in common both a quiet merger history (less than $\sim$20\% of their baryonic mass brought by mergers) and a low gas accretion rate at high redshift. Both a more violent merger history or a higher gas accretion rate give birth to more prominent bulges.}
\label{fig:mu10}
\end{figure}

To summarize this Section on the contribution of mergers and gas accretion to the bulge growth of simulated spiral galaxies (16 galaxies with B/T$<0.3$):
\begin{itemize}
\item we find a correlation between the bulge fraction at $z = 0$ and the merger history, both quantified by the fraction of baryons brought in by major and intermediate mergers and by the mass ratio of the largest merger undergone after $z = 2$, 
\item we also find a correlation between the bulge fraction at z = 0 and the gas accretion history, particularly when we consider the gas accretion rate at $z>1$
\item the most disk-dominated galaxies both have an extremely quiet merger history (with in most cases only minor mergers after $z = 2$) and an extremely quiet gas accretion history (gas accreted at a low and constant rate, with an angular momentum vector always in the same direction). By contrast, more violent merger or gas accretion histories give birth to galaxies with more prominent bulges (Figure \ref{fig:mu10} highlights this result)
\item the galaxies with the highest bulge S\'{e}rsic index at z = 0 are not those with many mergers but those with intense gas accretion at $z = 1$ and either early bar formation or other disk instabilities. 
\end{itemize}

\section{Discussion}

\subsection{Spiral galaxy formation in simulations}

Spiral galaxies formed in modern cosmological simulations usually suffer from two main issues: a too high bulge fraction, and a too high stellar mass for a given halo mass. Both issues can be found to some extent in our simulations.

\subsubsection{The bulge content of simulated galaxies}
Zoom cosmological simulations have recently started to be able to produce spiral galaxies with realistic bulge fractions, reaching values of 0.2--0.3 \citep{Agertz2011,Guedes2011,Brook2011}. However, since gathering large samples of such simulations is extremely hard, it is still unknown whether these simulations can reproduce the observed range of bulge fractions.

In our case, even if we do form some galaxies with a very low bulge fraction, and even if most of them host pseudo-bulges instead of classical ones, the distribution of our bulge-to-total ratios is significantly offset from observed distributions: we do not form enough bulgeless galaxies.

This could in some cases be a consequence of the re-simulation technique we use, and in particular of the way we populate dark matter haloes with galaxies. Indeed, we assume that each halo only hosts one galaxy, so that when a satellite is accreted, satellites of this satellite are neglected. If the incoming stellar mass was distributed not only in the accreted satellite but also in smaller dwarfs around it, these dwarfs could be more easily stripped during the infall phase, and the overall impact on the main galactic disk would be weaker. Our technique thus maximizes the impact of mergers, and we thus probably slightly overestimate bulge formation due to this process. However, many of the simulated galaxies have a quiet merger history, so that this effect alone is unlikely to account for the lack of bulgeless galaxies.

This lack is possibly aggravated by the relatively low number of violently unstable disks at $z=1$--2 in the simulations, which could instead lead us to underestimate bulge formation. Indeed, we find that $\sim$25\% of the simulated galaxies show a ring morphology at some point between $z=1$ and 2. These rings correspond to m=0 instabilities, and in approximately half of the cases, the ring itself becomes unstable and fragments (quite similarly to what is shown in Figure \ref{fig:unstable_disk}), leading to bulge formation. In some other cases however, the rings do not fragment but dissolve slowly, and we also find cases of disks remaining very smooth and stable. If the stability of these disks were artificial, we would then underestimate bulge formation, and the lack of bulgeless galaxies would become even more stricking.

These disks could be artificially too stable because of resolution effects. To test this, we have re-run one of the simulations at twice higher resolution (see the results in Appendix \ref{app:res}). In the standard resolution run, the chosen galaxy has a ring which remains stable for several Gyrs before slowly dissolving, leading to the formation of a bulgeless galaxy. At higher resolution, we find that the ring does not stay stable for such a long time, becomes unstable earlier, and a bulge grows more quickly. This gives birth however to only a very low mass bulge, with B/T $\sim$ 0.02. We also find that the surface density profiles are very similar in the standard and high resolution runs.

It thus seems that resolution alone could not be enough to explain  the stability of some disks in our simulations. While it could also be due to our gas dynamics scheme, this seems similarly unlikely since such a scheme has already been used successfully to model gas rich unstable disks  (see for instance \citealp{Bournaud2007a}). It is also probably not a consequence of our implementation of supernova feedback (that could dissolve the clumps in some cases, see \citealp {Genel2010b}), because our feedback is rather weak. In turn, it might be linked with star formation, and with the overly rapid gas consumption that the  galaxies seem to experience between $z=2$ and $z=1$. At $z=1$, the simulated galaxies only have gas fractions of 0.20--0.40, which might not be enough for strong instabilities to develop.

A direct comparison with the observed fraction of clumpy high redshift disks is however hard to perform, particularly because of the different mass ranges explored.  A large fraction of our galaxies are small at $z=1$--2, and have low stellar masses and SFRs (in the range 3--30 M$_\odot$/yr), which would make them hard to observe at such high redshifts.

\subsubsection{The relation between stellar mass and halo mass}

In addition to forming galaxies with overly massive bulges, simulations also usually struggle to reproduce the observed relation between stellar mass and halo mass. 
Observationally, direct measures can be obtained from weak lensing \citep{Mandelbaum2006, VanUitert2011} or satellite kinematics \citep{More2010}.  Many recent studies have also used abundance matching, a more indirect technique where the observed stellar masses are associated with dark matter halos extracted from simulations by assuming that the most massive galaxies are assigned to the most massive (sub)halos \citep{Guo2010a,Moster2010a}. \cite{Behroozi2010} discuss the various uncertainties associated with this technique, arising mostly from errors in the stellar masses (mostly linked with stellar population modelling), with additional errors coming from the scatter in assigning galaxies to halos. Overall, most techniques give similar results within a factor of 2 \citep{More2010}, with possible differences as a function of galaxy Hubble type and color \citep{Mandelbaum2006, VanUitert2011}, and with additional variations between central and satellite galaxies \citep{Neistein2011}. It is interesting to note that at fixed stellar mass spiral galaxies appear to be living in lower mass halos compared to ellipticals \citep{VanUitert2011}. \cite{Comeron2011b} also point out that the stellar masses of spirals could be underestimated by 10 to 50 \% if the same stellar mass-to-light ratio is used for thin and thick disks, so that the stellar-to-halo mass ratio could be further decreased for these galaxies.

However the stellar mass in simulations seems in most cases to be a few times too high (see for instance the compilation presented in \citealp{Guo2010a}). Our simulations suffer from the same shortcoming: depending on the simulated galaxy, the fraction of the cosmic baryons found in stars (i.e., the galaxy formation efficiency) varies between 0.38 and 0.88, with an average of 0.62. This is 3 times higher than the observed fraction of $\sim 0.2$ (e.g., \citealp{Behroozi2010}). Observational uncertainties probably cannot account for such a discrepancy. Note however that the agreement between simulations and observations is usually improved when the stellar masses are measured in the simulations in the same way as observers do, i.e. when stellar masses are estimated from the spectral energy distribution of the simulated galaxies \citep{Oh2011,Guedes2011}.

Overly high stellar masses could be due to the way we take the cosmic accretion of baryons into account, and could happen if gas accretion were overestimated. Indeed, our simulation technique assumes that filaments contain the cosmic fraction of baryons, which might not be true, and one of the consequences of the sticky particle scheme is that we ignore hot halos and assume all gas is accreted cold. If less cold gas reached the galactic disk, this would help decrease the stellar content of the simulated galaxies.
However, the galaxy formation efficiencies that we find are extremely similar to what simulations using other gas modelling techniques find (e.g. with AMR, \citealp{Agertz2011}, or with  SPH, \citealp{Governato2007a,Scannapieco2009}, see also a comparison of different codes in \citealp{Scannapieco2011b}).
This discrepancy suggest a much deeper issue, found in most simulations, whatever the technique used. 

An exception has recently been published by \cite{Guedes2011}, whose disk galaxy does sit on the observed stellar-to-halo mass relation. However, they use a way of deriving stellar mass slightly different from what other simulations have used, and neglect metal-dependent cooling at high temperatures, thus probably maximizing the effect of supernovae. They argue that their success is due to a high star formation threshold, which they set at 5 atoms cm$^{-3}$ (this is 5 times higher than the threshold we use but it's unsure if this could be enough to cause the differences between the simulations). They present only one case however, and it would be interesting to see if their recipes also produce realistic galaxies over a wider range of masses and formation histories. Regardless, a high star formation threshold is probably not a universal solution by itself, since  \cite{Avila-Reese2011a} used a similarly high value but still find it hard to match the observed stellar to halo mass relation.

\subsubsection{Modelling the physics of baryons}

The fact that simulations struggle to reproduce the stellar masses and bulge fractions of observed galaxies suggests some deep issues with the current recipes used to model the physics of baryons. 

AGN feedback is probably not relevant for our study, given the halo mass range on which we focus, and given that black holes are probably not massive enough in late-type galaxies. In turn, star formation and supernova feedback parameters likely play a  much more important role. While we already mentioned possible effects of the chosen star formation threshold, the star formation efficiency itself strongly impacts the morphology of the simulated galaxies \citep{Agertz2011}. Calibrating this efficiency is difficult, especially since it could vary with redshift (as suggested by \citealp{Agertz2011}),  which would be the case if it depended on metallicity (\citealp{Krumholz2011} showed that taking into account the influence of metallicity suppresses star formation, especially in small galaxies at high redshift).
This calibration requires simulations at much higher resolution than what can currently be achieved in cosmological context, following the structure of the GMCs and capturing supersonic turbulence. Such simulations have indeed produced results differing significantly from lower resolution ones, showing for instance an increased star formation efficiency during mergers \citep{Teyssier2010}, or a different structure for the merger remnants \citep{Bournaud2011}.

Supernova feedback has been shown to be a key ingredient to produce realistic galaxies \citep{Dekel1986,Scannapieco2008,Governato2010,Piontek2011}. Supernova-driven winds cannot escape the gravitational potential of massive galaxies, which makes them inefficient at directly reducing the baryon fraction in Milky-Way type galaxies at low redshift. However, some authors argue that the winds and galactic fountains are essential for redistributing angular momentum and producing bulgeless galaxies, all the more so since most of the ejected material has a low angular momentum \citep{Maller2002,Brook2011b,Brook2011}. It seems unlikely that a stronger feedback would be the solution to the  issues we encountered in our simulations, both because of the relatively low SFR, even at high redshift, in which case outflows are not very powerful, and because a significant fraction of the mass growth of the simulated galaxies happens at relatively low redshift ($z<1$).

In addition to these standard physical prescriptions, it could also be that current simulations are missing key physical ingredients like radiative transfer. Indeed, while radiation is usually thought to play an important role in regulating the fraction of baryons in low mass galaxies at high redshift (see for instance the recent simulations by \citealp{Petkova2010}), it might also be important at lower redshifts and in massive galaxies. For instance, \cite{Cantalupo2010} argues that the cooling in a galactic halo can be regulated by the photoionizing radiation from the galaxy itself that would remove the main coolants from the hot gas, so that the cooling timescale would directly depend on the SFR of the host galaxy. This could be an efficient mechanism for regulating gas cooling and galaxy formation.

\subsection{The emergence of the Hubble sequence}
Observations have revealed significant differences between galaxy populations at high redshift and the $z=0$ Hubble sequence.

Already at $z \sim 0.5$ grand-design spirals are rare and the spiral structure of disk galaxies is often more chaotic  \citep{VandenBergh2000,VandenBergh2002}. Disk galaxies are still however observed up to $z \sim 1.5$, and at fixed stellar mass their size seems to undergo little change between $z=1$ and 0 \citep{Ravindranath2004,Barden2005}. Large blue disks have also been found at $z=2$ (some with spiral arms), but these object are extremely rare, and probably correspond only to the most massive galaxies at that redshift \citep{Szomoru2011}.

Irregular and interacting galaxies become increasingly common with increasing redshift, with the population of such galaxies already significant at $z \sim 0.5$. In the Hubble Deep Field, \cite{Abraham1996} find a fraction of irregular galaxies of 40 \%, ten times more than in the Local Universe. In the UDF \cite{Elmegreen2005c} find only 31\% of spirals and 13\% of ellipticals, the other galaxies being classified as chains, doubles, tadpoles or clump clusters. While most of these disturbed galaxies are probably disks (undergoing a phase of violent instabilities), they could not fit anywhere on the standard $z=0$ Hubble sequence.

Most surveys conclude that the Hubble sequence disappears at $z>1.5$--2, and most galaxies at that redshift seem to be irregular or compact \citep{Daddi2004,Conselice2005,Papovich2005}.

Our simulations support this picture. At $z=2$, the majority of our simulated galaxies have a  stellar mass between $10^9$ and $10^{10}$  M$_{\odot}$. None of these galaxies has a spiral-like morphology: some have very thick disks, some are ellipticals and there is a large fraction of galaxies undergoing interactions. Their kinematics classifies them as dispersion-dominated systems and no clear Hubble sequence could be defined. This may be a consequence of the mass range we explore, and is consistent with observations showing that more massive galaxies are less irregular (with for instance some cases of massive spirals as observed at $=2$ by  \citealp{Szomoru2011}).
By $z=1$, a number of our simulated galaxies have acquired a more standard spiral morphology, with a few cases of galaxies with spiral arms, and some with a bar. The majority of disks however show no signs of either a bar or spiral arms. Most galaxies contain several distinct dynamical components, with in some cases bulges and disks that can be separated from their kinematics.
By $z=0.5$ bars are much more common and the final structure is more or less in place even if spiral arms often appear fragmented. Between $z=0.5$ and $z=0$, spiral arms become better defined and bars continue their growth. Note that our simulations suggest bars to be long lived structures, which will be investigated in more detail in a future paper. 

We thus confirm that a sequence of disks and ellipticals is found at $z=1$, but we find no one-to-one correspondence with $z=0$. Most galaxies in our sample acquire their final morphology at $z \sim 0.5$ (or even later if the details of bars and spiral arms are taken into account).

\subsection{Mergers and  bulge formation in Milky Way type galaxies}

We notice that the most disk-dominated galaxies of our sample share extremely quiet merger histories, with only very minor mergers after $z=2$. While this could seem quite obvious, it also contradicts recent claims that mergers could actually help disk survival by triggering central starbursts and subsequent strong outflows of low angular momentum material \citep{Brook2011b,Guedes2011}. 

Our simulated galaxies show however that galaxies with a major merger at $z=1$--2 can end up at $z=0$ with a relatively low bulge fraction of 0.15--0.3. In fact, we find that 6 out of 16 (i.e. 37\%) of simulated galaxies with B/T$<0.3$ have a major merger between $z=2$ and $z=1$. This fraction is significantly higher than the values quoted in \cite{Weinzirl2009}, where the different models tested conclude that less than 10\% of galaxies with such a bulge fraction have a major merger at $z<4$, which is a very strong constraint.
Our simulations give much weaker constraints on the merger histories of spiral galaxies, provided that major mergers happen at $z>1$. 

This result is a combination of some mergers having disky remnants (all of them have a S\'{e}rsic index lower than 2) and of a significant amount of disk growth after $z=1$. This significant disk growth at low redshift both comes from the accretion of fresh gas (although galaxies with mergers tend to accrete gas at relatively low rates--- see middle panel in Figure \ref{fig:rate_accretion}) and from gas recycling due to stellar mass loss \citep{Martig2010}.

Regarding mergers and bulge formation, merger remnants can be significantly affected by the simulation resolution : \cite{Bois2010} show that a resolution of $\sim 32$~pc (compared to our resolution of 150 pc) is required to follow the evacuation of angular momentum necessary for the formation of slow rotators, and that the effect is even stronger for mergers of gas-rich galaxies. \cite{Bournaud2011} also showed that simulations resolving the turbulent and fragmented nature of the ISM produced merger remnants that were more compact and with higher S\'{e}rsic indices. The low S\'{e}rsic indices of the bulges formed in our simulations could thus be a consequence of their limited resolution.

Independent to the detailed properties of the bulges, we find mergers to make a significant contribution to bulge formation at high redshift, all the more so since violent disk instabilities are relatively rare in our simulations. 
Whether this is realistic or not can be tested with studies of thick disks. Both the mass, shape and kinematics of thick disks are indeed tracers of their formation mechanism \citep{Bournaud2009, Sales2009}. The fact that the edges of thick disks are not flared (see for instance \citealp{Comeron2011b}) was used by \cite{Bournaud2009} to show that they most likely did not form because of minor mergers, but rather are the relics of gas-rich clumpy disks at high redshift.
A future paper will be dedicated to thick disks in our simulations, and if they are flared, this would indicate that we probably overestimate the importance of mergers for bulge formation at high redshift.

\vspace{2cm}

\section{Conclusion}
We present the results of a series of cosmological simulations targetting 33 galaxies in an isolated environment with halo masses between  $2.7 \times 10^{11} $ and $2 \times 10^{12}$ M$_{\odot}$ . We study the evolution from $z=2$ to $z=0$ of their morphology with techniques based both on their kinematics and on 2D  decomposition with GALFIT. 

We find at $z=0$ galaxies with a large range of Hubble types, from bulgeless disks to bulge-dominated galaxies, although the fraction of bulgeless galaxies may be lower than the observed fraction. Most of the galaxies host pseudo-bulges, with a S\'{e}rsic index lower than 2, and 70\% of them have a bar.

At $z=2$, the simulated galaxies are very perturbed, and if there is a disk, it is usually thick, and sometimes unstable and clumpy.  By contrast, it is much easier to identify disks and spheroids at $z=1$, even if spiral structure is usually absent, and bars are rare (except for galaxies with very high gas accretion rates at early times, in which case a bar can already have formed by $z=1$). We find that spiral arms and bars are usually in place and are much more common by $z=0.5$.

Even if a Hubble Sequence could be defined at $z=1$, we do not find any correlation between the morphology at $z=1$ and at $z=0$,  with a whole range of possibilities for the $z=1$ progenitors of spiral galaxies (interacting galaxies, bulge-dominated systems, pure disks, unstable disks...). There is a much better morphological correlation between $z=0.5$ and 0.

Focussing on the formation histories of galaxies with B/T$<0.3$ (typically Sb and later types, and corresponding to about half the galaxies in our sample), we find a correlation between the bulge fraction at $z=0$ and the merger history, both in terms of the mass ratio of the largest merger undergone after $z=2$ and in terms of the fraction of baryons brought in by major and intermediate mergers. We also find a correlation with the gas accretion rate at $z>1$. We note that the most disk-dominated of these galaxies both have an extremely quiet merger history (with in most cases only minor mergers after $z=2$) and an extremely quiet gas accretion history: they accrete their gas at a low and constant rate, with an angular momentum vector always in the same direction. 

By contrast, more violent merger or gas accretion histories give birth to galaxies with more prominent bulges. Interestingly, $\sim 40$\% of the galaxies with B/T$<0.3$ undergo a major merger between $z=2$ and $z=1$ (none of these galaxies has a major merger at $z<1$). Their disk-dominated nature at $z=0$ is the consequence of both the relatively disky nature of the merger remnants and a significant disk growth at $z<1$.

Finally, we find that the galaxies with the highest bulge S\'{e}rsic index at $z=0$ are not those with many mergers but those with intense gas accretion at $z=1$ and either early bar formation or other disk instabilities. The relation between old bars and high S\'{e}rsic index bulges will be investigated in a  future paper, but it seems consistent with observations of local Sb-Sc galaxies by \cite{Durbala2008} showing that most galaxies with a bulge S\'{e}rsic index greater than 1.7 also have a bar.

While our simulations are successful in producing galaxies with a large range of Hubble types, and even a few nearly bulgeless galaxies, we still find that we probably overestimate bulge formation as well as the stellar mass at fixed halo mass. These issues, that are common to many cosmological simulations, are very likely related to the modelling of the physics of baryons. A better calibration of the efficiency of star formation (with a possible dependency on metallicity, see \citealp{Krumholz2011}), and a better treatment of stellar feedback (from supernovae explosions but also for instance the radiation pressure from young massive stars) are promising paths to follow in order to resolve these issues. In particular, the formation of disk-dominated galaxies seems to be easier in simulations where star formation is significantly delayed \citep{Scannapieco2011b}. This effect should be convolved with our study to get a complete picture on how galaxies get their morphology.

\acknowledgements
We thank the referee for valuable comments and Mark Sargent for useful discussions. MM and DC acknowledge funding from a QEII Fellowship awarded by the Australian government. This work has been partly supported by ISF grant 6/08, by GIF grant G-1052-104.7/2009, and by a DIP grant. 
Simulations were carried out at CEA/CCRT computing center.

\appendix
\section{Resolution study} \label{app:res}

At high redshift, a number of our simulated galaxies show disks that remain extremely smooth, with some cases of rings that do not fragment. Rings could be a frequent evolutionary stage in high redshift galaxies \citep{Genzel2008, Aumer2010}, but they are usually thought to fragment into clumps, that can then migrate to the center and form a bulge. Since we are interested in bulge formation, it is particularly important to make sure that our disks are not kept artificially too stable, in which case we would strongly underestimate the bulge content of our galaxies.

The non-fragmentation of our rings is probably not an artifact of the sticky-particle technique, which has already been successfully used to model gravitationally unstable disks (see for instance \citealp{Bournaud2007a}). It could however be due to the limited resolution of our simulations. To test this, we have re-run one of the simulations with a twice higher spatial resolution (corresponding to 65 pc), and a mass resolution increased by a factor of 6 (the mass of a gas particle is then 2500 M$_{\odot}$). 

\begin{figure*}
\centering 
\includegraphics[height=5cm]{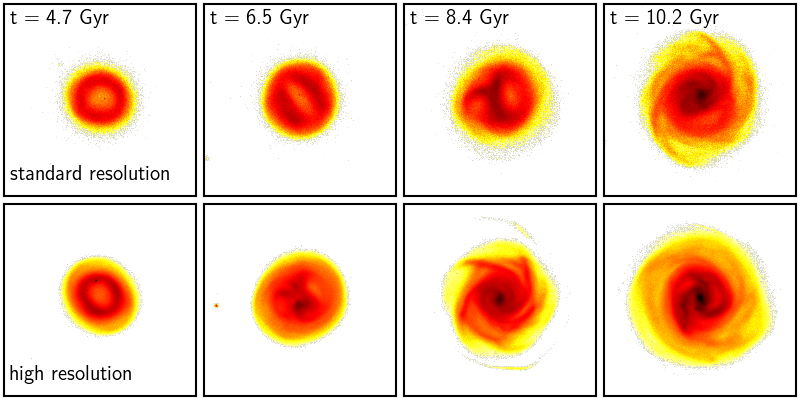}
\includegraphics[height=5cm]{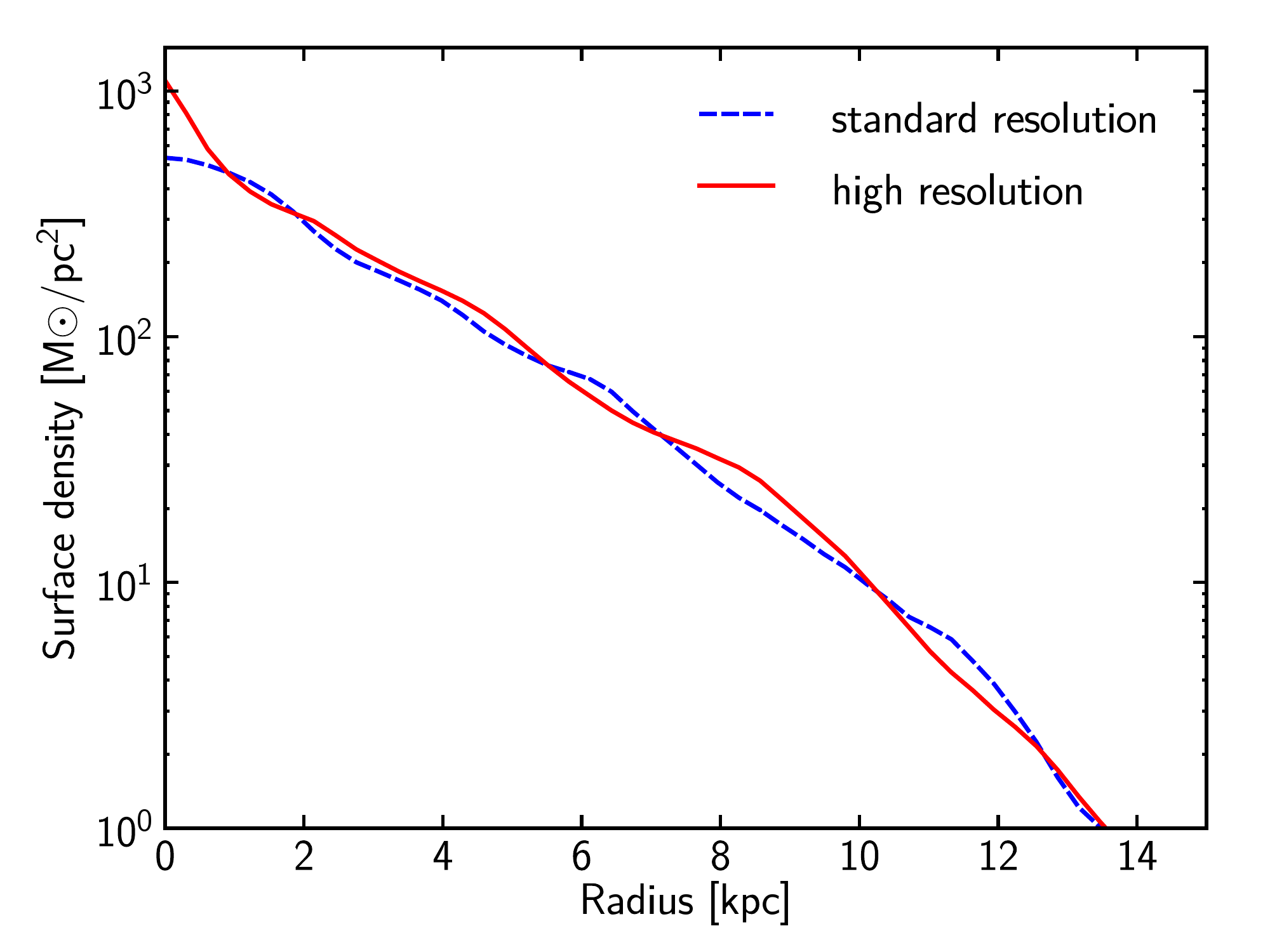}
\caption{Comparison of the stellar structure of a simulated galaxy in a standard run and at twice higher resolution. The left panels compare stellar surface density maps (30 x 30 kpc, identical colormap in all cases) for different times of the simulations, while the right panel shows the stellar surface density profile at t=10.2 Gyr. At both resolutions a ring structure first forms, but this structure is less stable at high resolution, and a bulge forms much earlier. In the end however, at both resolutions galaxies have a very similar structure, both in terms of visual morphology and stellar density profile.}
\label{Compar_resolution}
\end{figure*}

We compare the structure of the simulated galaxy in the standard run and in the high resolution run. Figure \ref{Compar_resolution} shows that in both cases a stellar ring forms at high redshift. This ring is more unstable at higher resolution, it dissolves more quickly and a bulge is formed earlier.
However, if the details of the time evolution are different, the final galaxies are very similar at both resolutions. At t=10.2 Gyr, both galaxies have acquired a spiral structure, they have similar radii and stellar masses: the total stellar mass within the inner 8 kpc is $2.17 \times 10^{10}$ M$_{\odot}$ in the standard run, and $2.28 \times 10^{10}$ M$_{\odot}$ at higher resolution. Furthermore, the surface density profiles are very similar, with the only difference being that the low resolution galaxy is bulgeless, while a small bulge has formed at high resolution. The bulge fraction at high resolution is however only 0.02.

This test thus confirms that the stability of the rings can be an artifact of the resolution we standardly use. Increasing the resolution however only leads to a minor increase of the bulge fraction, so that most of our results should still be valid.

\section{The sample of simulated galaxies}\label{app:sample}
\begin{figure*}
\centering 
\includegraphics[width=0.9\textwidth]{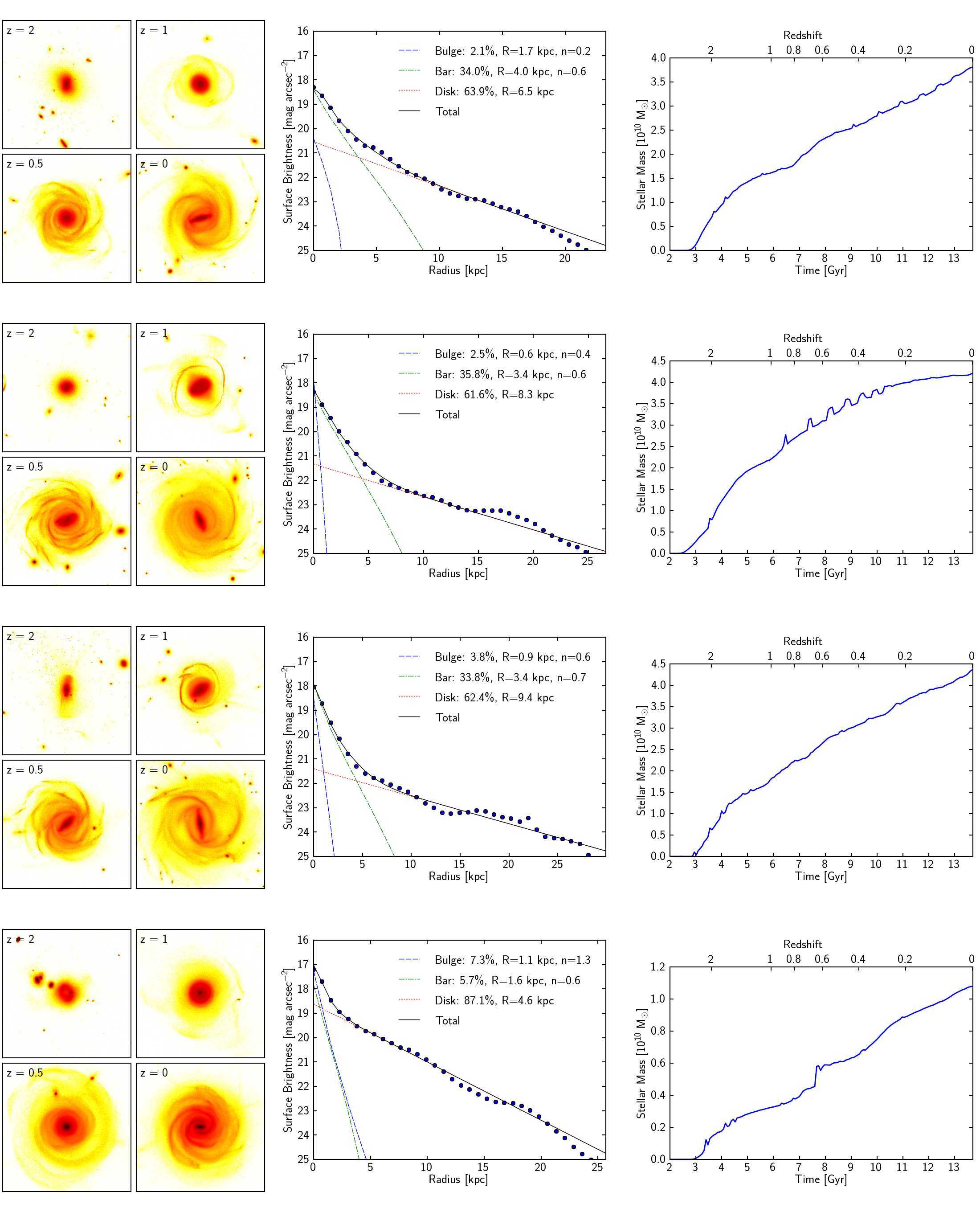}
\caption{Galaxies with a bulge fraction from 0.02 to 0.10}
\label{fig:AppA1}
\end{figure*}

\begin{figure*}
\centering 
\includegraphics[width=0.9\textwidth]{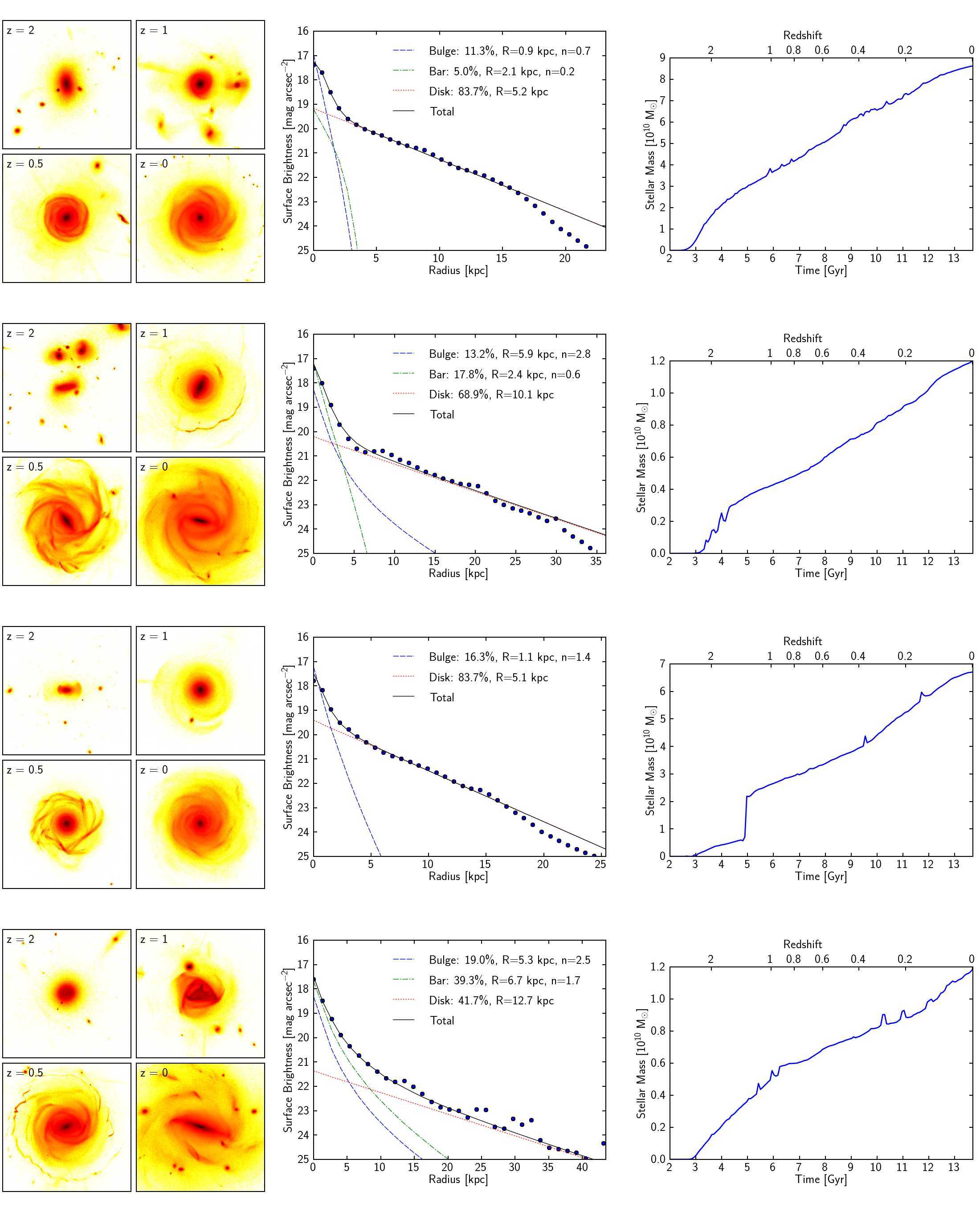}
\caption{Galaxies with a bulge fraction from 0.10 to 0.19}
\label{fig:AppA2}
\end{figure*}

\begin{figure*}
\centering 
\includegraphics[width=0.9\textwidth]{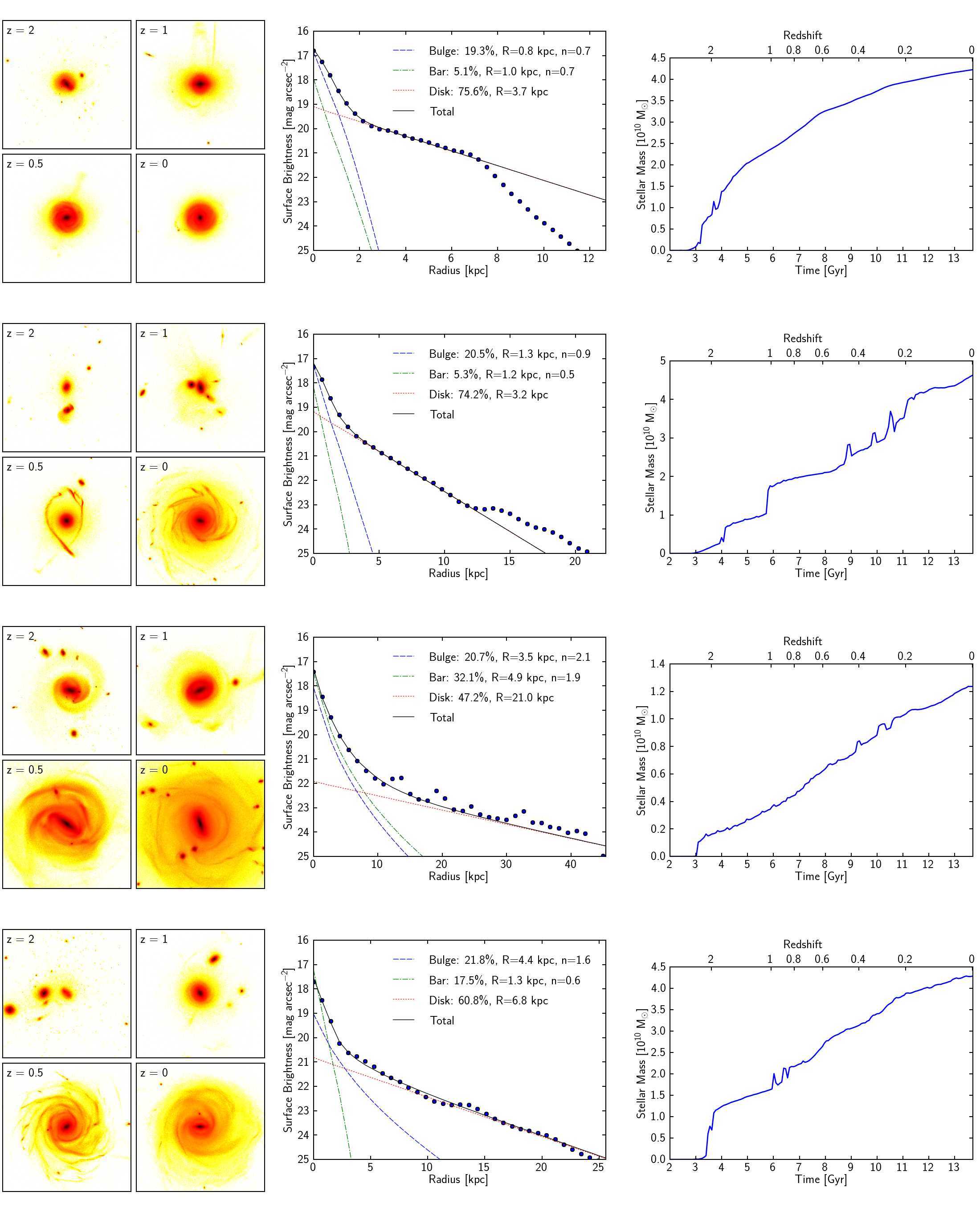}
\caption{Galaxies with a bulge fraction from 0.19 to 0.22}
\label{fig:AppA3}
\end{figure*}

\begin{figure*}
\centering 
\includegraphics[width=0.9\textwidth]{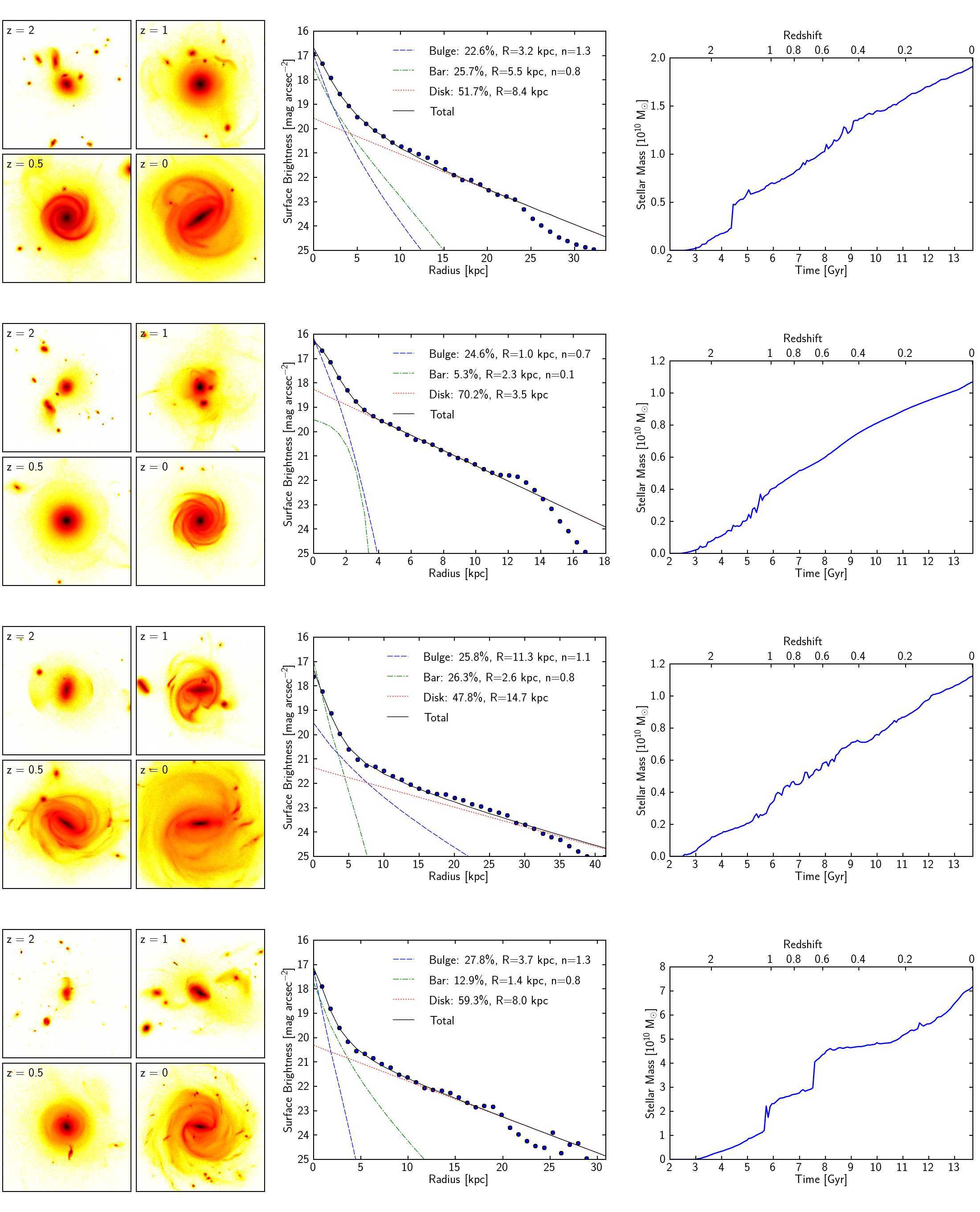}
\caption{Galaxies with a bulge fraction from 0.22 to 0.30}
\label{fig:AppA4}
\end{figure*}

\begin{figure*}
\centering 
\includegraphics[width=0.9\textwidth]{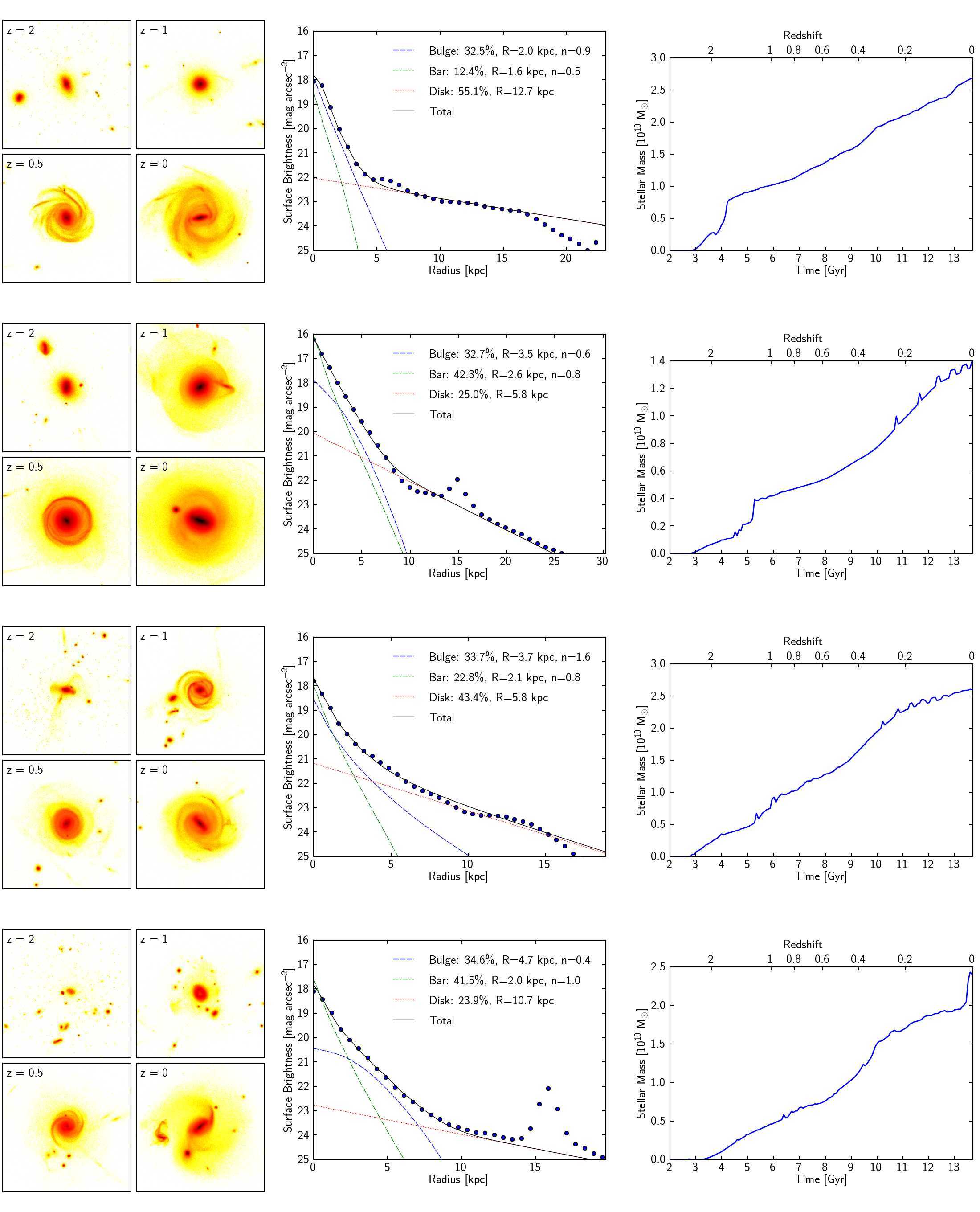}
\caption{Galaxies with a bulge fraction from 0.30 to 0.35}
\label{fig:AppA5}
\end{figure*}

\begin{figure*}
\centering 
\includegraphics[width=0.9\textwidth]{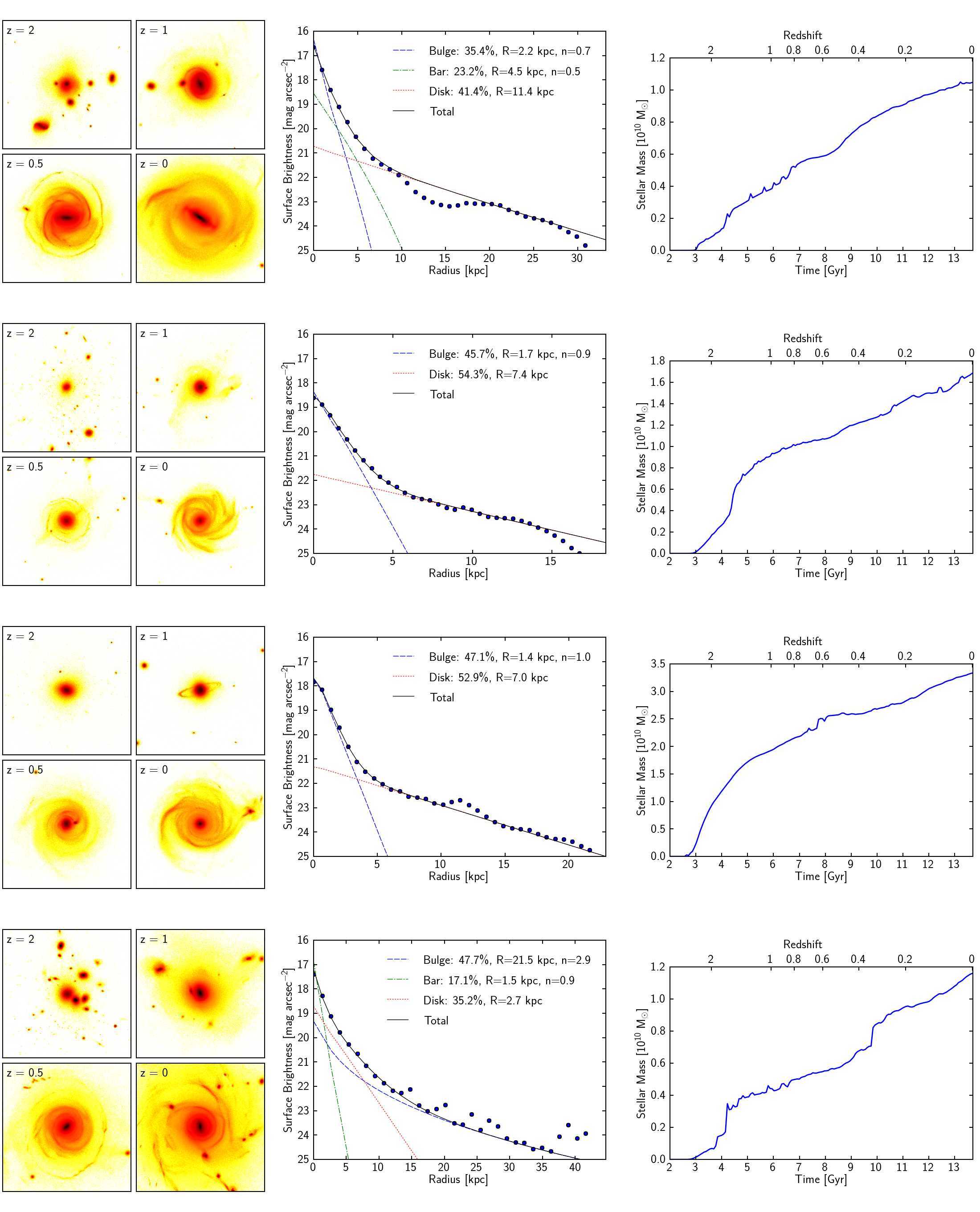}
\caption{Galaxies with a bulge fraction from 0.35 to 0.50} 
\label{fig:AppA6}
\end{figure*}

\begin{figure*}
\centering 
\includegraphics[width=0.9\textwidth]{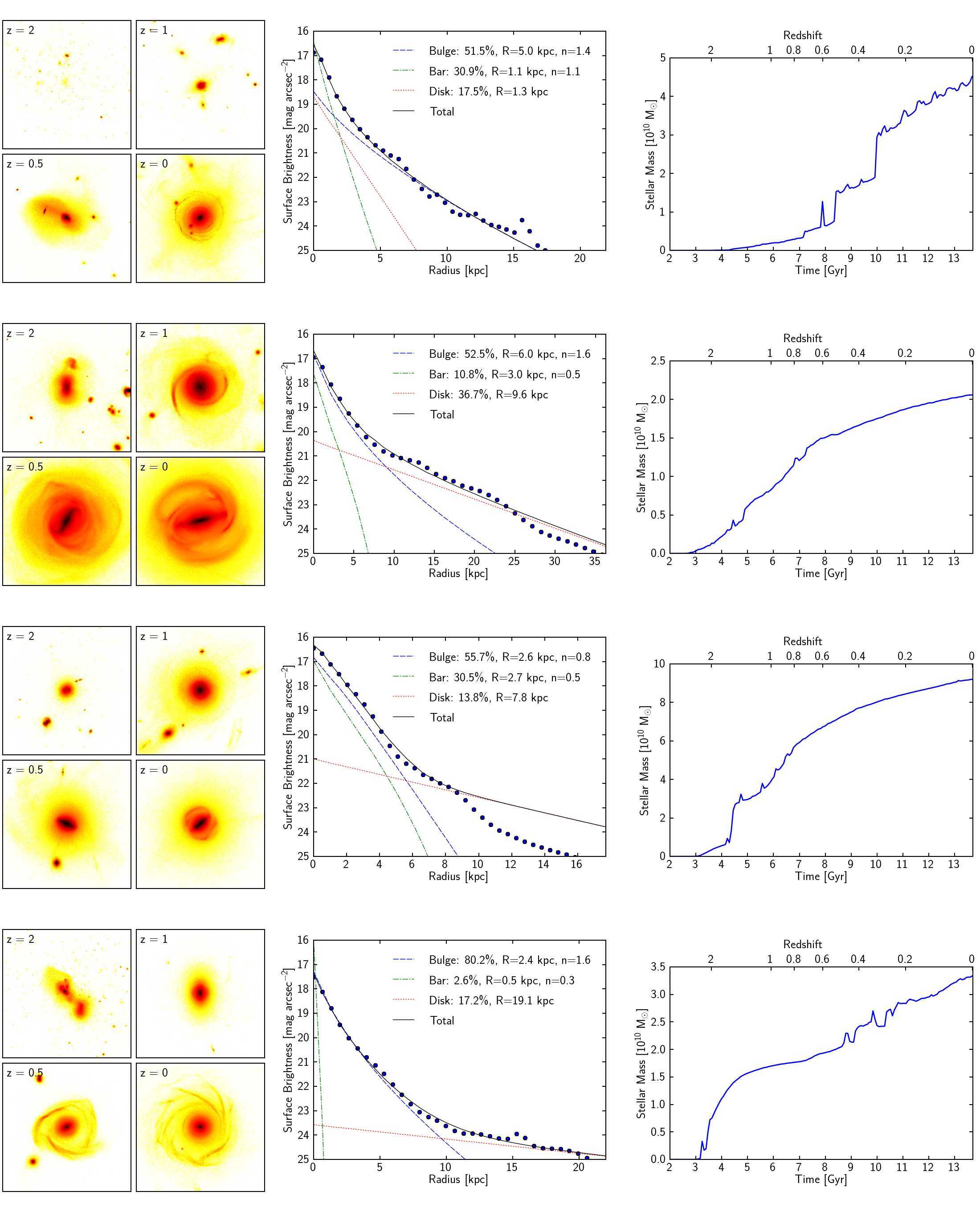}
\caption{Galaxies  with a bulge fraction from 0.50 to 0.80} 
\label{fig:AppA7}
\end{figure*}

\begin{figure*}
\centering 
\includegraphics[width=0.9\textwidth]{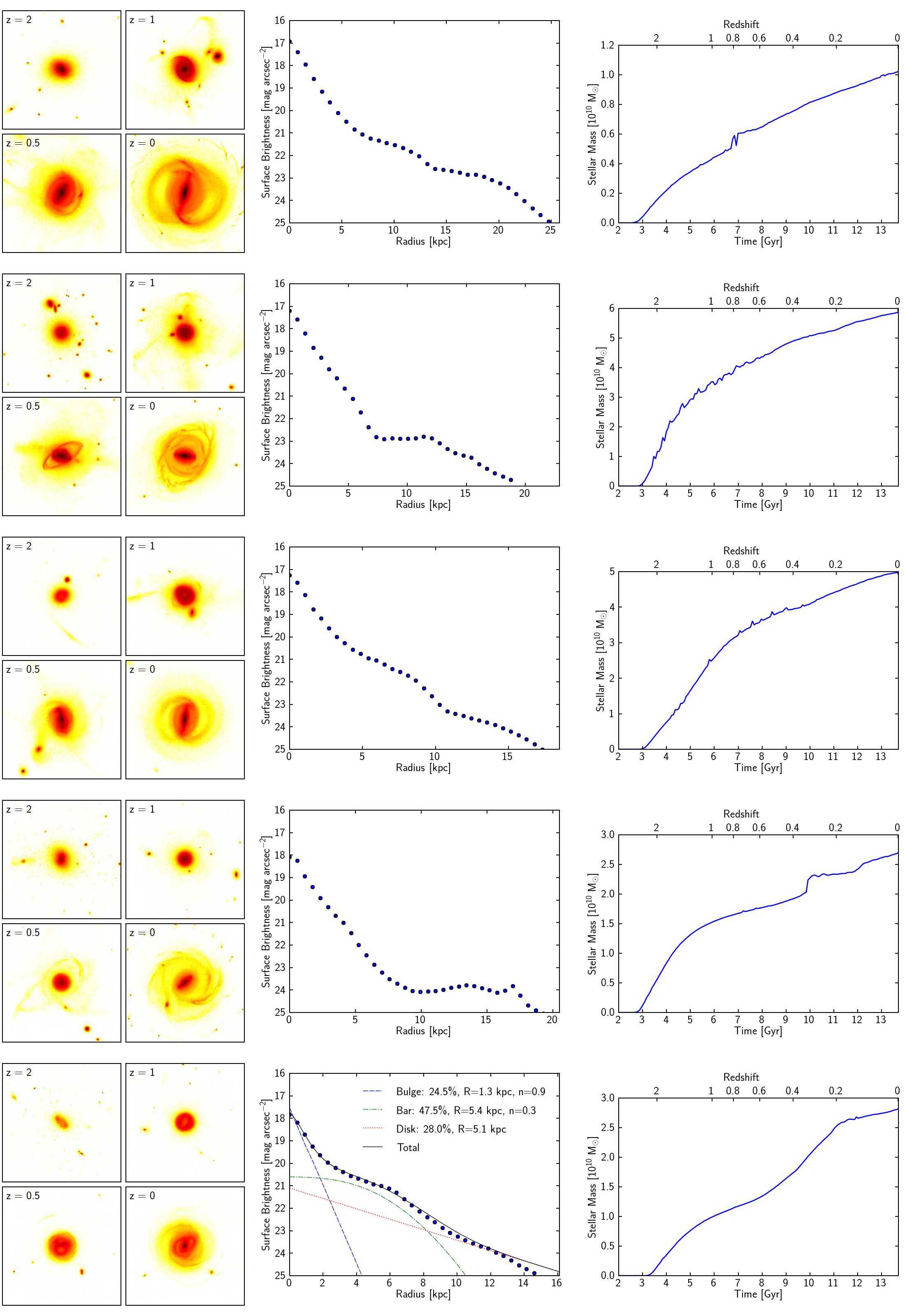}
\caption{Galaxies for which no GALFIT decomposition has been achieved}
\label{fig:AppA8}
\end{figure*}

In Figures \ref{fig:AppA1} to \ref{fig:AppA8} we present for each simulated galaxy:
\begin{itemize}
\item  i-band face-on images (70x70kpc, color scale from 16 to 27~mag~arcsec$^{-2}$) at $z=2$, 1, 0.5 and 0
\item  i-band surface brightness profiles at $z=0$ and the corresponding GALFIT bulge+bar+disk decomposition
\item  stellar mass evolution with time  (stellar mass within the optical radius)
\end{itemize}

\bibliographystyle{apj}

\end{document}